\documentclass{jfm-per}
\pdfoutput=1
\usepackage{graphicx,amsmath}
\usepackage{epstopdf, epsfig}
\renewcommand{\v}{{\boldsymbol{v}}}
\newcommand{\w}{{\boldsymbol{w}}}
\newcommand{\q}{{\boldsymbol{q}}}
\newcommand{\f}{{\boldsymbol{s}}}
\newcommand{\0}{{\boldsymbol{0}}}
\newcommand{\h}{{\boldsymbol{h}}}
\newcommand{\D}{{\boldsymbol{D}}}
\newcommand{\nuhat}{{\hat{\boldsymbol{\nu}}}}

\renewcommand{\u}{{\boldsymbol{u}}}
\newcommand{\x}{{\boldsymbol{x}}}
\newcommand\bdot{{\boldsymbol{\cdot}}}
\newcommand\del{{\boldsymbol{\nabla}}}
\newcommand\delsq {\nabla^2}
\newcommand\bcol{{\boldsymbol{:}}}
\renewcommand{\r}{{\boldsymbol{r}}}

\newcommand{\nhat}{{\hat{\boldsymbol{n}}}}
\newcommand{\J}{{\boldsymbol{J}}}
\renewcommand{\d}{{\rm d}}
\renewcommand{\p}{{\boldsymbol{p}}}
\newcommand{\bsigma}{{\boldsymbol{\sigma}}}
\newcommand{\bepsilon}{{\boldsymbol{\epsilon}}}
\renewcommand{\S}{{\mathcal S}}
\newcommand{\Q}{{\mathsfbi{Q}}}
\renewcommand{\D}{{\mathsfbi{D}}}
\newcommand{\bOmega}{{\boldsymbol{\Omega}}}
\newcommand{\I}{{\mathsfbi{I}}}
\DeclareMathOperator{\Tr}{Tr}

\shorttitle{Binary Fluid Mixtures}
\shortauthor{M. E. Cates and E. Tjhung}

\title{Theories of Binary Fluid Mixtures: \\ From Phase-Separation Kinetics to 
Active Emulsions}

\author{Michael E. Cates\aff{1}
  \corresp{\email{m.e.cates@damtp.cam.ac.uk}} and Elsen Tjhung\aff{1}\corresp{\email{et405@cam.ac.uk}}}

\affiliation{\aff{1}DAMTP, University of Cambridge, Centre for Mathematical Sciences,
Wilberforce Road, Cambridge CB3 0WA, UK}

\begin{document}

\maketitle

\begin{abstract}
Binary fluid mixtures are examples of complex fluids whose microstructure and flow are strongly coupled. For pairs of simple fluids, the microstructure consists of droplets or bicontinuous demixed domains and the physics is controlled by the interfaces between these domains. At continuum level, the structure is defined by a composition field whose gradients -- which are steep near interfaces -- drive its diffusive current. These gradients also cause thermodynamic stresses which can drive fluid flow. Fluid flow in turn advects the composition field, while thermal noise creates additional random fluxes that allow the system to explore its configuration space and move towards the Boltzmann distribution. This article introduces continuum models of binary fluids, first covering some well-studied areas such as the thermodynamics and kinetics of phase separation, and emulsion stability. We then address cases where one of the fluid components has anisotropic structure at mesoscopic scales creating nematic (or polar) liquid-crystalline order; this can be described through an additional tensor (or vector) order parameter field. We conclude by outlining a thriving area of current research, namely active emulsions, in which one of the binary components consists of living or synthetic material that is continuously converting chemical energy into mechanical work. Such activity can be modelled with judicious additional terms in the equations of motion for simple or liquid-crystalline binary fluids. Throughout, the emphasis of the article is on presenting the theoretical tools needed to address a wide range of physical phenomena. Examples include the kinetics of fluid-fluid demixing from an initially uniform state; the result of imposing a steady macroscopic shear flow on this demixing process; and the diffusive coarsening, Brownian motion and coalescence of emulsion droplets. We discuss strategies to create long-lived emulsions by adding trapped species, solid particles, or surfactants; to address the latter we outline the theory of bending energy for interfacial films.  In emulsions where one of the components is liquid crystalline, `anchoring' terms can create preferential orientation tangential or normal to the fluid-fluid interface. These allow droplets of an isotropic fluid in a liquid crystal (or {\em vice versa}) to support a variety of topological defects, which we describe, altering their interactions and stability. Addition of active terms to the equations of motion for binary simple fluids creates a model of `motility-induced' phase separation, where demixing stems from self-propulsion of particles rather than their interaction forces, altering the relation between interfacial structure and fluid stress. Coupling activity to binary liquid crystal dynamics creates models of active liquid-crystalline emulsion droplets. Such droplets show various modes of locomotion, some of which strikingly resemble the swimming or crawling motions of biological cells.
\end{abstract}

\section{Introduction}\label{Intro}
A binary fluid mixture contains two types of molecule, A and B (oil and water, for instance). In many cases these molecules have an energetic preference to be surrounded by others of the same type. At high temperatures, this is overcome by entropy and the fluid remains well mixed at a molecular scale. At low temperatures, it undergoes phase separation into A-rich and B-rich domains.  (In a few systems, particularly those with hydrogen bonding, this temperature dependence is inverted so the system demixes upon raising temperature instead.) A sudden change in temperature, known as a `quench',  initiates phase separation \citep{Chaikin95}. 

Usually the resulting fluid domains grow indefinitely in time so that phase separation goes to completion \citep{Bray94,Onuki02}. But in many cases one wants to avoid or arrest this process. One strategy is to steadily stir the fluid, in the hope of remixing the domains so that they can never become large.  Another is to introduce additional (molecular or colloidal) species that inhibit the growth of domains. The resulting finely divided mixtures, called emulsions, are generally not thermodynamically stable, but can be long-lived. They have many applications ranging from foods via agrochemicals and pharmaceuticals, to display device materials \citep{Bibette02}. Partly because of these applications, there is increasing interest in emulsions where at least one of the components is not a simple fluid but has its own microstructure: for instance, a liquid crystal in which rod-like molecules align along a common axis. In the confined geometry of an emulsion droplet, liquid crystals can show complex behaviour caused by an interplay between boundary conditions and bulk energy minimization \citep{Poulin99}.

Another growth area for binary fluids research concerns cases where one (or both) of the two fluids escapes the laws of conventional thermodynamics by continually consuming fuel. This allows continuous fluxes of energy, momentum and particles through the system, whereas in thermal equilibrium steady-state fluxes are prohibited by the time-reversal symmetry of the microscopic laws of motion \citep{Marchetti13}. In these so-called `active fluids' thermal equilibrium is only reached when the fuel has run out; prior to that, the system can show steady-state behaviour without microscopic time reversal symmetry, leading to new effects. Many models of active fluids also address liquid crystallinity of the active component; such models were first developed to describe a system of active fibres such as those present in the cytoskeleton of eukaryotic cells. The cytoskeleton, which is responsible for shape changes and locomotion of these cells, contains locally aligned rod-like structures that are tugged lengthwise towards one another by active molecular motors; the rods can also actively move along their own length by adding protein subunits at one end and dropping them from the other \citep{Marchetti13}. (Both of these processes are fuelled by adenosine triphosphate, ATP.)  An interesting question in then whether the emergent dynamics of such cells can be understood in terms of simple physical models: can a cell be viewed as an active liquid crystal emulsion droplet? 

This Perspectives article explains some of the theoretical tools and approaches that can be used to investigate quantitatively all the above issues. The focus is on theoretical concepts rather than quantitative prediction, particularly in the later sections. Although in many cases the ideas presented have been amply confirmed by experiments, the corroborating evidence will not be much discussed. In addition, we will often consider simplified or asymptotic regimes for which the main testing ground of theoretical ideas is provided by computer simulations, in which an appropriate numerical methodology (such as the Lattice Boltzmann method \citep{Kendon01, Cates09}) is used to solve the equations of simplified models of the type presented below. Such simulations are vital in checking our theoretical beliefs about how such a model should behave. They can also tell us whether the resulting behaviour is close to that seen experimentally. If it is, we have evidence that the simplified model captures the dominant mechanisms in the experimental system, allowing us to better identify what those mechanisms are. 

\section{Order parameters for complex fluids}\label{OPs}
We first consider an isothermal, incompressible, simple fluid with Newtonian viscosity $\eta$ and density $\rho$. This obeys the Navier Stokes equation (NSE)
\begin{equation}
\rho(\dot \v + \v\bdot\del\v) = \eta\delsq \v - \del P , \label{NSE}
\end{equation}
where the pressure field $P$ must be chosen to enforce the incompressibility condition 
\begin{equation}
\del\bdot\v = 0 .\label{continuity}
\end{equation}
One generic approach to complex fluids is to consider a simple fluid obeying the NSE, coupled to a set of coarse-grained internal variables $\psi(\r,t)$, each a function of position and time. These variables are generally called order parameter fields or simply `order parameters'. Obviously they are not parameters of the model but its dynamical variables, and indeed $\v(\r,t)$ is itself an order parameter since it describes a local average of random molecular velocities. Other order parameters encountered below include the following:

(i) A scalar field $\phi$ that describes the local molecular composition of a binary fluid mixture. We define it at each instant as
\begin{equation}
\phi(\r) = \frac{\langle n_A-n_B \rangle_{\rm meso}}{\langle n_A+n_B \rangle_{\rm meso}}.
\label{phidef}
\end{equation}
Here $n_{A,B}$ denotes the number of A,B molecules per unit volume locally; the mesoscopic average $\langle\cdot\rangle_{\rm meso}$ is taken over a large enough (but still small) local volume so that $\phi(\r)$ is smooth. For notational simplicity we have assumed that A and B molecules have the same (constant) molecular volume. Given the incompressibility condition (\ref{continuity}), the denominator in (\ref{phidef}) is a constant, and our composition variable obeys $-1\le\phi\le 1$ with $\phi=1$ in a fluid of pure A. 

(ii) A vector field $\p$, describing the mean orientation of rodlike molecules:
\begin{equation}
\p(\r) = \langle \nuhat\rangle_{\rm meso},\label{pdef}
\end{equation}
with  $\nuhat$ a unit vector along the axis of a single molecule. A material of nonzero $\p$ is called a polar liquid crystal. This order parameter makes sense only for molecules that have one end different from the other. Even in that case $\p$ vanishes when molecules are oriented but not aligned, in the sense that they point preferentially along some axis but are equally likely to point up that axis as down it. 

(iii) To describe cases with orientation but not alignment (in the sense just defined), we need a second rank tensor
 \begin{equation}
\Q(\r) = \langle \nuhat\nuhat\rangle_{\rm meso}-\I/d,
\label{Qdef}
\end{equation}
where $\nuhat\nuhat$ is a dyadic product (and independent of which way the unit vector points along the molecule); $\I$ is the unit tensor, and $d$ is the dimension of space. The resulting tensor is traceless by construction and therefore vanishes if the rods are isotropically distributed. A fluid in which $\Q$ is finite but $\p$ is zero is called a nematic liquid crystal.

In general the density and viscosity in (\ref{NSE}) should depend directly on our chosen set of order parameters $\psi(\r,t)$. For example, in a binary fluid A and B molecules may have the same volume but different masses, and pure A and pure B fluids might have different viscosities.  However a big simplification, which does not affect much the conceptual physics discussed below, is to assume these dependences are negligible. The remaining effect of the structural order parameters $\psi(\r,t)$ is then to create an additional thermodynamic stress $\bsigma$ that enters the NSE as
\begin{equation}
\rho(\dot \v + \v\bdot\del\v) = \eta\delsq \v - \del P 
+\del\bdot\bsigma[\psi].
\label{NSE2}
\end{equation}
(This is more properly now called a Cauchy equation; but we call it the NSE in this article.) The stress term, which is a functional of the order parameters  can alternatively be viewed as a force density ${\f} = \del\bdot\bsigma$ exerted by the order parameter fields on the fluid continuum. Note that any isotropic contribution to $\bsigma$ can be absorbed into $P$. 

To fully specify the dynamics of our system, we need two further things. The first is a set of equations of motion for the order parameters themselves. In general these must allow for their advection by the fluid flow $\v$; this enters alongside whatever physics would describe the system at rest. More precisely, for a composition variable $\phi$, with $\v=0$ one has $\dot\phi = \del\bdot\J$ where $\J$ is a diffusive current;  this form reflects the fact that $\phi$  is a conserved quantity that cannot be created or destroyed locally. In contrast, $\p$ and $\Q$ are not conserved and can relax directly towards their thermodynamic equilibrium state. In all three cases, the equations of motion involve derivatives of a functional $F[\psi]$ which gives the (Helmholtz) free energy in terms of the  order parameter fields. The second thing we need is a recipe for calculating the stress $\bsigma[\psi]$ from the (instantaneous) order parameter configuration $\psi(\r)$. This calculation is nontrivial, particularly for liquid crystals \citep{Beris94}, but is unambiguous so long as the system is not too far from thermodynamic equilibrium locally. In active systems, this does not apply and additional terms arise in both the equations of motion and the stress, whose form is less rigorously known but can be selected empirically. In what follows we consider both the order parameter evolution equations and the stress expression on a case by case basis.

\section{The symmetric binary fluid}
In the simplest model of a binary fluid, the AA and BB interactions are the same but there is an additional repulsive energy, say $E_{AB}$, between adjacent molecules of A and B. Combined with our previous assumptions, the system is now completely symmetric at a molecular level.  At high temperatures, $T>T_C\simeq E_{AB}/k_B$ (with $k_B$ Boltzmann's constant) the repulsive interactions are overcome by mixing entropy and the two fluids remain completely miscible. At lower $T$ however, the A-B repulsion causes demixing into two co-existing phases, one rich in A, one rich in B. Entropy ensures that there is always a small amount of the other type of molecule present in each phase; close to the critical temperature $T_C$ the two phases differ only slightly in $\phi$, merging at $\phi  = \phi_C = 0$. 

\begin{figure}
	\centering
	\includegraphics[width=5cm]{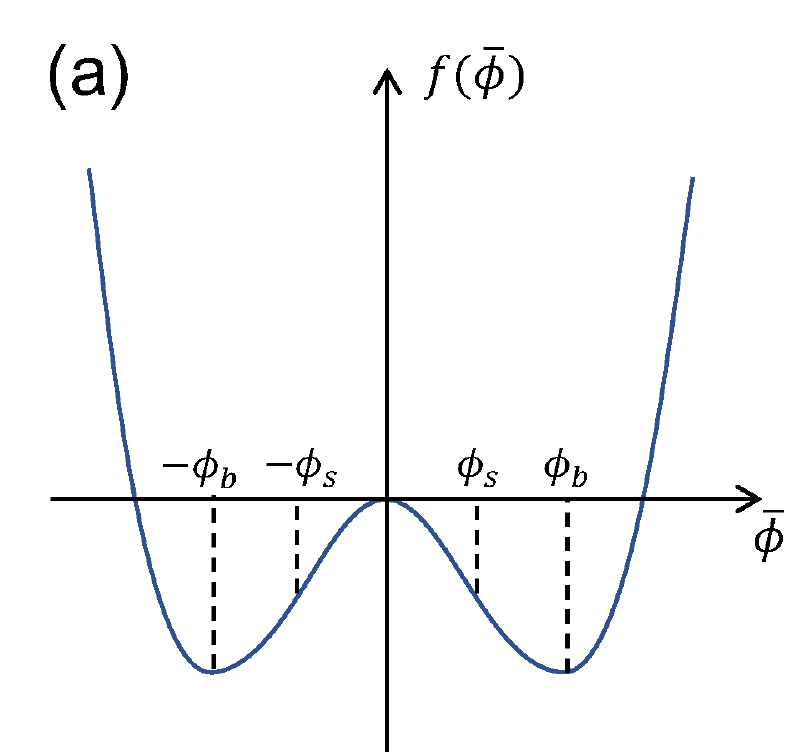}
	\includegraphics[width=5.7cm]{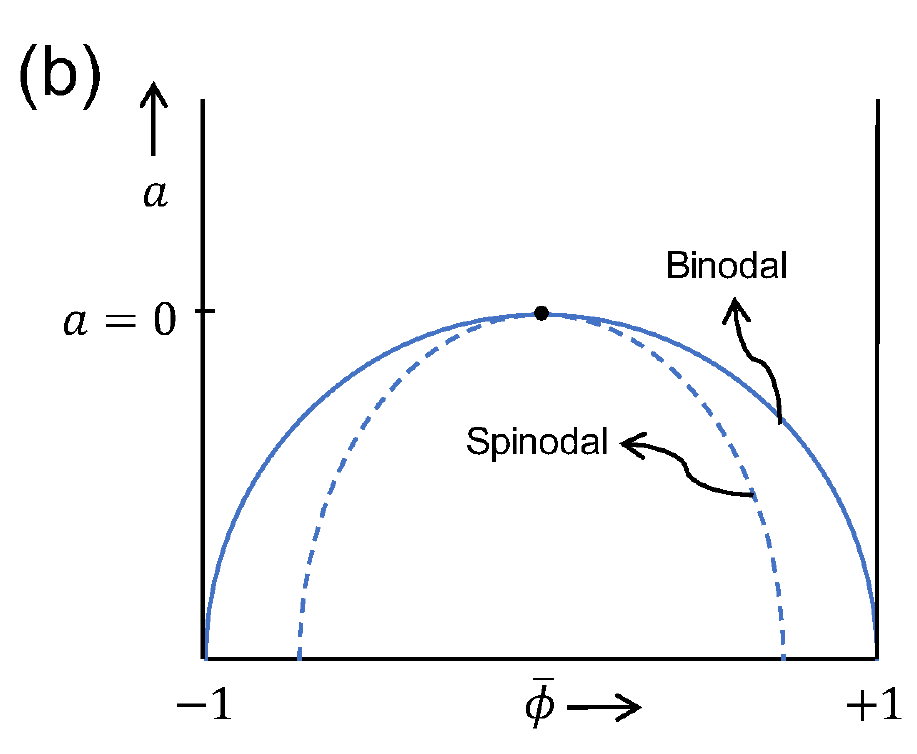}
	\caption[]{Mean-field free energy density (a) and phase diagram (b) of a symmetric binary fluid mixture. In (b), the dot at the top of the binodal curve is the critical point, where two coexisting phases become identical.}
	\label{one}
\end{figure}

A schematic phase diagram for the symmetric binary fluid  is shown in figure \ref{one}.
The locus of coexisting compositions $\phi = \pm \phi_b(T)$ is called the binodal curve; for global compositions $\bar\phi = \int \phi(\r)\d\r$ within the binodal, the equilibrium state comprises two phases of composition $\pm\phi_b$. The volumes occupied by the A-rich and B-rich phases, $V_{A,B}$, obey $V_A+V_B = V$ where $V$ is the overall volume of the system and
\begin{equation}
(V_A-V_B)\phi_b = \bar\phi.
\label{phasevol}
\end{equation}
Thus the `phase volume' of the A-rich phase, $\Phi_A\equiv V_A/V$, evolves from zero to one as the overall composition $\bar\phi$ is swept across the miscibility gap from $-\phi_b$ to $\phi_b$. The dotted line on the phase diagram is the spinodal, $\bar\phi = \pm \phi_s(T)$, within which the globally uniform state $\phi(\r) = \bar\phi$ is locally unstable. Between the spinodal and the binodal ($\phi_s\le|\bar\phi|\le\phi_b$) the uniform state is metastable; to get started, phase separation requires nucleation of a large enough droplet. This is a random rare event, driven by thermal noise.

\subsection{Free energy functional and mean-field theory}
The simplest next step is to postulate the following free energy functional
\begin{equation}
F[\phi] = \int \left(\frac{a}{2}\phi^2+\frac{b}{4}\phi^4 +\frac{\kappa}{2}(\del\phi)^2\right)\d\r,
\label{functional}
\end{equation}
where $a = a(T)$ while $b$ and $\kappa$ are positive and (for simplicity) independent of temperature.
The bulk free energy density of a uniform state, $f(\phi) = \frac{a}{2}\phi^2+\frac{b}{4}\phi^4$, is an approximation, inspired by a Taylor expansion in small $\phi$ for weakly demixed states; for a symmetric fluid this contains only even powers. (Note that at given $\bar\phi$ any linear term merely adds a constant to $F$. Were asymmetric interactions present, any cubic term could also be eliminated by an additive shift, $\phi \to \phi-\phi_C$.)
This expansion breaks down at large negative $a$ where it overshoots the saturating asymptote $\phi_b-1\sim \exp[-u/k_BT]$, with $u$ a solubilization energy, set by ideal (dilute) solution thermodynamics. A more accurate form is $f(\phi)= -u\phi^2/2 -k_BT[ \phi\ln\phi+(1-\phi)\ln(1-\phi)]$, but the quartic approximation is sufficient for our purposes, and much easier to use in calculations of both the phase diagram and the interfacial tension between phases. 

These calculations can be tackled without further approximation by addressing (\ref{functional}) using field theoretic methods, including the renormalization group theory which is essential to understanding the behaviour very close to the critical point at $T_C$ \citep{Chaikin95}. Such methods take averages over the Boltzmann weight $\exp[-{F}/k_BT]$ of our fluctuating order parameter(s); we do not pursue them here. A simpler alternative is mean-field theory which considers only the most probable states  found
by minimizing $F$ at fixed global composition $\bar\phi$. 

We first consider states of uniform $\phi(\r) = \bar\phi$. For such states
\begin{equation}
\frac{F}{V} = \frac{a}{2}\bar\phi^2+\frac{b}{4}\bar\phi^4 = f(\bar\phi). \label{uniform}
\end{equation} 
For $a>0$ this has a single minimum at $\bar\phi = 0$, with positive curvature everywhere.  The latter means that whatever $\bar\phi$ is chosen, one cannot lower the free energy by introducing a phase separation. On the other hand, for $a<0$, $f$ has negative curvature between the spinodals $\pm\phi_s$ where $\phi_s = (-a/3b)^{1/2}$ (see figure \ref{one}). Also it has two symmetric minima at $\bar\phi = \pm\phi_b$ with $\phi_b = (-a/b)^{1/2}$. For $|\bar\phi|<\phi_b$, $F$ is minimized by demixing the uniform state at $\bar\phi$ into two coexisting states at $\phi = \pm\phi_b$. A price must be paid to create an interface between these, but the interfacial area scales as $V^{1-1/d} \ll V$ so that for a large enough system, this price is always worth paying. 

Although so far restricted to {\em symmetric} fluid pairs, this calculation is more general than it first appears. For an asymmetric fluid, one expects (to quartic order) additional linear and cubic terms in $f(\phi)$. However, any linear term in $F$ is of the form $\int \phi \d\r =\bar\phi V$ which is simply a constant set by the global composition. Such an additive constant to $F$ has no physical effects. In contrast,  a cubic term $\int (c\phi^3/3) \d\r$ creates an asymmetric phase diagram, which is useful in fitting the model to real fluid pairs for which some asymmetry is always present. However, it is a simple exercise then to show this cubic contribution can be absorbed by shifts $a\to a-a_C$ and $\phi\to\phi-\phi_C$ with $a_C = c^2/3b$ and $\phi_C = -c/3b$. In other words, at our chosen level of treating $f(\phi)$ as a quartic polynomial, the cubic term merely shifts the mean-field critical point to a new position on the phase diagram; measuring $\phi$ and $a$ relative to this new position, nothing has changed.

\subsection{Interfacial profile and tension}\label{profiletension}
In equilibrium, our two bulk phases will minimize their mutual surface area; in most geometries, this requires the interface to be flat. To calculate its interfacial tension, we take the surface normal along the $x$ direction so that $\phi(\r) = \phi(x)$. The boundary conditions are that $\phi(x)$ approaches $\pm\phi_b$ at $x=\pm\infty$. To find the profile, we minimize $F[\phi]-\lambda\int\phi \d \r$ with these boundary conditions. (Here $\lambda$ is a Lagrange multiplier that holds  the global composition fixed during the minimization.) 
The resulting condition,
\begin{equation}
\frac{\delta}{\delta\phi}\left[F-\lambda\int \phi \d\r\right] = 0, \label{minim}
\end{equation}
involves the functional derivative of $F$ which we denote by
\begin{equation}
\mu(x) \equiv \frac{\delta F}{\delta\phi} =  a\phi + b\phi^3 -\kappa\delsq \phi .\label{chempotdef}
\end{equation}
This is called the chemical potential (or more properly the exchange chemical potential) because, up to a factor of molecular volume, it gives the free energy change on replacing an A molecule with a B molecule locally such that if $\phi$ is incremented by $\delta\phi(\r)$, then the free energy change is $\delta F = \int \mu\, \delta \phi \,\d\r$.

Equation (\ref{minim}) requires that $\mu =\lambda$ which is constant in space. 
For our symmetric choice of $F[\phi]$ we have $\mu = df/d\phi = 0$ in the two bulk phases at density $\pm\phi_b$, so it follows that $\lambda = 0$.
It is then a good exercise \citep{Chaikin95} to show that, with the boundary conditions already given, the solution for $\phi(x)$ of the ODE $\mu(x)=0$ is
\begin{equation}
\phi(x) = \pm \phi_0(x) \equiv \pm\phi_b\tanh\left(\frac{x-x_0}{\xi_0}\right). \label{profile}
\end{equation}
Here $\xi_0 = (-\kappa/2a)^{1/2}$ is an interfacial width parameter, $x_0$ marks the midpoint of the interface, and the overall sign choice depends on whether the A-rich or B-rich phase occupies the region at large positive $x$. 
The interfacial profile is fixed by a trade-off between the penalty for sharp gradients (set by $\kappa$) and the purely local free energy terms which, on their own, would be minimized by a profile $\phi(x)$ that jumps discontinuously from one binodal value to the other. A further exercise is to show that the equilibrium interfacial tension $\gamma_0$, defined as the excess free energy per unit area of a flat interface, obeys \citep{Chaikin95}
 \begin{equation}
 \gamma_0 = \int \kappa(\partial_x\phi_0(x))^2 dx =  \left(\frac{-8\kappa a^3}{9b^2}\right)^{1/2}.
 \end{equation} 

\subsection{Stress tensor \label{sec:stress}}
If the interfacial profile departs from the equilibrium one, a thermodynamic stress $\bsigma$ will act on the fluid. An important example is when the interface is not flat but curved; under these conditions $\mu$ cannot be zero everywhere.
For use in the NSE we require not the stress tensor directly but the thermodynamic force density ${\f} \equiv \del\bdot\bsigma$. Consider now a small incompressible displacement field $\u$: that is $\r\to\r+\u(\r)$ with $\del\bdot\u = 0$. Advection of the $\phi$ field by this displacement induces the change $\phi(\r) \to \phi(\r-\u)$.  To linear order this gives the increment $\delta\phi  = -\u\bdot\del\phi$ from which we find the free energy increment as 
\begin{equation}
\delta F = \int\frac{\delta F}{\delta \phi}\delta\phi\,\d\r=-\int \mu\u\bdot\del\phi\, \d\r = \int (\phi\del\mu)\bdot\u\,\d\r,\label{stress1}
\end{equation}
where the final form follows by partial integration and incompressibility. (We consider periodic boundary conditions without loss of generality; this eliminates the boundary term.) This result can be compared with the free energy increment caused by a strain tensor $\bepsilon = \del\u$
\begin{equation}
\delta F = \int \bsigma\bcol(\del\u)\,\d\r = - \int \f\bdot\u\,\d\r ,\label{stress2}
\end{equation}
where the second form again follows by partial integration. Comparison of (\ref{stress1},\ref{stress2}) shows that $ \del\bdot\bsigma\equiv\f = -\phi\del\mu$. 
This form, which is all we need to know about the stress tensor for the purposes of solving the NSE, could also have been derived from the following expression for the stress tensor itself:
\begin{equation}
\sigma_{ij} = -\Pi\delta_{ij} -\kappa(\partial_i\phi)(\partial_j\phi).\label{stress3}
\end{equation}
Here $\Pi = \phi\mu - {\mathbb F}$ is the order parameter contribution to the (local) pressure, better known as the osmotic pressure; we have defined ${\mathbb F} = f(\phi)+\kappa(\del\phi)^2/2$ which is the local free energy density, {\em i.e.,} $\int {\mathbb{F}}\, d\r = F$. The non-gradient part of $\Pi$ is in turn $\Pi_{\rm bulk} =\phi \mu_{\rm bulk}-f$ with $\mu_{\rm bulk} \equiv df/d\phi$; this is the usual thermodynamic relation between pressure and free energy density \citep{Chaikin95}. In some cases it is possible to think of the flow resulting from the force density $\f$ as loosely arising from ``a gradient of osmotic pressure". This can be useful for estimating the size of terms in the NSE, but cannot be interpreted too literally, since a term $-\del P$ already appears there to enforce incompressibility. The irrotational term in the order parameter force density $\f = -\del \Pi$ can be absorbed into this and so has, strictly speaking, no role in the dynamics.

\section{Model H and Model B}\label{ModH}
Above we have determined the force density $\f = -\phi\del\mu$ that appears in the NSE for a binary fluid mixture, as a result of spatial variations in composition $\phi(\r)$. Next we need an equation of motion for $\phi$ itself. This takes the form
\begin{equation}
\dot\phi +\v\bdot\del\phi = - \del\bdot{\bf J}, \label{current1}
\end{equation}
where the left-hand side is the co-moving derivative of $\phi$. This derivative must be the divergence of a current, because A and B particles are not created or destroyed and thus $\phi$ is a conserved field. The form for the compositional current is
\begin{equation}
{\bf J}  = -M\del\mu ,\label{current2}
\end{equation}
where $M$ could in principle depend locally (or indeed nonlocally) on composition, but is here chosen constant for simplicity. 
The collective mobility $M$ describes, under conditions of fixed total particle density, how fast A and B molecules can move down their (equal and opposite) chemical potential gradients to relax the composition field. The linear relation between flux and chemical potential gradient assumes that gradient is small enough for the system to remain locally close to thermal equilibrium everywhere.

Combining (\ref{current1},\ref{current2}) with our earlier results for the chemical potential and the NSE, we arrive at a closed set of equations for an isothermal, binary fluid mixture: 
\begin{eqnarray}
\rho(\dot\v+\v\bdot\del\v) &=& \eta\delsq \v-\del P - \phi\del\mu +\del\bdot\bsigma^n,\label{HNSE}\\
\del\bdot\v &=& 0, \label{incomp}\\
\dot\phi + \v\bdot\del\phi &=& -\del\bdot(-M\del\mu+\J^n),\label{Hphi}\\
\mu(\r) &=& a\phi+ b\phi^3 -\kappa\delsq \phi.\label{mumodelH}
\end{eqnarray}
Here the quantities $\bsigma^n$ and $\J^n$ represent noise terms, discussed below.
As derived so far, however, these equations are at the level of a deterministic hydrodynamic description for which both such terms are zero. 

The noise terms are important in several situations. One is near the critical point (not addressed in this article) where thermal fluctuations play a dominant role in the statistics of $\phi$: the mean-field theory implicit in the noise-free treatment then breaks down. Another case where noise matters is when there are well-separated fluid droplets suspended in a continuous fluid phase. These objects will move by Brownian motion, but without noise terms (specifically $\bsigma^n$) droplets of one fluid in another cannot diffuse and, in a quiescent system, would never meet and coalesce. The lifetime of such a droplet emulsion is thus noise-controlled. Also, if a uniform system is prepared with $\phi_s<\bar\phi<\phi_b$ (between the spinodal and the binodal on the phase diagram), a noise-free treatment would predict this to remain uniform indefinitely, whereas in practice the system phase separates after an accumulation of noise (entering via $\J^n$) has driven it across a nucleation barrier involving formation of a droplet of the phase on the opposite binodal.

The noise terms can be determined using the fluctuation dissipation theorem which stems from the requirement for Boltzmann equilibrium in steady state \citep{Chaikin95}. The order parameter current ${J}_i^n$ (Cartesian component $i$; superscript $n$ for noise) is a zero-mean Gaussian variable of the following statistics:
\begin{equation}
\langle J^n_i(\r,t)J^n_j(\r',t')\rangle = 2k_BTM\delta_{ij}\delta(\r-\r')\delta(t-t'),\label{HJnoise}
\end{equation}
where $\langle \cdot \rangle$ denotes an average over the unresolved microscopic dynamics responsible for the noise.
Meanwhile the random stress in the NSE is likewise a zero-mean Gaussian thermal stress whose statistics obey \citep{Landau59}
\begin{equation}
\langle \sigma^n_{ij}(\r,t)\sigma^n_{kl}(\r',t')\rangle = 2k_BT\eta[\delta_{ik}\delta_{jl}+\delta_{il}\delta_{jk}]\delta(\r-\r')\delta(t-t').\label{Hvnoise}
\end{equation}
Note that due to incompressibility the isotropic (pressure-like) part of this noise stress is optional, and often explicitly removed in the literature. The resulting noisy NSE is fundamental to the fluid mechanics of thermal systems, although in some cases  -- notably suspensions of Brownian spheres -- it can be impersonated by adding a set of correlated noise forces direct to the equations of motion of individual suspended objects \citep{Brady88}. That approach does not generalize to fluid domains of variable shape and is therefore not helpful in the binary fluid context.

With these noise terms duly added, equations (\ref{HNSE}--\ref{mumodelH}) are lifted from a {\em mean-field} dynamics of the  simplest binary fluid, whose free energy functional is the chosen $F[\phi]$ and whose mobility and viscosity are $\phi$-independent, to a {\em complete} dynamical description of the same simplified model. Accordingly, in a quiescent system with no external forcing, the model ultimately achieves in steady state the Boltzmann distribution ${\mathcal P}[\phi,\v]\propto\exp\left(-\beta (F[\phi]+K[\v])\right)$, with $\beta\equiv(k_BT)^{-1}$ and where $K= \int ({v^2}{/2\rho})\,\d\r$ is the kinetic energy of the fluid, which is subject to the incompressibility constraint $\del\bdot\v = 0$.

The above dynamical equations are generally known as `Model H' \citep{Chaikin95}. An important limiting case is when there is no fluid flow; this can be viewed as the limit of infinite viscosity $\eta$. In this limit one has simply
\begin{eqnarray}
\dot\phi &=& -\del.\J = -\del. \left(-M\del\mu +\J^n\right), \label{ModelB}\\
\mu&=& a\phi + b\phi^3 -\kappa\delsq \phi, \label{ModelBmu}
\end{eqnarray}
which is called `Model B'. This model can be used to describe systems where molecular diffusion is much more efficient than viscous flow at relaxing structure (true for some binary mixtures of polymeric fluids) and also applies to system where there is no appreciable flow because of a momentum sink, such as a solid wall in contact with a two dimensional film of the binary mixture. Model B is also the proper coarse-grained description for interacting Brownian particles in the absence of hydrodynamic interactions, which is a widely used simulation model, as well as for interacting A and B particles hopping randomly on a lattice representing a metallic alloy. Its simplicity means that often insight can be gained by first understanding how Model B behaves, before turning to the more complicated scenario of Model H. 

The unhelpful names of these models have been embedded in the physics literature since the much-cited review of \cite{Hohenberg77} which also features Models A, C, D, E, F, G and J (each describing a different combination of broken symmetries and conservation laws). Consider however a systems in which $\phi$ represents a coarse-grained density of colloidal particles in solution. In such a system one can either treat the particles as undergoing Brownian motion with independent thermal noises (so that the only coupling between them comes from the enthalpic interaction forces encoded in $F$), in which case momentum is not conserved, or interacting additionally via hydrodynamic interactions, so that momentum is conserved (and the noise forces on different particles are not independent).  A mnemonic thereby emerges: the corresponding coarse-grained models are then B for Brownian, and H for hydrodynamic, respectively.

\section{Phase-separation kinetics} \label{PSK}
Having assembled the necessary conceptual and mathematical tools, we now examine some of the dynamics of phase separation in binary simple fluids \citep{Bray94}.

\subsection{Spinodal decomposition}
We start by considering the spinodal instability. Ignoring advection and noise initially, we write (\ref{Hphi}) as 
\begin{eqnarray}
\dot\phi &=& \del\bdot(M\del\mu)\\
&=& \del\bdot\left(M\del\left[f'(\phi)-\kappa\delsq \phi\right]\right)\\
&=& \del\bdot\left(M\left[f''(\phi)\del\phi-\kappa\delsq \phi\right]\right).
\end{eqnarray}
Here $f(\phi)$ is the local free energy density of a uniform state as defined in (\ref{uniform}), and primes denote differentiation of this function with respect to $\phi$. 
Next we linearize this equation about a uniform initial composition $\bar\phi$, and Fourier transform, to give
\begin{equation}
\dot\phi_q = -Mq^2\left[f''(\bar\phi)+\kappa q^2\right]\phi_q
\equiv -r(q)\phi_q,
\end{equation}
where we have defined a wave-vector dependent decay rate $r(q)$. For $f''(\bar\phi)>0$ this is positive for all $q$: all Fourier modes decay and the initial state is stable. In contrast for $f''(\bar\phi)<0$, the system is unstable, with $r(q)$ negative at small and intermediate wavevectors. (Stability is restored at high enough $q$ by the $\kappa$ term.)
Solving $dr/dq = 0$ for $q$ identifies the fastest growing instability to be at $q^* = -f''(\bar\phi)/2\kappa$. Because coupling to fluid flow was neglected this is technically a Model B result, but in fact such coupling (Model H) does not change the result of this linear stability analysis.

Even neglecting the thermal noise in the dynamics (\ref{HJnoise},\ref{Hvnoise}), the initial condition can be assumed to have some fluctuations. Those whose wavenumbers lie near $q^*$ grow exponentially faster than the rest, so that the time-dependent composition correlator $S_q(t) \equiv\langle\phi_\q(t)\phi_{-\q}(t)\rangle$ soon develops a peak of height growing as $\exp[-r(q^*)t]$ around $q^*$. Hence during this `early stage' of spinodal decomposition a local domain morphology is created by compositional interdiffusion with a well defined length-scale set by $\pi/q^*$. The amplitude of these compositional fluctuations grows until local values approach the binodals $\pm\phi_b$. From this emerges a domain pattern, still initially with the same length scale, consisting of phases in local coexistence, separated by relatively sharp interfaces of width $\xi_0$ and interfacial tension $\gamma_0$ as found from the equilibrium interfacial profile (\ref{profile}).

The next stage of the phase separation depends crucially on the topology of the newly formed fluid domains. This is controlled mainly by the phase volumes $\Phi_{A,B}$ of the A-rich and B-rich phases. Roughly speaking, if $0.3\le\Phi_A\le 0.7$ the domains remain bicontinuous: one can trace a path through the A-rich phase from one side of the sample to the other, and likewise for the B-rich phase. Outside this window, the structure instead has droplets of A in B ($\Phi_A<0.3$) or B in A ($\Phi_A>0.7$). The values $0.3$ and $0.7$ are rule-of-thumb figures only, with details depending on many other factors (including any asymmetry in viscosity or mobility) that we do not consider here. Note also that the window of bicontinuity shrinks to zero width in two dimensions, where the slightest asymmetry in phase volume and/or material properties will generally result in a droplet geometry in which only one phase is continuous. We next assemble the tools needed to understand the late stages of phase separation before discussing their dynamics.

\subsection{Laplace pressure of curved interfaces}
Once the interfaces are sharp, their geometry becomes a dominant factor in the time evolution.
Unless they are perfectly flat, interfaces exert forces on the fluid via the $\f=-\phi\del\mu$ term in (\ref{HNSE}), in response to which the fluid may or may not be set in motion. (No motion is implied if $\f$ remains irrotational, as discussed previously.) The physics of this term, for interfaces that are locally equilibrated but not flat, is that of Laplace pressure. Consider for example a spherical droplet of one fluid in another with radius $R$ and interfacial tension $\gamma$. Suppose that that the pressure inside the droplet is greater than that outside by an amount $\Delta P$. The total force on the upper half of the droplet exerted by the bottom half is then (in three dimensions)
\begin{equation}
\pi R^2\Delta P -2\pi \gamma R = 0,
\end{equation}
which must vanish if the droplet is not moving. The first term comes from the vertical component of the extra pressure acting across the equatorial disc, and the second is the tension acting across its perimeter. The resulting pressure excess, known as `Laplace pressure', is $\Delta P = 2\gamma/R$. More generally in three dimensions one has 
$
\Delta P = \gamma H = \gamma \left({R_1}^{-1}+{R_2}^{-1}\right)
$
where $H$ is the mean curvature and $R_1$ and $R_2$ are the principal radii of curvature. Note also that for a (hyper-)sphere in $d$ dimensions, $\Delta P = (d-1)\gamma/R$. In all these expressions we can set $\gamma = \gamma_0$ as calculated in section~\ref{profiletension}, so long as curvature is weak ($H\xi_0\ll 1$) and there is local thermodynamic equilibrium across the interface. 

A closely related result concerns the chemical potential $\mu$ at a representative point on a weakly curved interface. Local equlibrium fixes the interfacial profile as $\phi(\r) \simeq \phi_0(w) \equiv \phi_b\tanh(w/\xi_0)$ where $w$ is a Cartesian coordinate normal to the interface that vanishes at the midpoint of the profile. Then $\del\phi = \partial_w\phi(w)\hat\w$ with $\hat\w$ the unit normal (pointing from low to high $\phi$). It follows that $\nabla^2\phi = \partial_w^2\phi + \partial_w\phi\del\bdot\hat\w$. From the definition of $\mu$ as $\delta F/\delta\phi = f'(\phi)-\kappa\nabla^2 \phi$ we have
\begin{equation}
\mu = f'(\phi)-\kappa\partial_w^2\phi -\kappa\partial_w\phi\,\del\bdot\hat\w.
\end{equation}
For the local equilibrium profile $\phi(\r) = \phi_0(w) = \phi_b\tanh(w/\xi_0)$ the first two terms on the right cancel. Moreover, local equilibrium requires that $\mu$ is almost constant on the scale of the interfacial thickness $\xi_0$. Denoting this locally constant value interfacial by $ \mu_I$, multiplying by $\partial_w\phi$ and integrating through the interface  gives $2\phi_b\mu_I =  H\kappa\int(\partial_w\phi)^2\,\d w$ where $H = -\del\bdot\hat\w$ is the mean curvature, defined as above to be positive when the interface curves towards the high $\phi$ phase. Using the previously quoted result $\gamma_0 = \int \kappa(\partial_x\phi_0(x))^2 dx$ we find that $\mu_I = \gamma_0 H/2\phi_B$ \citep{Bray94}.

Thus for a large spherical droplet of the high $\phi$ phase in the low $\phi$ phase (so that $H=2/R$) the chemical potential $\mu_I$ at the interface has a small positive value proportional to the Laplace pressure. If the chemical potential in the surrounding phase takes its equilibrium value far away ($\mu = 0$ at coexistence) we requires a gradient of $\mu$ and thus a constant outward flux of $\phi$ from the droplet surface. Stable equilibrium of a finite droplet surrounded by the opposite phase is possible, but only in a finite system that has $\mu(\r) =  \mu_I$ everywhere so that $\J = -M\del\mu = \0$. This arises when the phase volumes $\Phi_{A,B}$  are sufficiently asymmetric that the interfacial area of a droplet configuration is less than that of a slab geometry, which depends on the shape of the container and its boundary conditions; even if so, $R\to \infty$ and $\mu_I \to 0$ in the large system limit. In all other situations (nonspherical domains, spherical droplets of unequal sizes, or a droplet in an infinite bath) there is a flux onto or off the interface proportional to the local curvature. 

To quantify this, consider the state of stable equilibrium for a droplet of A-rich phase ($\phi\simeq +\phi_b$) in a finite domain of the B-rich phase ($\phi\simeq -\phi_b$). After expanding $f(\phi) = a\phi^2/2+b\phi^4/4$ for small deviations about $\pm\phi_b$, equating the Laplace pressure $\Delta P$ to the difference in osmotic pressure $\phi\mu(\phi)-f(\phi)$ of the A-rich and B-rich bulk phases  gives the result that $\phi_{A,B} = \pm\phi_b+\delta$, with both phases equally shifted upwards in composition $\phi$ relative to  the equivalent state of bulk coexistence with a flat interface. The upward shift obeys $\delta = \gamma_0/(\alpha\phi_bR)$ where $\alpha \equiv f''(\pm\phi_b) = -2a$ and $R$ is the droplet radius. (This $\delta$ would reverse in sign for a B-rich droplet in an A-rich phase.) The result can alternatively be derived by setting $\mu_I = \alpha \delta$ in the equation $\mu_I = \gamma_0 H/2\phi_B$ quoted above.

We next assume that these local equilibrium conditions apply across the curved droplet interface in an infinite B-rich reservoir whose composition far away is $-\phi_b+\epsilon$ where $\epsilon < \delta$. Here the far-field parameter $\epsilon$ is known as the ambient supersaturation. We then seek a spherically symmetric quasi-static ($\dot\phi = -\del\bdot\J = 0$) exterior solution $\phi(r) = -\phi_b+\tilde\phi(r)$ where $\tilde\phi(R) = \delta$ and $\tilde\phi(\infty) = \epsilon$. With $\J = - M\del\mu$ this solution obeys $\mu = \alpha\tilde\phi$ where $\nabla^2\tilde\phi = 0$; note that the square gradient contribution to $\mu$ vanishes in this geometry. With the given boundary conditions, the result is $\tilde\phi = \epsilon +(\delta-\epsilon)R/r$. If we now calculate the radial current $J = |\J|$ exterior to the droplet, we have $J = -\alpha M\partial\tilde\phi/\partial r = \alpha M(\delta-\epsilon)R/r^2$; just outside the droplet itself this becomes $-\alpha M(\delta-\epsilon)R$. By mass conservation (given the compositional jump $\Delta\phi = 2\phi_b$ across the interface) we then find $\dot R = -J(R)/2\phi_b$ or
\begin{equation}
\dot R = \frac{1}{2\phi_b}\left[\frac{\alpha M}{R}\left(\epsilon - \delta(R)\right)\right] = \frac{1}{2\phi_b}\left[\frac{\alpha M}{R}\left(\epsilon - \frac{\gamma_0}{\alpha\phi_bR}\right)\right].
\label{evaporation}\end{equation}
This result will be used in section \ref{Ostwald} below to discuss the process of competitive growth and shrinkage of droplets known as Ostwald ripening.

\subsection{Coalescence of droplet states}

If the post-spinodal structure is that of droplets, each relaxes rapidly to minimize its area at fixed volume, resulting in a spherical shape. For well separated droplets the thermodynamic force $\f$ in the neighbourhood of each droplet has radial symmetry and hence is the gradient of a scalar (pressure) contribution. This has no consequences for fluid flow since all such terms are subsumed into an overall pressure set by the incompressibility constraint. Accordingly there is no net fluid motion in the absence of noise. Thermal noise allows droplets to move around by Brownian motion, resulting in collisions that cause the mean droplet radius $R$ to increase by coalescence. To estimate the time dependence of $R$ we observe that for A droplets in B, the mean inter-droplet distance $L$ is of order $R\Phi_A^{-1/3}$ at small $\Phi_A$; more generally, $L/R$ is a function of
$\Phi_A$ only. Each droplet will collide with another in a time $\Delta t$ of order $L^2/D$ where $D \simeq k_BT/\eta R$ is the droplet diffusivity. Upon collision, two droplets of radius $R$ make a new one of radius approximately $2^{1/3}R$ causing an increment $\Delta \ln R \simeq (\ln 2)/3$. This gives
\begin{equation}
\frac{\Delta \ln R}{\Delta t} \propto \frac{k_BT}{\eta R^3},
\end{equation}
where the precise factor of $2^{1/3}$ has ceased to matter, and  the left hand side can be approximated as $d\ln R/dt = \dot R/R$. By integration we then obtain the scaling law
\begin{equation}
R(t) \sim\left(\frac{k_BTt}{\eta}\right)^{1/3}. \label{coalescence}
\end{equation}
This argument assumes that coalescence is diffusion-limited, and shows that in this case Brownian motion will cause indefinite growth of the mean droplet size, culminating in total phase separation. 

The assumption of independently diffusing spherical droplets should be reliable at low enough phase volumes $\Phi_A$ of the dispersed phase, where it is the dominant coarsening mechanism so long as the Ostwald process (described next) is suppressed.  But at larger $\Phi_A$ more complicated routes to coalescence, some involving droplet-scale or macroscopic fluid flow, are possible. 
One of these is so-called `coalescence-induced coalescence' where the shape relaxation post-collision of a pair of droplets creates enough flow to cause another coalescence nearby \citep{Wagner01}. This gives a new scaling ($R\sim \gamma_0 t/\eta$) which coincides with one of the regimes described later for the coarsening of bicontinuous structures (section \ref{VH} below). Indeed it stems from the same balance of forces as will be discussed for that case.
Quite recently a new picture has emerged of hydrodynamic coarsening in moderately concentrated droplet suspensions ($\Phi_A> 0.2$), which supersedes a longstanding view that equation (\ref{coalescence}) still holds in this regime. This picture involves mechanical (Marangoni) forces resulting from departures of the interfacial tension from its equilibrium value ($\gamma \neq \gamma_0$). These departures arise because the presence of neighbours breaks rotational symmetry around any given droplet so that the thermodynamic force density is no longer a pure pressure gradient \citep{Tanaka15}.

In many droplet emulsions it is possible to inhibit the coalescence step, so that this entire route to phase separation is effectively blocked. For instance, adding charged surfactants can stabilize oil droplets in water against coalescence by creating a Coulombic barrier opposing the close approach of droplet surfaces. (Surfactants are amphiphilic molecules that have a polar or charged head-group and an oily tail; they are widely used to stabilize oil-in-water droplet emulsions.) Steric interactions between surfactant tails can likewise stabilize water-in-oil emulsions. If the rupture of a thin film of the continuous phase between droplets has a high enough free energy barrier, coalescence rates can be reduced to a manageable if not negligible level in both cases \citep{Bibette02}.

\subsection{Ostwald ripening}\label{Ostwald}
\begin{figure}
	\centering
	\includegraphics[width=5cm]{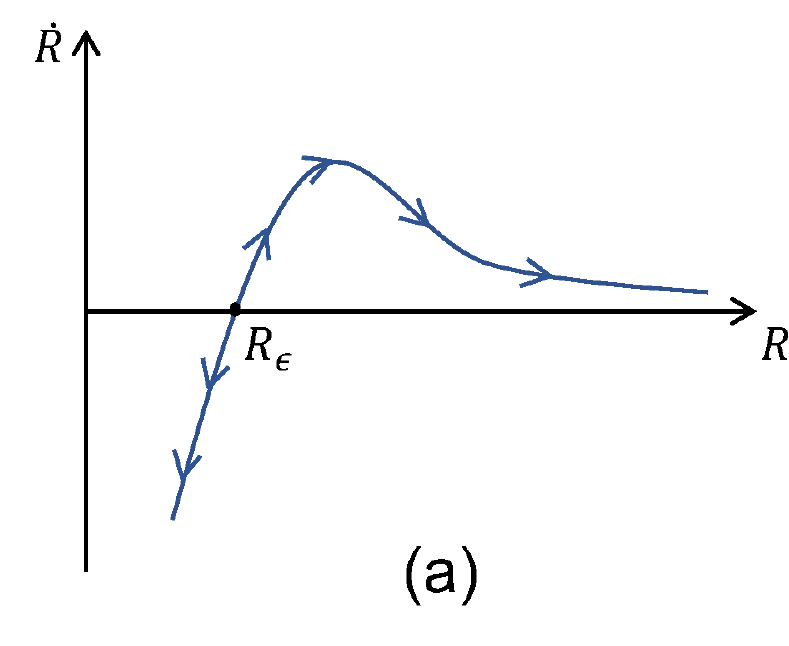}
	\includegraphics[width=5cm]{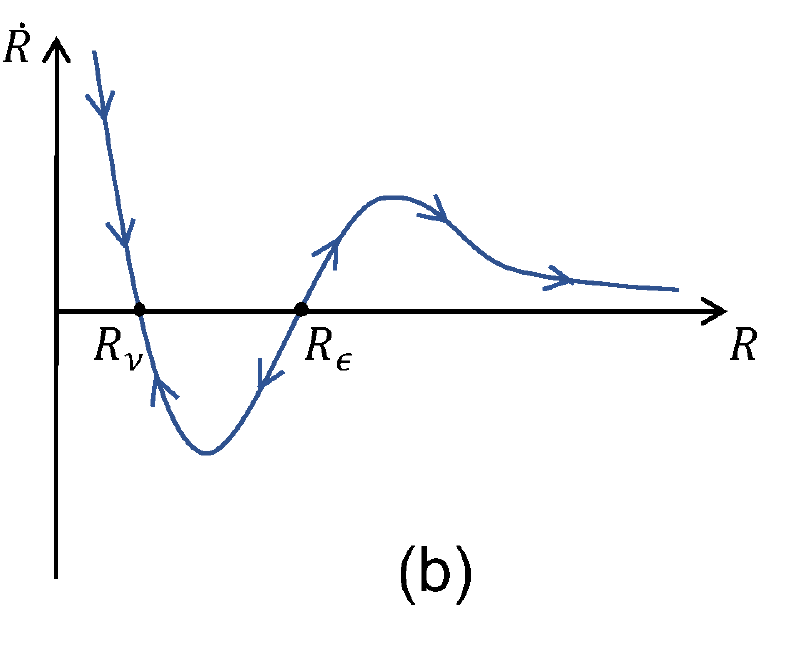}
	\caption[]{Growth rate of droplet as a function of size during the Ostwald process (a) without and (b) with trapped species.}
	\label{three}
\end{figure}

Sadly however, switching off coalescence is not enough to prevent macroscopic phase separation of droplet emulsions. This is because of a process called Ostwald ripening, in which material is transported from small droplets to large ones by molecular diffusion across the intervening continuous phase. 
The physics of this process follows directly from equation (\ref{evaporation}). Assuming that the system has not already reached its end-point of full phase separation, the ambient supersaturation $\epsilon$ in that equation remains finite. The function $\dot R(R)$ is shown in figure \ref{three}. This exhibits an unstable fixed point at 
\begin{equation}
R = R_\epsilon(t)\equiv \frac{\gamma_0}{\alpha\phi_b\epsilon}.
\end{equation}
Droplets bigger than this grow, those smaller, shrink. This is because the Laplace pressure in each droplet raises the chemical potential at its outer surface by an amount inversely proportional to its radius. This creates chemical potential gradients from small to large droplets which shrink and grow in response to the fluxes set up by those gradients.

We can now find the scaling of the typical droplet size $\bar R$ by assuming this to be comparable (but not exactly equal) to $R_\epsilon$ and conversely that the ambient supersaturation $\epsilon$ is comparable (but not exactly equal) to the local supersaturation near a typical drop: $\epsilon \simeq \delta(\bar R) = \gamma_0/(\alpha \bar R\phi_b)$. Substituting in (\ref{evaporation}) gives
\begin{equation}
\dot {\bar R} \simeq \frac{M\gamma_0}{\phi_b^2\bar R^2},
\end{equation}
which results in the scaling law
\begin{equation}
\bar R(t) \simeq \left(\frac{M\gamma_0 t}{\phi_b^2}\right)^{1/3}
\sim t^{1/3.}\label{LSlaw}\end{equation}
A more complete theory of Ostwald ripening, known as the Lifshitz-Slyov-Wagner theory, is reviewed by \cite{Onuki02}. This not only confirms these scalings but gives detailed information on the droplet size distribution.
Note that (\ref{LSlaw}) has similar time dependence to (\ref{coalescence}) for coalescence; this stems from the fact that both mechanisms are ultimately diffusive. However, the nature of the diffusing species (droplet in one case, molecule in the other) is quite different, resulting in prefactors that involve unrelated material properties for the two mechanisms.

\subsection{Preventing Ostwald ripening}\label{ts}
It follow from (\ref{LSlaw}) that the Ostwald process can be slowed by reducing the interfacial tension $\gamma_0$, but so long as this remains positive it cannot be stopped entirely. A more effective approach is to include within the A phase a modest concentration of a some species that is effectively insoluble in B. This might be a polymer or, if A is water and B oil, a simple salt. The idea is that the trapped species in the A droplets creates an osmotic pressure which rises as $R$ falls, hence opposing the Laplace pressure. Treating the trapped species as an ideal solution in A,  (\ref{evaporation}) is replaced by \citep{Webster98}
\begin{equation}
\dot R =  \frac{1}{2\phi_b}\left[\frac{\alpha M}{R}\left(\epsilon - \frac{\gamma_0}{\alpha\phi_bR}+ \frac{\hat\eta \nu}{R^3}\right)\right],
\label{webster}\end{equation}
where $\hat\eta$ (unrelated to the fluid viscosity $\eta$) is a combination of molecular parameters and physical constants, and $\nu$ is the number of trapped molecules in the droplet. The final term reflects the extra osmotic pressure of the trapped material. 
There is now a stable fixed point, even for zero ambient supersaturation, at finite droplet size
\begin{equation}
R_\nu = \left(\frac{\hat\eta \nu \alpha\phi_b}{\gamma_0}\right)^{1/2}.
\end{equation}
If we start from a uniform set of droplets of size $R_0$, then $\nu = c 4\pi R_0^3/3$ with $c$ the initial concentration of the added species in the A-rich phase.  Droplets that have shrunk to a size $R_\nu$ can coexist with a bulk A-rich phase without shrinking further: the Laplace pressure is balance by the trapped species' osmotic pressure. Moreover, if the initial size obeys $R_0<R_\nu$ the Ostwald process is reversed: if an emulsion of such droplets placed in contact with a bulk A-rich phase, they will expand up to size $R_\nu$. By adding trapped species one can thus make robust `mini-emulsions' \citep{Landfester03}
or `nano-emulsions' \citep{Fryd12}  which permanently resist coarsening by the Ostwald process. However these are still metastable: so long as the tension $\gamma_0$ is positive, the free energy can always be reduced by coalescing droplets to reduce the interfacial area.

\subsection{Coarsening of bicontinuous states}
For roughly equal volumes of the two coexisting phases, say $0.3\le\Phi_A\le 0.7$ the domains of A-rich and B-rich coexisting fluids remain bicontinuous. (This applies in three dimensions; the two-dimensional case is special and considered later.) This allows coarsening by a process in which the interfacial stresses -- or loosely speaking, Laplace pressure gradients -- pump fluid from one place to another. In all but the most viscous fluids, this process is ultimately faster than either coalescence or Ostwald ripening: we will see below that it allows $L(t)$ to grow with a larger power of the time. Here $L(t$) is now defined as the characteristic length scale of the bicontinuous structure, for example as the inverse of its interfacial area per unit volume. The flow-mediated coarsening is captured by Model H, but absent in Model B. In the latter, domain growth is controlled by a version of the Ostwald process in which diffusive fluxes in both phases act to flatten out interfaces (reducing their curvature) and move material from thinner to fatter regions of the bicontinuous structure. This leads again to (\ref{LSlaw}) for the characteristic domain size, which is now unaffected by adding insoluble species which can move diffusively throughout the relevant phase. In both cases 
the driving force is interfacial tension.

To describe flow-mediated coarsening, we assume that $L(t)$ is much larger than the interfacial width, and is the only relevant length in the problem, so that in estimating terms in the NSE (\ref{HNSE}) we may write $\del\sim 1/L$. The forcing term $\phi\del\mu$ is then of order $\gamma_0/L^2$ where we have used the result found above, $\mu_I = \gamma_0 H/2\phi_B$, for the chemical potential on an interface of curvature $H\sim 1/L$. The fluid velocity is of order $\dot L$ so the viscous term scales as $\eta \dot L/L^2$. The inertial terms are $\rho\dot\v \sim \rho \ddot L$ and $\rho \v.\del\v \sim \rho (\dot L)^2/L$. The $\del P$ term, enforcing incompressibility, is slave to the other terms.

Importantly, once sharp interfaces are present so that $\phi\del\mu\sim \gamma_0/L^2$, the NSE (\ref{HNSE}) involves only three parameters, $\rho, \gamma_0$ and $\eta$. From these three quantities one can make only one length, $L_0 = \eta^2/\rho\gamma_0$, and one time $t_0 = \eta^3/\rho\gamma_0^2$. This means that the domain scale $L(t)$ must obey \citep{Siggia79,Furukawa85}
\begin{equation}
\frac{L(t)}{L_0} = f\left(\frac{t}{t_0}\right).\label{Scalinglaw}
\end{equation} 
where, for given phase volumes, $f(x)$ is a function common to all fully symmetric binary fluid pairs. This scaling applies only to bicontinuous states; as previously described, in states comprising spherical droplets the forcing term in the NSE is instead subsumed by the pressure term at leading order.

In systems showing fluid-mediated coarsening via (\ref{Scalinglaw}), we expect different behaviour according to whether $L/L_0$ is large or small. For $L/L_0$ small it is simple to confirm that the inertial terms in (\ref{HNSE}) are negligible. (Note also that, in any regime where $f(x)$ is a power law, the two inertial terms have the same scaling.) The primary balance in the NSE is then $\eta \dot L/L^2 \sim \gamma_0/L^2$ resulting in the scaling law $L(t) \sim \gamma_0 t/\eta$ so that
\begin{equation}
f(x) \propto x\;\;;\;\; x\ll x^*. \label{VH}
\end{equation} 
This is called the viscous hydrodynamic (VH) regime and holds below some crossover value $x = x^*$. In contrast, for $x\gg x^*$ the primary balance in the NSE is between the interfacial and inertial terms. It is a simple exercise then to show that $L(t) \sim (\gamma_0/\rho)^{1/3}t^{2/3}$ so that 
\begin{equation}
f(x) \propto x^{2/3}\;\;;\;\; x\gg x^* .\label{IH}
\end{equation} 
This is called the inertial hydrodynamic (IH) regime. 
In practice this crossover from VH to IH is several decades wide, and the crossover value rather high: $x^*\simeq 10^4$ \citep{Kendon01}. The high crossover point is less surprising if one calculates a domain-scale Reynolds number
\begin{equation}
{\Rey} = \frac{\rho L \dot L}{\eta} = f(x)\frac{df}{dx}.
\label{Re}\end{equation} 
The crossover value of $\Rey$ then turns out to be of order 10 \citep{Kendon01}, and the largeness of $x^*$ is found to stem from a modest constant of proportionality in (\ref{VH}). It means that in practice a clean observation of the inertial hydrodynamic regime has only been achieved in computer simulation: in terrestrial laboratory experiments the domains are by then so large (millimetres to centimetres for typical fluid pairs) that the slightest density difference between A and B causes gravitational terms to dominate. These terms give a body force pulling the two fluids apart along the gravitational axis, greatly speeding phase separation.

Note finally that in two dimensions, bicontinuity is exceptional. For fully symmetric fluid pairs it arises only when the phase volumes $\Phi_{A,B}$ are also symmetric, allowing percolating paths through both fluids to cross the container (but only just). More generally, it arises at only one special phase volume whose value is set by the asymmetry of the two fluids; even if one could create this state, it would typically not be sustained during coarsening (since the meticulous dynamical balance required varies with the scale parameter $x$). Alongside Ostwald and coalescence regimes, one can find, at least in simulations, various complex structures involving cascades of nested droplets whose details may be noise-dependent \citep{Gonnella99}.

\section{Emulsification by shearing}
Suppose we wish to stop the coarsening of a bicontinuous fluid mixture by the flow-mediated mechanism just described. In a processing context, it is often enough to temporarily maintain a well-mixed, emulsified state merely by stirring the system. Industrial stirring is complicated, with nonuniform flows that often combine very different local flow geometries (extension or shear) within a single device. Here we restrict attention to the effects of a simple shear flow, focusing on scaling issues and the question of whether or not such a flow can actually stop the coarsening process. 

We consider uniform shearing with macroscopic fluid velocity along $x$, and its gradient along $y$; $z$ is then the neutral (or vorticity) direction. In simulations one can use boundary conditions with one static wall at $y=0$, another sliding one at $y = \Lambda$ and periodic BCs in $x,z$. In practice there are ways to introduce periodic BCs also in $y$; nonetheless, the system size in that direction, $\Lambda$, is important in what follows. The top plate moves with speed $\Lambda/t_s$ where $1/t_s$ is the shear rate: $\v_{\rm macro} = y\hat\x/t_s$. 

For the coarsening to be fully arrested by shear, the system must reach a nonequilibrium steady state time-reversal symmetryin which the fluid domains have finite length scales $L_{x,y,z}$ in all three directions. (These can be defined as the inverse of the length of interface per unit area on a plane perpendicular to each axis.) The simplest hypothesis is that these steady-state lengths, if they exist, all have similar scaling: $L_{x,y,z}\sim L$ \citep{Doi91}. Moreover, given the preceding discussion of terms in the NSE, we expect in steady state that $L/L_0$ is now a function, not of $t/t_0$, but of $t_s/t_0$. That is, in steady state the previous dependence on time is replaced by a dependence on the inverse shear rate. The functional form of this dependence could in principle be anything at all, but the simplest scaling ansatz is that the system coarsens as usual until $t\sim t_s$, whereupon the shearing takes over and $L$ stops increasing. If so, 
\begin{equation}
\frac{L}{L_0} \simeq f\left(\frac{t_s}{t_0}\right),
\end{equation}
where $f(x)$ is approximately the same function as introduced previously, for which (\ref{VH},\ref{IH}) hold. If so, for $t_s/t_0 \ll x^*$ we have $L/L_0 \sim t_s/t_0$ and for $t_s/t_0 \gg x^*$ we have $L/L_0\sim (t_s/t_0)^{2/3}$. In the first of these regimes, the force balance in NSE between viscous and interfacial stresses corresponds to setting a suitably defined capillary number (on the scale of $L$) to an order-unity value. (This is a criterion for the maximum size of droplets in dilute emulsions also.)
Note however that the Reynolds number obeys $\Rey\sim f(x)df/dx$ as in (\ref{Re}). In contrast to what happens for any problem involving shear flow around objects of fixed geometry, $Re$ is now {\em small} when the shear rate is {\em large} and vice versa. This is because at small shear rates, very large domains are formed.

\begin{figure}
	\centering
	\includegraphics[width=10cm]{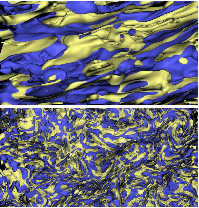}
	\caption[]{Upper panel: Simulation snapshot for Model H of binary fluid domains under shear for a symmetric binary fluid in 3D. The image shows only the interface between phases, with the two sides coloured blue and yellow; the macroscopic flow is roughly laminar. (This is a 3D view into the sample from an arbitrary $(x,y)$ plane cut through it.)
	In this particular case the flow eventually transitions to a turbulent mixing regime (lower panel). At lower $\Rey_\Lambda$ states resembling the upper panel persist indefinitely.
	The imposed flow axis $\hat\x$ is horizontal with the velocity gradient vertical; the upper part of the system is moving to the right. (Images courtesy of Kevin Stratford.)}
	\label{four}
\end{figure}

There are several possible complications to this picture. First, in principle $L_{x,y,z}$ could all have different scalings. The resulting anisotropies could spoil any clear separation of the viscous hydrodynamic and inertial hydrodynamic regimes; with three-way force balance in the NSE, there is no reason to expect clean power laws for any of these quantities. Secondly, at high shear rates the system-size Reynolds number $\Rey_\Lambda \sim \rho \Lambda^2/\eta t_s$ becomes large. The presence of a complex microstructure could promote or suppress the transition to conventional fluid turbulence expected in this regime. In practice, experimental studies on sheared symmetric binary fluids are sparse \citep{Onuki02}. Gravity complicates matters, as does viscosity asymmetry (unavoidable in practice) between phases. However, relatively clean tests are possible via computer simulation \citep{Stansell06,Stratford07}. In 2D one finds data fitted by $L_x/L_0\sim (t_s/t_0)^{2/3}$ and $L_y/L_0\sim (t_s/t_0)^{3/4}$ over a fairly wide range of length and timescales, most of which are however in the crossover region around $x^*$ \citep{Stansell06}. The fitted exponents change slightly if instead of the flow and gradient direction one uses the principal axes of the distorted density patterns, but they remain unequal. There is no theory for these exponents as yet, but it is credible that the length-scales along and across the stretched domains have different scalings with shear rate. (Indeed if the dynamics resembled stretching of droplets of fixed volume, the two exponents would add to zero in 2D.)
In 3D, where the simulations require very large computations, the inertial hydrodynamic (2/3 power) scaling has been observed within numerical error for all three length scales within the range of accessible domain-scale Reynolds numbers, defined as $\Rey_L\sim \rho L^2/\eta t_s$, which lies
between 200 and 2000 \citep{Stratford07}. These measurements are however limited by the onset of a macroscopic instability to turbulent mixing at $\Rey_\Lambda\simeq 20,000$. Snapshots of the highly distorted domain structures seen in computer simulations of sheared binary fluids are shown in figure \ref{four}. 

The above simple arguments suggest that in both 2D and 3D, a nonequilibrium steady state can be achieved in which the force balance in (\ref{HNSE}) is entirely between viscous and interfacial stresses, with the inertial terms negligible, as holds in the viscous hydrodynamic regime for the coarsening process. If so, this should be the situation at low enough $\Rey_L$, that is, large enough shear rate. Evidence that things are yet more complicated has been presented by \cite{Fielding08}, who showed that if inertial terms are omitted altogether from (\ref{HNSE}), coarsening in two dimensions appears to proceed indefinitely. This suggests that a three-way balance of viscous, inertial and interfacial terms ultimately controls the formation of the nonequilibrium steady state.

\section{Thermodynamic emulsification}
Surfactant molecules, comprising small amphiphilic models with a polar or charged head group and an apolar tail, are well known to reduce the interfacial tension between coexisting phases of oil and water (as well as other pairs of apolar and polar fluids). They generally have fast exchange kinetics between the interface and at least one bulk phase in which they are soluble; this means that the interface remains locally in equilibrium. Use of surfactants can thus be viewed as a thermodynamic route to the partial or complete stabilization of interfacial structures. This is a separate effect from their role in creating a kinetic barrier to coalescence discussed previously.

\subsection{Interfacial tension in the presence of surfactant}
Consider first a solution of surfactant molecules each carrying a unit vector $\nuhat_i$ denoting its head/tail orientation;  local coarse graining creates a smooth field $\p(\r) = \langle\nuhat\rangle_{\rm meso}$. In the absence of interfaces, this fluctuating mesoscopic field is zero on average, and its small (hence Gaussian) fluctuations are governed by a free energy contribution
\begin{equation}
F_{s} = \int \left(\frac{|\p(\r)|^2}{2\chi}\right)\d\r.\label{chidef}
\end{equation}
This is defined such that the variance of $\p$ is set by the `osmotic compressibility' $\chi(\mu_s)$ which is an increasing function of the surfactant chemical potential $\mu_s(c_s)$, which is itself a function of surfactant concentration $c_s$. We can add this to the binary fluid free energy  (\ref{functional}), alongside a coupling term to represent the reduction in free energy caused when a surfactant molecule resides at the A-B interface with its orientation suitably aligned along the composition gradient $\del\phi$:
\begin{equation}
F[\phi,\p] = \int \left(\frac{a}{2}\phi^2+\frac{b}{4}\phi^4 +\frac{\kappa}{2}(\del\phi)^2+\frac{1}{2\chi}|\p|^2 + \omega\p.\del\phi\right)\d\r.
\label{functional2}
\end{equation}
It is a simple exercise to minimize this over $\p(\r)$ at fixed $\phi(\r)$, and a slightly more complicated one to explicitly integrate over the fluctuating $\p$ field by Gaussian integration
to obtain $e^{-\beta F[\phi]} = \int e^{-\beta F[\phi,\p]}{\mathcal D}\p$. The result of either calculation is to recover the original Model H free energy $F[\phi]$ as in (\ref{functional}),  but with a renormalized square gradient coefficient
$\kappa_r = \kappa - \omega^2\chi$.
From this it follows that the interfacial tension is reduced by adding surfactant, according to
\begin{equation}
\gamma_0(c_s) = \left(\frac{-8a^3(\kappa-\omega^2\chi)}{9b^2}\right)^{1/2}. \label{gammasurf}
\end{equation}
This calculation is based on a Gaussian approximation (\ref{chidef}) that assumes $\p(\r)$ to deviate only mildly from zero everywhere. It is not very realistic -- for instance the interfacial width diverges as $\gamma_0$ becomes small (whereas in practice this width is set by the size of a surfactant molecule). However it captures the main physical effects of interest here; for a fuller discussion, see \cite{Gompper94}. Note that the sign of $\omega$, which determines whether $\p$ points up or down the interfacial $\phi$ gradient, is irrelevant. 

The simplest possible assumption is that the surfactant molecules form an ideal solution, with no interactions between them. In this case $\chi = \tilde\chi c_s$ with $\tilde\chi$ a constant. (Linearity in $c_s$ of the variance parameter $\chi$ then follows from the Poisson statistics of randomly located and oriented molecules.) In this case the interfacial tension vanishes, with infinite slope, at $c_s = \tilde c = \kappa/\omega^2\tilde\chi$. But in fact, surfactant solutions are far from ideal, due to a phenomenon called `micellization' in which individual molecules self-assemble into micelles, which contain several tens of molecules. (Micelles are typically spherical; in water they have the polar heads at the exterior surface and the apolar tails in the interior of the sphere. In oil this structure is reversed.)  

The effect of micellization is to cause the surfactant chemical potential $\mu_s$, and therefore $\chi$, to rapidly saturate above a so-called `critical micelle concentration' $c_s = c^*$. Any surfactant added beyond this level becomes sequestered into micellar aggregates. The concentration $c_1$ of `free' molecules remains very close to $c^*$ at all higher concentrations and, 
to a good approximation, $\chi$ saturates at $\tilde\chi c^*$, so that $\gamma_0(c_s)$ does not fall further. For a more detailed discussion of micellization, see \cite{Safran03} or \cite{Cates12}.

The outcome is to have two general classes of behaviour, depending on whether $c^*$ lies below or above $\tilde c$. In case 1, $c^*<\tilde c$, so that $\gamma_0(c_s)$ follows (\ref{gammasurf}) so long as $c_s<c^*$ but then abruptly stops decreasing as micellization intervenes.  In case 2, $c^*>\tilde c$ so that $\gamma_0(c_s)$ hits zero before micelles are formed. At this point, if water and oil are both present in bulk quantities, the system can minimize its free energy by creating a macroscopic amount of interface on which the surfactant can reside. When this happens $\mu_s$ again saturates: adding further surfactant simply creates more surface at fixed $\mu_s$. Hence $\gamma_0$ does not become finitely negative but remains stuck at an effectively zero value. This analysis is grossly simplified, ignoring (among other things) free energy contributions from interfacial curvature which can allow interfaces to proliferate even for small positive tension. Nonetheless, the broad distinction between case 1 and case 2 is a useful one.

\subsection{Finite tension: metastable emulsions and biliquid foams}
Case 1 is the more common: the typical effect of surfactant is to reduce interfacial tension to half or a third of its previous value. The global free energy minimum then comprises complete phase separation just as it does without surfactant; if emulsions are formed (for instance by stirring) they are at most metastable. Since fluid-mediated coarsening is always present in bicontinuous states, metastability generally requires a droplet geometry; as already mentioned, surfactants can help prevent their coalescence by inhibiting film rupture. 

Typically such emulsions have spherical droplets (of A in B, say) but by evaporation or drainage under gravity, for instance in a centrifuge, much of the continuous B phase can often be expelled to create a so-called biliquid foam \citep{Bibette02}, in which polyhedral droplets of A (say) are separated by thin films of B. In many cases, biliquid foams can persist for hours or days, and sometimes longer. To achieve this one must suppress not only the rupture of thin films but also the Ostwald process which, despite the more complicated geometry, still drives diffusion of A from small (few-sided) to large (many-sided) polyhedral droplets \citep{Weaire99}. To achieve this with a trapped species requires an especially low level of solubility in the B phase so as to ensure negligible diffusion even across the thin B films present in the foam structure. So long as they remain metastable against rupture and coarsening, biliquid foams, like soap froths, are solid materials (generally amorphous, though ordered examples can be made). As such they have an elastic modulus, and also a yield stress, both of which scale as $G\sim\gamma_0/R$ with $R$ the mean droplet size. This is an interesting example of a solid behaviour emerging solely from the spatial organization of locally fluid components -- for even the surfactant on the interface is (normally) a 2D fluid film. 

\subsection{Near-zero tension: stable microemulsions}
\label{tensionless}
In case 2, added surfactant can reduce $\gamma_0$ to negligible levels for $c_s\ge\tilde c$.  This can lead to thermodynamically stable emulsions, generally called ``microemulsions", in which enough A-B interface is created to accommodate all surplus surfactant, of which the concentration is $c_s-\tilde c$. There are two broad approaches to describing the resulting structures. One avenue is to base a description on the $\phi,\p$ order parameters already introduced, addressing (\ref{functional2}) in the case where $\kappa_r = \kappa - \omega^2\chi$ so that the thermodynamic interfacial tension is negative. Unsurprisingly, this model is unstable unless further terms are added to prevent interfacial proliferation; the most natural addition to $F$ is a term in $\int(\del\phi)^4\d\r$. The competition of this term with the effectively negative square gradient coefficient $\kappa_r$ sets a characteristic length scale for stable domains of A-rich and B-rich fluids. This approach leads to many insights \citep{Gompper94} but is mainly appropriate for systems in which `weak' surfactants (whose adsorption energy at an interface is not much larger than the thermal energy $k_BT$) are present at high concentration so that the domain size $L$ and interfacial width $\xi$ are comparable. The structure of the microemulsion can then be viewed as a smooth spatial modulation of composition $\phi(\r)$ rather than a system of well separated, surfactant-coated interfaces. 

For strong surfactants, which have small $\tilde c$, one can instead treat almost all the surfactant as interfacial. The interfacial area $\S$ of the fluid film then obeys
\begin{equation}
\frac{\S}{V}  =  (c_s-\tilde c)\Sigma \simeq c_s\Sigma = \frac{\phi_s}{v_s}\Sigma. \label{area}
\end{equation}
Here $\Sigma$ is a preferred area per molecule; $\phi_s$ is the volume fraction of surfactant and $v_s$ its molecular volume. 
For the soluble surfactants normally used for emulsification, the area per molecule is maintained very close to $\Sigma$ by rapid adsorption and desorption at the interface; the specific interfacial area $\S/V$ is then fixed directly by $\phi_s$ via (\ref{area}). 

With interfacial tension negligible, the energetics of a given structure in case 2 is determined by the cost of {\em bending}  the interfacial surfactant film at fixed area. We treat this by a leading order harmonic expansion about a state of preferred curvature set by molecular geometry. By a theorem of differential geometry \citep{David04}, at each point on the A-B interface one can uniquely define two principal radii of curvature, $R_1, R_2$ which we take positive for curvature towards A. (For a spherical droplet of A of radius $R$, we have $R_1=R_2 = R$ whereas for a cylinder of radius $R$, we have $R_1 = R$ and $R_2 = \infty$. A saddle shape has $R_{1}$and $R_{2}$ of opposite signs.)
The harmonic bending energy for a fluid film then reads
\begin{equation}
F_{\rm bend} = \int  \left[\frac{K}{2}\left(\frac{1}{R_1}+\frac{1}{R_2}-\frac{2}{R_0}\right)^2 +\frac{\bar K}{R_1R_2}\right] \d\S. \label{bending}
\end{equation}
There are 3 material parameters: the elastic constants $K$ and $\bar K$ (both with dimensions of energy), and $R_0$, a length defining the preferred radius of mean curvature, whose relation to the molecular geometry of surfactants is discussed by \cite{Safran03}. The integral in (\ref{bending}) is over an interface $\S$ between phases that can have disconnected parts (droplets) but must be orientable so that A is enclosed by $\S$ and B excluded. Therefore, its enclosed volume $V_{\rm in}$ must obey
\begin{equation}
\frac{V_{\rm in}}{V} = \Phi_A+\frac{\phi_s}{2} \equiv \Phi, \label{Psi}
\end{equation}
where we have partitioned the surfactant equally between A and B to allow us to define the volume $V_{\rm in}$ as enclosed by a mathematical surface of no thickness. The phase volume of $V_{\rm in}$ is then $\Phi$ (with $V_{\rm out} = 1-\Phi$), and a completely symmetric state has $\Phi = 1/2$.   

\subsection{The physics of bending energy}
The statistics of the A-B interface in case 2 is, by the above arguments, determined by the Boltzmann distribution ${\cal P}[\S] \propto \exp[-\beta F_{\rm bend}]$. Performing averages over this distribution is intractable, in general. However, some key concepts can be identified that allow the problem to be understood qualitatively.

\subsubsection{Gauss-Bonnet theorem and emulsification failure}
The Gauss-Bonnet theorem states that
\begin{equation}
\int \frac{1}{R_1R_2}d\S = 4\pi[N_c-N_h].
\end{equation} 
Here $N_c$ is the number of components of our surface (where a component means a disconnected piece such as a sphere) and $N_h$ is the number of handles. A handle is a doughnut-like connection between one part of the surface and another. Thus for a sphere $N_c = 1$ and $N_h = 0$ whereas for a torus, $N_c = 1$ and $N_h=1$. Accordingly the bending energy term governed by $\bar K$ in (\ref{bending}) vanishes for a torus but not a sphere. More generally, this term does not care about the local deformations of the surface, only its topology.

\begin{figure}
	\centering
	\includegraphics[width=11cm]{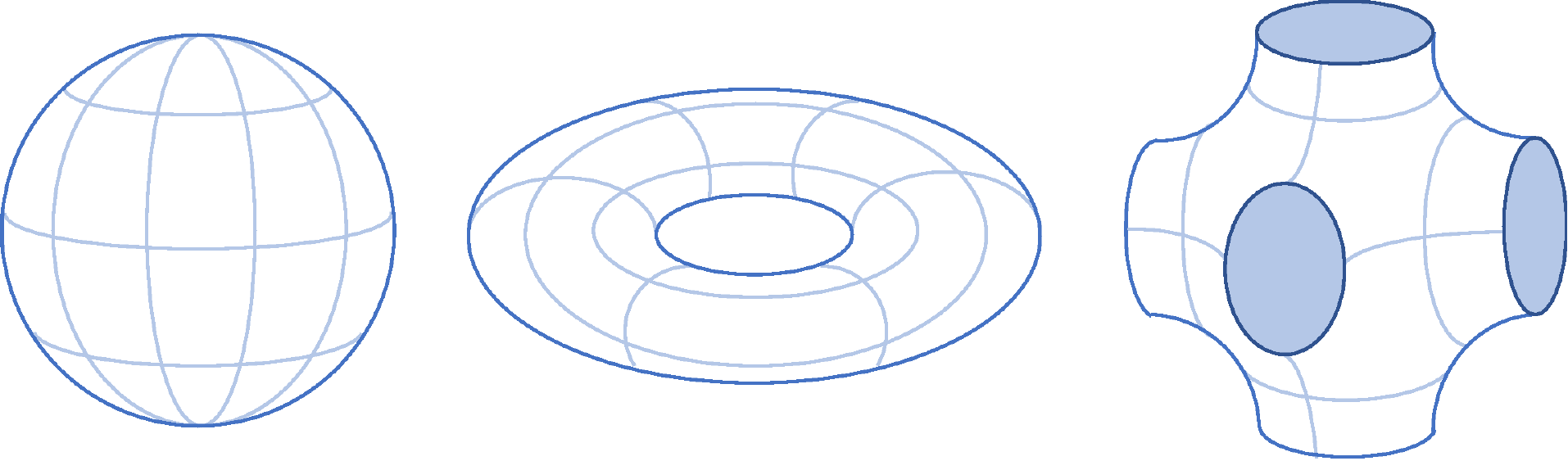}
	\caption[]{A sphere, a torus, and a sketch of the unit cell of a periodic surface of constant (approximately zero) mean curvature. The hole through the torus is a handle. The grey discs on the periodic surface are cuts across it at the junction points between unit cells. Gluing a pair of these discs together at the faces of the unit cell creates one handle. Thus the final periodic structure has three handles per unit cell, but only one component in total. }
	\label{eight}
\end{figure}

As well as spheres and tori, one can devise extended surfaces of constant mean curvature, comprising a periodic surface element (figure \ref{eight}) which connects with identical copies of itself in neighbouring unit cells to create a structure with only one global component, but several handles per unit cell. Choosing the mean curvature to be $1/R_0$, the $K$ term in the bending free energy (\ref{bending}) vanishes everywhere. If $\bar K$ is positive, favouring handles, $F_{\rm bend}$ is unbounded below for a periodic state with an infinitesimal unit cell. In practice the unit cell is small and finite, due to anharmonic terms omitted from (\ref{bending}); 
the result is a bicontinuous cubic liquid crystal \citep{Safran03}. This phase has large $\phi_s$ (typically tens of percent) and so, unless the global mean of $\phi_s$ is similarly large, can occupy only a small part of the total system volume, meaning that emulsification has failed, giving coexistence of bulk A and B phases. Likewise, if  $2K+\bar K <0$, the bending energy of a sphere with $R \ll R_0$ is negative and (modulo anharmonic corrections) almost independent of $R$. For similar reasons one then expects a proliferation of tiny spheres containing only a negligible amount of A: again, emulsification has failed. 

Successful emulsification thus requires both $\bar K<0$ and $2K+\bar K > 0$. This is not sufficient, however. For values of $R_0$ that are less than a thermal persistence length $\xi_K$ introduced below in (\ref{xiK}), we can neglect the entropic shape fluctuations of the interface and need only minimize $F_{\rm bend}$ at fixed total area $\S$ and fixed $V_{\rm in}/V = \Phi$ to find the thermodynamic equilibrium state of the system. 
The computation of $F_{\rm bend}$ for spheres, cylinders and lamellae (infinite flat sheets) is straightforward and for simplicity we limit attention only to these geometries. We have
\begin{equation}
F_{\rm bend} = 4\pi\left(2K\left[1-\frac{R}{R_0}\right]^2+\bar K\right)
\end{equation}
for a sphere of radius $R$; 
\begin{equation}
F_{\rm bend} = \frac{\pi K L}{R_0}\left[1-\frac{2R}{R_0}\right]^2
\end{equation}
for a cylinder of radius $R$ and length $L$; and $F_{\rm bend} = 2KA/R_0^2$ for a flat sheets of area $A$. 
It is also a simple exercise to show, by equating the enclosed volume to $V\Phi$ and the interfacial area to $\S$, that the droplet size $R$ for spheres is
\begin{equation}
R_s = \frac{3\Phi v_s}{\phi_s\Sigma}.
\end{equation}
If $R_s<R_0$, then (modulo a transition to cylinders at small negative $\bar K$) the droplet phase is stable. However, if surfactant is removed, or the internal phase volume fraction $\Phi$ is increased, to the point where $R_s$ exceeds $R_0$, the system is not obliged to pay the additional bending cost of having droplets larger than their preferred curvature radius. Instead, the droplets remain of size $R_0$, and coexist with a bulk phase of leftover A-rich fluid: emulsification has once again failed \citep{Safran03,Safran83}. 
Given the various routes to emulsification failure described above, formulators generally aim to avoid any intrinsic tendency to strong curvature of the surfactant film.

\subsubsection{Persistence length and thermal softening of $K$}
Reflecting that strategy, we now set $R_0 = \infty$ so a flat interface is preferred. The bending energy can then be evaluated for small fluctuations in shape described by a height field $h(x,y)$ above a flat reference plane. One finds 
\begin{equation}
F_{bend} = \int \left(\frac{K}{2}(\delsq h)^2\right) dx\,dy \simeq \frac{K}{2}\sum_q q^4|h_q|^2, \label{monge}
\end{equation}
where in the first expression $\delsq $ is defined with respect to the $x$ and $y$ coordinates and in the second we have taken a Fourier transform of the height field. 
From this it is a simple exercise in statistical mechanics \citep{Safran03} to show that (with $r^2 = x^2+y^2$)
\begin{equation}
\langle|\del h(r)-\del h(0)|^2\rangle \propto \frac{k_BT}{2\pi K}\ln\left(\frac{r}{\ell}\right),
\end{equation}
where $\ell$ is a microscopic cutoff length of order the film thickness and $\langle\cdot\rangle$ denotes a thermal average. When this logarithmic deviation in orientation becomes large, the expansion underlying (\ref{monge}) breaks down. 

Generally we only want to know the interface's coarse-grained properties on some scale $\lambda$ set by, for instance, the size of emulsion droplets. Under coarse graining we replace an entropically wiggly interface by a smooth one on the scale $\lambda$. By carefully integrating out the thermal undulations, one can show \citep{David04} that their perturbative effect is to soften the elastic constant:
\begin{equation}
K_{\rm eff}(\lambda)  = K - \frac{3k_BT}{4\pi}\ln\left(\frac{\lambda}{\ell}\right).
\label{Keff}\end{equation}
Thus there is very little resistance to bending at scales beyond a `persistence length'
\begin{equation}
\xi_K \simeq \ell \exp\left[\frac{4\pi K}{3 k_BT}\right]
\label{xiK}.\end{equation} 
Note that $\xi_K$ is exponentially dependent on $K$: for $\ell = 1$ nm, $\xi_K \simeq 1\mu$m when $K = 1.65k_BT$. For $K= 3k_BT$, we already have $\xi_k\ge 300\mu$m; and $\xi_K$ is irrelevantly large, for our purposes, once $K$ is much larger than this.

\subsection{Bicontinuous microemulsions} 
\label{Bicon} In so-called `balanced' microemulsions, the spontaneous curvature is tuned to be small, so that $R_0 \gg \xi_K$. For simplicity we set $R_0 \to \infty$ as above. We also assume $\ell \ll \xi_K \ll 100\mu$m, so that the entropy of the interface (including the renormalization of $K$) cannot be ignored. Assuming $\Phi$ of order 0.5 (roughly symmetric amounts of A- and B-rich fluid) we can introduce a structural length scale $\lambda$ which is then set by $\phi_s$. Specifically for a lamellar phase, comprising a 1D stack of alternating layers of A and B with relatively flat interfaces between these, one has
a layer spacing $\lambda$ between adjacent surfactant films set by
$\phi_s \simeq \frac{\ell}{\lambda+\ell}$.
When $\lambda\simeq \ell$ the system has no option but to fill space with flat parallel layers. As $\phi_s$ is reduced ($\lambda$ raised) the layer spacing $\lambda$ becomes comparable to $\xi_K$. For $\lambda/\xi_K \le 1/3$ (or so), the lamellar phase fluctuates but remains stable. 

On the other hand, if $\phi_s$ is then decreased further so that $\lambda \simeq \xi_K$, these layers melt into an isotropic phase comprising (for $\Phi\simeq 0.5$) bicontinuous domains of A and B fluids separated by a fluctuating surfactant film. This is the {\em bicontinuous microemulsion} and effectively represents a thermodynamic route to prevent coarsening of the transient bicontinuous structures encountered in section~\ref{PSK} above. If $\Phi$ now deviates strongly from $0.5$, then (just as found there) the structure depercolates, forming a droplet phase. This differs from the one discussed above for $R_0\ll\xi_K$, since this one is stabilized by entropy and fluctuations, not by a preferred curvature of the droplets. 
Theories of the bicontinous microemulsion were initiated by \cite{DeGennes82}. Some of these theories use coarse grained lattice models in which fluid domains are placed at random on a lattice of some scale $\xi$; the bending energy and area of the resulting interface can be estimated and used to calculate a phase diagram \citep{Andelman87}. One specific feature is the appearance of three-phase coexistence in which a  `middle phase' microemulsion coexists with excess phases of both oil and water: this roughly corresponds to a `double-sided' emulsification failure in which both oil and water are expelled.

\subsection{The sponge phase and vesicles} 
Closely related to the bicontinuous microemulsion is the `sponge phase'. This arises when there is a huge phase volume aysmmetry between A-rich and B-rich fluids, but only in case-2 systems where the surfactant has a molecular preference to form a flat film rather than highly curved structures such as micelles. The interfacial structure that forms spontaneously at $c_s = \tilde c$ is, in the almost complete absence of B, necessarily now a bilayer with A (usually water) on both sides and a thin B layer in the middle. The lamellar state now consists of flattish bilayers alternating with A domains; for small volume fractions of bilayer such that their separation exceeds their persistence length, this structure melts, just as described above for the microemulsion. The result is subtly different though: we now have a bilayer film that separates two randomly interpenetrating domains containing the same solvent A. This is called the `sponge phase'. It is not directly useful for A/B emulsification, since the minority B phase has negligible phase volume. However, the sponge phase does have remarkable phase transitions associated with the `in-out' symmetry between the two A domains, which can be broken spontaneously \citep{Huse88,Roux92}.  When the symmetry is strongly broken, one has discrete droplets of A separated from a continuous A phase by a bilayer. This can be viewed as a thermodynamically stabilized A-in-A emulsion (typically water-in-water), which can be useful for encapsulation. Such structures, called vesicles, can also exist in the complete absence of B, so that the bilayer contains only surfactant.

Similar vesicles can also be formed from lipids, which are biological molecules closely related to surfactants. The main difference is that for lipids $\tilde c$ is {\em extremely} small compared to typical values for surfactants. This renders them effectively insoluble in water, since any attempt to increase $c_s$ above this small value leads instead to the creation of more interface on which the lipid resides (typically organized as a lamellar phase). This insolubility means that although in local equilibrium $\gamma_0$ is effectively zero, there is no rapid equilibration of the surface area per molecule $\Sigma$ by molecular exchange between interface and bulk. As a result, if a lipid vesicle is mechanically stretched a tension soon develops. Such tension can also arise by swelling the vesicle to an inflated spherical shape rather than a relaxed, floppy shape of lesser volume. Only in the absence of such stretching can a lipid bilayer can be described by the bending free energy (\ref{bending}) in which typically $R_0=\infty$ because there is no preferred curvature by symmetry; in this case there are large shape fluctuations so long as $K/ k_BT$ is not large.  The spontaneous curvature radius $R_0$ can however be finite if the composition of the lipid bilayer is different on its two faces. This is common in biology where a bilayer of mixed lipids separates the interior and exterior of a cell. Note that, because of their insoluble character, the global distribution of lipids to form vesicles is almost never in equilibrium: the size of each vesicle is set by the amount of lipid present at its surface, not by thermodynamics. 

\section{Particle-stabilized emulsions}\label{PE}
A typical soluble surfactant has an energy of attachment to the A-B (oil-water) interface of between $5k_BT$ and $15k_BT$. This is high enough to alter interfacial properties but low enough to maintain local equilibrium by exchange of molecules with one or both bulk phases. Larger amphiphilic species range from lipids as just discussed, via diblock copolymers (in which two polymer chains of different chemistry are covalently bonded together) and globular proteins, to colloidal `Janus beads'. The latter are colloidal spheres, up to a micron in size, with surface chemistry that favours water on one hemisphere and oil on the other. For typical solid-fluid interfacial tensions ($\simeq 0.01$ Nm$^{-2}$) Janus beads have attachment energies of order $10^7k_BT$ or larger. Such species are adsorbed irreversibly at the A-B interface, in the sense that Brownian motion alone will almost never lead to detachment. 

\subsection{Colloidal particles at fluid-fluid interfaces}
\begin{figure}
	\centering
	\includegraphics[width=10cm]{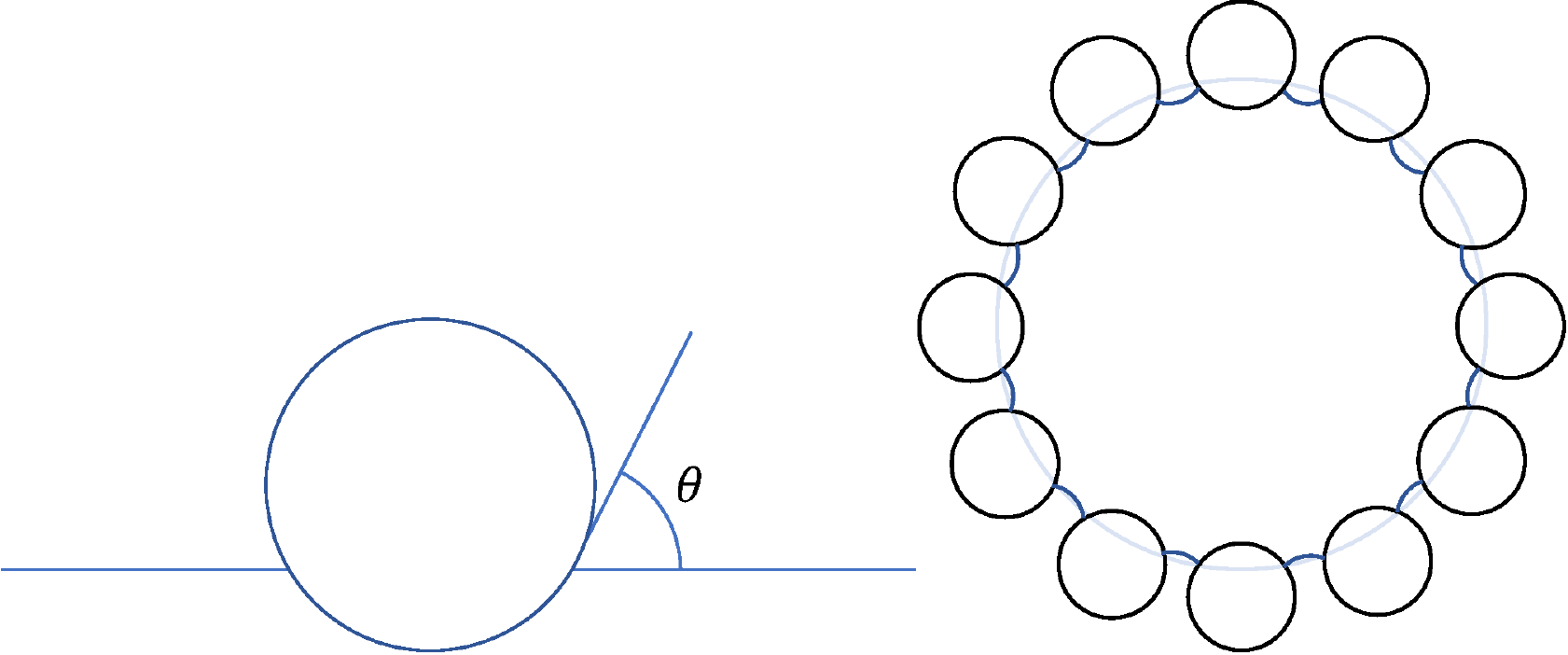}
	\caption[]{Left: Geometry of a partially wet colloidal particle at the fluid fluid interface; $\theta$ is the contact angle. This is usually measured through the polar phase, so the upper fluid is water and the lower oil as drawn here.
	Right: In light grey is the initial locus of a fluid interfaces into which are inserted the particles shown in darker grey. These are jammed in a 2D layer but have interstitial fluid regions as shown between them. If the volume of fluid in the droplet is now reduced, to maintain a fixed contact angle with particles that cannot move, the curvature of the interface is reversed to give the final fluid locus (black). This creates a negative Laplace pressure which can switch off the Ostwald process.}
	\label{ten}\label{eleven}
\end{figure}

Although Janus beads are often studied \citep{Lattuada11}, they are rarely used for emulsion stabilization, because it is much cheaper to use ordinary colloidal spheres of homogeneous surface chemistry. Perhaps surprisingly, so long as the colloid has roughly equal affinity to the two fluids A and B, interfacial attachment energies remain vastly larger than $k_BT$. The simplest case is when the two solid--fluid interfacial tensions, $\gamma_{SA}$ and $\gamma_{SB}$, are the same. The energy of such a  particle (of radius $a$) is then independent of where it resides, but the energy of the A-B interface is reduced by $\pi a^2 \gamma_0$ if the particle is placed there, because a disc of interface is now covered up by the colloid. Typically $\gamma_0 = 0.01$Nm$^{-2}$, giving an attachment energy of order $10^7k_BT$ for $a=1\mu$m and $10k_BT$ for $a=1$nm. Similar remarks apply for unequal solid-fluid tensions so long as the contact angle $\theta$, defined via $\gamma_0 \cos \theta = \gamma_{SB}-\gamma_{SA}$ for the case of partial wetting (figure \ref{ten}), is not too close to zero or $\pi$, whereupon the particle becomes fully wetted by one or other phase. 

Consider now placing a number of spherical colloidal particles on the surface of a spherical droplet of fluid A in fluid B, or vice versa. Each colloid can be accommodated with the required contact angle $\theta$ by cutting a small spherical cap out of the interface and slotting the particle into place there. The contact line is a perfect circle, as required for tangency at fixed angle to a sphere. To conserve the enclosed volume the droplet radius changes slightly, but it remains perfectly spherical. There is minimal change in Laplace pressure, and the interfacial energy is independent of {\em where} the colloids are placed. Accordingly there is no capillary force between them. These statements can only change if particles become jammed so they interact directly via particle-particle forces. 

The physics is very different for spherical colloids residing on a hypothetical cylinder of fluid A in B (or vice versa). It is not possible now to insert a spherical particle onto the surface of this cylinder at fixed contact angle $\theta$ unless the fluid interface becomes deformed. This deformation costs extra interfacial energy, and can be minimized by placing two particles close together. Accordingly, there is a capillary attraction between the colloids, which will have a strong tendency to aggregate. Similar arguments apply to nonspherical particles such as ellipsoids, even on flat surfaces \citep{Cavallaro11}. 

In all these cases, particle adsorption to the interface is thermally irreversible for particles bigger than a few tens of nanometres. This nonequilibrium interfacial physics allows stabilization not only of spherical droplets (classically known as `Pickering' or `Ramsey' emulsions) but also of more complex structures with frozen shapes maintained by a layer of interfacial particles clamped together by interfacial tension \citep{Binks06,Cates08}. Closely related structures can also be made with air as one of the two fluids \citep{Subramanian05}. Unless Janus particles are used \citep{Aveyard12}, these emulsions are {\em always metastable}: their minimum free energy state comprises bulk A and B phases, separated by a flat interface, with as much of this interface as possible covered by particles, and the remainder distributed randomly in one or both solvents. However the vast detachment energies allow the metastable state to survive almost indefinitely in many cases.

\subsection{Resistance to coalescence and Ostwald ripening}
One route to Pickering emulsions is to create numerous small droplets by applying a strong flow that mixes fluids A and B. The flow sweeps particles onto the interface whose initial surface area is much larger than they can cover. Coalescence initially proceeds as normal (perhaps assisted by maintaining a lower flow rate).  Coalescence decreases the surface-to-volume ratio of droplets at fixed numbers of attached colloids. This proceeds until the surface particle density is high enough to prevent further coalescence. This requires a coverage of comparable to, but somewhat below, that of a densely packed 2D amorphous film. Hence the resulting particle layer is usually not  jammed, and droplets can relax to a spherical shape. In some cases, though, for instance if droplets remained stretched by flow as they coalesce, the final droplet is arrested in an aspherical jammed state, with particles clamped in position by interfacial tension \citep{Cates08}.  

For both spherical and aspherical structures, these clamped layers offer very strong stability against coalescence. They also resist Ostwald ripening, for the following reason. Recall that a close packed monolayer of particles (whether ordered or amorphous) can be placed on the surface of a spherical fluid droplet of radius $R$ (say) without altering its interfacial geometry. Imagine such a droplet with the particles just in contact with one another. There is still a fluid-fluid interface at the interstices between particles, and this has the same curvature, and hence Laplace pressure $\gamma_0/R$, as the original drop. 

Suppose now that this droplet is in diffusive equilibrium with one or more larger ones. According to the Ostwald mechanism, it will start to shrink. However, if the particles are already in contact they cannot follow the droplet surface inwards as this happens. Moreover, each particle demands an unchanged contact angle with the interface, which is effectively now pinned to the particle layer. It is easy to see (figure \ref{eleven}) that even a small loss of volume of the droplet under these conditions will reduce the Laplace pressure to zero and then negative values. Once zero is reached, the droplet is fully resistant to Ostwald ripening. The mechanism is similar to that described in section \ref{ts} using a trapped species; indeed, due to their high detachment energies, interfacial particles are effectively such a species.

\subsection{Morphologies of particle-stabilized emulsions}
\begin{figure}
	\centering
	\includegraphics[width=12cm]{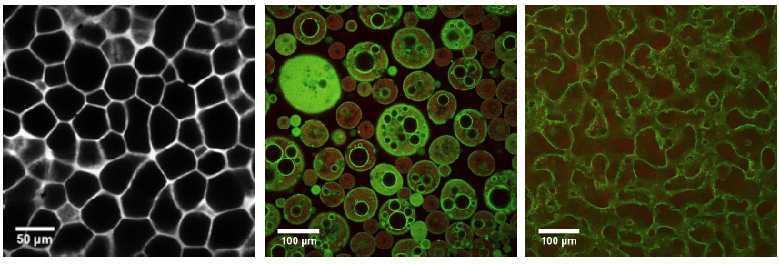}
	\caption[]{Various particle-stabilized interfacial structures between fluids A and B, as imaged by confocal microscopy (which effectively views a thin slice through the material). Left: A biliquid foam with thin B films separating polyhedral A droplets; the bright regions are layers of fluorescently labelled particles and appear as lines or regions according to whether these lie oblique to the confocal plane or within it. Centre: A multiple emulsion.
The brightest regions (light green in online colour version) are again particle-rich. Fluid A is dark grey and fluid B is mid-grey (dull red in online colour version). Right: A bijel; the fluorescent labelling is similar to the multiple emulsion except that fluid B (dull red in online colour) is now the darker of the two in grey scale. (Images courtesy of Paul Clegg.)}
	\label{twelve}
\end{figure}

Alongside conventional spherical emulsion droplets, with particle stabilization a number of alternative morphologies are available (figure \ref{twelve}). First,
drainage or centrifugation of a Pickering emulsion leads to a compressed foam structure that is like the biliquid foams discussed previously. These can be very stable thanks to the combined resistance to coalescence and Ostwald ripening provided by the particles. 
Second, simple manual agitation of binary immiscible solvents containing partially wettable particles often results in droplet-within-droplet structures known as multiple emulsions. Such structures require stability against both coalescence and Ostwald ripening for A droplets in B and B droplets in A simultaneously. This is relatively difficult for surfactant formulations but seemingly quite easy with particle-stabilized ones \citep{Clegg16}.

A third interesting morphology is that of ``bijels'', or bicontinuous interfacially jammed emulsion gels, which are metastable analogues of the bicontinuous microemulsion: a particle layer resides at the interface between interpenetrating domains of A and B. This structure was predicted first computationally by Model H simulations with added colloids \citep{Stratford05}; it was confirmed in the laboratory by \cite{Herzig07}. In a bijel, the interfacial film of non-detachable particles is clamped by tension into a 2D jammed layer, which imparts solidity to the whole 3D structure. This robustness can be improved further by having an interaction potential between particles with a steep barrier and then an attractive minimum at short distances. The interfacial tension pushes particles over the barrier creating a strongly bonded interfacial film \citep{Sanz09}.

One recipe for making a bijel is to choose a fluid pair A+B that are miscible at high temperature \citep{Cates08}. The colloidal particles are then dispersed within the single-phase mixture. On dropping the temperature, the fluids separate and particles are swept onto the interface. The coarsening process arrests when a jammed monolayer is formed, creating the bijel. The final structural domain size obeys $L\simeq a/\phi_p$ with $a,\phi_p$ the particle size and volume fraction, and the elastic modulus of the solid bijel scales as $G\sim\gamma_0/L$. Bijels are currently being explored for various applications in materials design \citep{Lee13}. 

%
%

\section{Liquid-crystalline emulsions \label{sec:LC}}

\begin{figure}
\centering
\includegraphics[height=3cm]{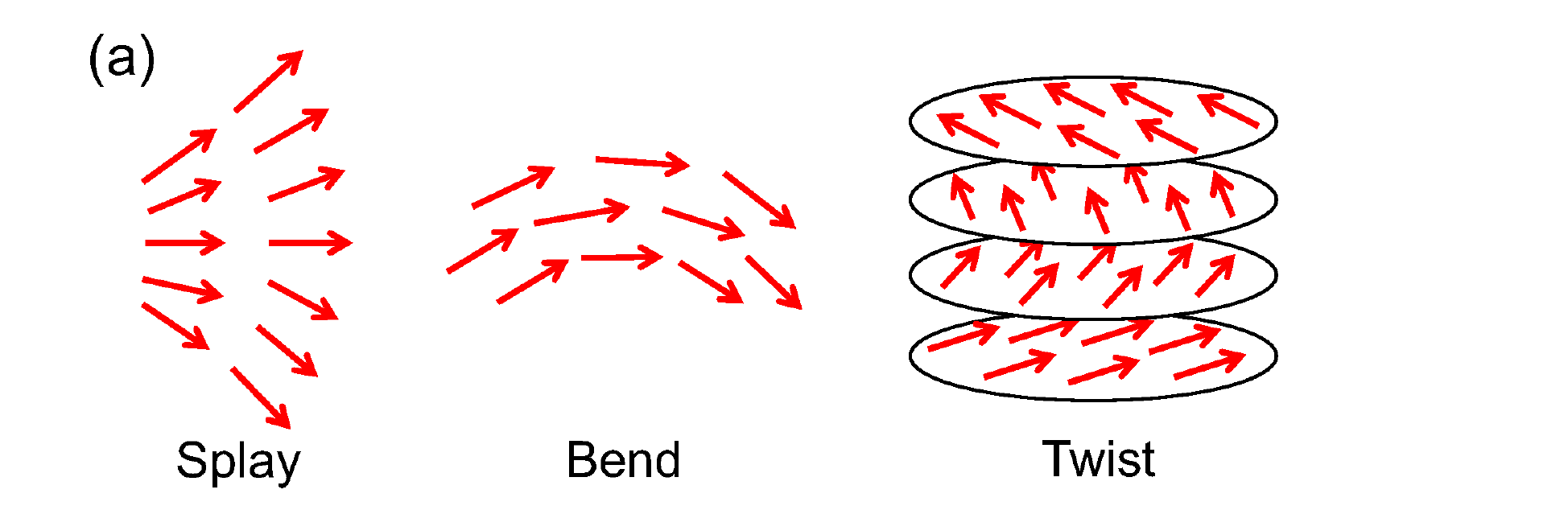}
\includegraphics[height=3cm]{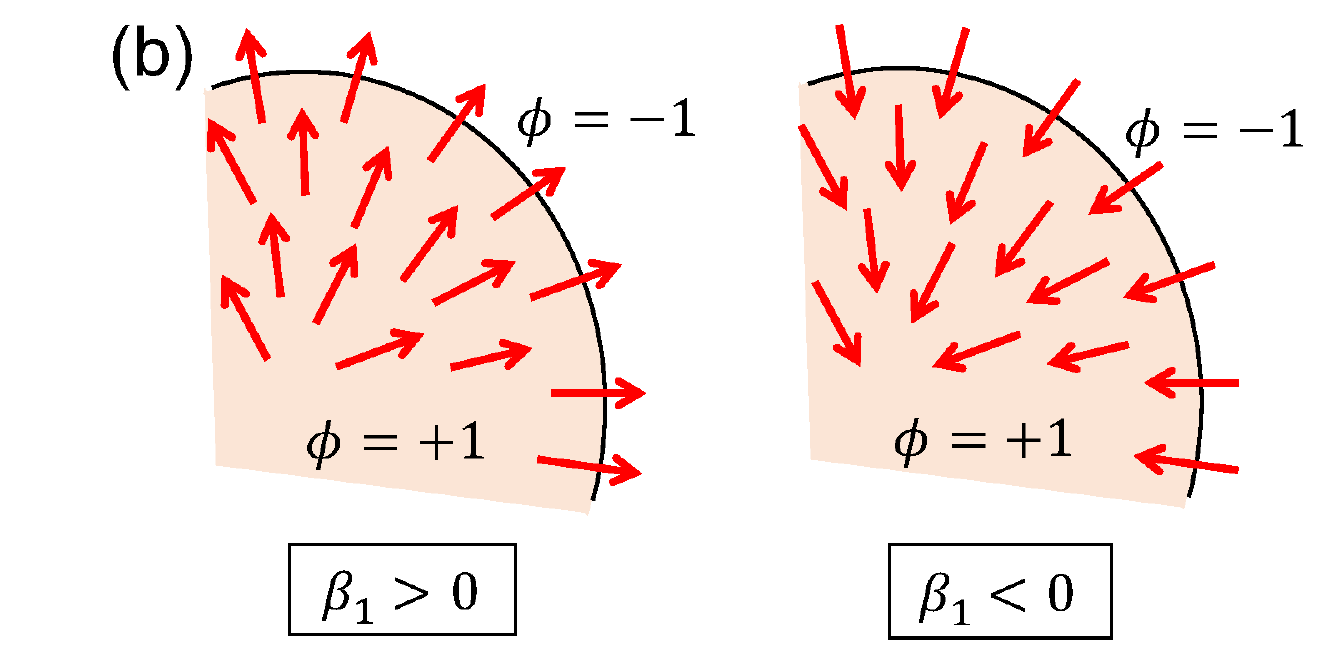}
\includegraphics[height=3cm]{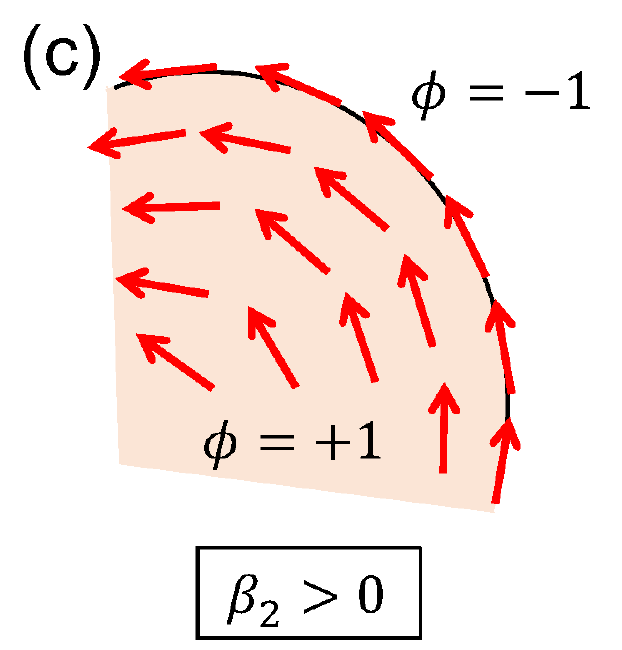}
\caption{(a) Elastic modes in polar liquid crystals: splay, bend and twist.
Red arrows indicates the average direction of the particles, $\p(\r)$.
In the case of nematic liquid crystals, the vector field $\p$ is replaced by a headless unit vector $\nhat$.
(b) The effect of an anchoring term $\beta_1$ in the free energy;
$\p$ tends to align perpendicularly outwards ($\beta_1>0$) or inwards ($\beta_1<0$) at the interface.
(c) The effect of another anchoring term $\beta_2>0$ in the free energy;
$\p$ tends to align parallel to the interface.
The black line indicates the interface between the isotropic phase ($\phi=-1$) and the liquid-crystalline phase ($\phi=+1$).}
\label{fig:elastic-modes}
\end{figure}

Nematic and polar liquid crystals are classes of materials with long range orientational order but without positional order.
Normally they are composed of rod-shaped particles which develop such order at high enough density. (A third class of liquid crystal are smectic phases, comprising a stack of fluid sheets with periodic order in one dimension; we do not address these here, but note that the lamellar phase referred to in section \ref{Bicon} above is an example.) 
Nematic and polar liquid crystals can flow like liquids but at the same time retain an elastic, solid-like response to deformations in their orientational order \citep{DeGennes02}.
The three independent elastic modes are splay, bend and twist deformation (see figure \ref{fig:elastic-modes}(a)).

Recently, there has been considerable interest in multiphase mixtures of liquid crystals and isotropic fluids, collectively termed liquid-crystalline emulsions.
This includes liquid-crystalline droplets in a bulk isotropic fluid or isotropic droplets in bulk liquid crystals.
Theoretically, such system can be described by two order parameters.
The first is a scalar composition field $\phi(\r,t)$ which distinguishes the liquid-crystalline fluid (say $\phi=+1$) from the isotropic fluid (say $\phi=-1$).
The second order parameter describes the orientational order of the liquid-crystalline phase.
As mentioned in section \ref{OPs}, there are two basic types of orientational order for rodlike particles: \emph{polar} order, described by a vector field $\p(\r,t)$; and \emph{nematic} order, described by a second rank traceless symmetric tensor field ${\Q}(\r,t)$. 

Polarity stems from a head-tail asymmetry of the particles, described by a unit vector $\nuhat$ which is embedded in each particle and points from the tail to the head. The order parameter $\p(\r,t)$ is defined in (\ref{pdef}) to be the average of these unit vectors inside some mesoscopic volume. In the isotropic phase, particles point in random directions such that $|\p|=0$, whereas
in a polar phase, particles point in a preferred direction (on average) such that $0\le |\p| \le 1$. 
Molecular liquid crystals rarely form a macroscopic polar phase because such molecules typically have a finite electric and/or magnetic dipole moment. In a polar phase, even if small, this dipole moment gives rise to macroscopic ferroelectric or ferromagnetic order whose long-range field energies tend to destabilize the phase at large scales \citep{DeGennes02}.
On the other hand, many examples of polar liquid crystals are not molecular but biological in nature, for example, 
dense swarms of rod-like bacteria such as \emph{E. coli}.
Here the bacteria also swim in the direction of $\nuhat$ and force the system to be out of equilibrium.
We return to such `active' liquid crystals in section~\ref{active}. 
Polar emulsions (especially active ones) have been the subject of several theoretical and numerical studies in recent years. For instance, as we shall see in section~\ref{sec:active-LC}, a droplet of polar active fluid on a solid substrate can capture some of the physics of cell crawling \citep{Ziebert13, Tjhung15}.

In nematic liquid crystals, the particles are again oriented along a preferred axis, but lie
parallel or anti-parallel to that axis with equal probability such that 
$\left<\nuhat\right>$ is zero but $\left<\nuhat\nuhat\right>$ is not zero.
The nematic order parameter $\Q(\r,t)$ is then defined as in (\ref{Qdef}) as a second rank tensor that is traceless and vanishes in the isotropic phase. Note that $\Q$ and $\p$ are linearly independent: while typically $\Q$ does not vanish in a polar phase it can do so. (A concrete example is given in section \ref{ASILC} below.)
We define the director field $\nhat(\r,t)$ as a unit eigenvector of ${\Q}$ corresponding to its eigenvalue of largest magnitude. 
Here $\nhat$ and $-\nhat$ contain the same information about the average orientation of the particles, and pictorially $\nhat$ is often represented as a ``headless'' vector field. For uniaxial nematics (in which rotational symmetry about this preferred axis is unbroken) we can then write
${\Q}(\r,t)=S(\r,t)(\nhat\nhat-{\I}/{d})$ where the local eigenvalue $S$ tells us the strength of nematic order. For rodlike particles with a preference for parallel alignment this is positive ($0\le S \le 1$)
but $S$ can become negative in some situations (e.g. close to walls) where it describes a `pancake-like' rather than `sausage-like' orientational distribution function. Importantly, in three dimensions there is no symmetry relating states of opposite $S$. This causes the phase transition from isotropic to nematic to be generically discontinuous \citep{DeGennes02}.
Emulsions of a nematic phase in an isotropic fluid, or vice versa, can give rise to interesting and exotic new states of organization.
For instance, isotropic droplets can form long chains which are stabilized by the surrounding nematic \citep{Loudet00}); they can also
form a hexagonal crystalline lattice \citep{Nazarenko2001}. These self-assembling properties can be attributed to the formation of topological defects (described below) around each droplet which mediate new interactions between them \citep{Poulin97}.

\subsection{Free energy of liquid-crystalline emulsions}

In a polar liquid-crystalline emulsion,
the hydrodynamic variables are the composition variable $\phi(\r,t)$, the polarization $\p(\r,t)$ and the fluid velocity $\v(\r,t)$.
These can be defined such that $\phi \simeq 1$ in the bulk liquid crystal where $|\p|>0$, and $\phi\simeq -1$ in the bulk isotropic fluid where $|\p| = 0$.
The free energy of such a system can be written as a sum of two contributions: $F[\phi,\p] = F_{\phi}[\phi] + F_{\p}[\phi,\p]$.
The first is the contribution from a simple binary fluid, similar to (\ref{functional})
\begin{equation}
F_{\phi}[\phi] = \int{\mathbb{F}}_\phi\,\d\r = \int\left( -\frac{a}{2}\phi^2 + \frac{a}{4}\phi^4 + \frac{\kappa}{2}|\nabla\phi|^2 \right) \d\r.
\label{eq:F-phi}
\end{equation}
Here $a>0$ and $\kappa>0$ are chosen such that $F$ is minimized by bulk phase separation into states with $\phi\simeq\pm1$.
The second contribution to the free energy stems from the liquid crystallinity which can be written as \citep{DeGennes02,Tjhung12}
\begin{equation}
\begin{split}
F_{\p}[\phi,\p] = \int{\mathbb{F}}_\p\,d\r =\int\Big( &\frac{\gamma(\phi)}{2} |\p|^2
			+ \frac{\alpha}{4} |\p|^4
			+ \frac{K_1}{2} ( \del\bdot\p )^2
			+ \frac{K_2}{2} ( \p\bdot\del\times\p )^2 \\
			&+ \frac{K_3}{2} | \p\times\del\times\p |^2 
		    + \beta_1 (\del\phi)\bdot\p
		      + \beta_2 ((\del\phi)\bdot\p)^2
			\Big) \d\r.
\end{split}
\label{Fp}
\end{equation}
Here $\gamma(\phi)$ is a thermodynamic parameter which controls the isotropic to polar transition:
$\gamma(\phi)<0$ in the polar phase and $\gamma(\phi)>0$ in the isotropic phase. (This notation is conventional and should be distinguishable by context from our previous use of $\gamma$ to denote interfacial tension.) For simplicity we can
assume $\gamma(\phi)=-\alpha\phi$ to obtain a bulk polar phase whenever $\phi>0$ and an isotropic phase ($|\p| = 0$) when $\phi<0$. With this choice, $|\p| = 1$ when $\phi = 1$, in effect setting the units for $\p$. We require the quartic coefficient $\alpha$ to be positive for thermodynamic stability.
The gradient terms involve $K_1$, $K_2$, and $K_3$ which are the splay, twist and bend elastic constants (all positive); see figure~\ref{fig:elastic-modes}(a).
In the literature, one often finds the approximation $K_1=K_2= K_3\equiv K$,
such that all three elastic terms combine into the simpler form $\frac{K}{2}(\partial_i p_j)^2$.
The remaining terms in (\ref{Fp}) describe anchoring effects. The $\beta_1$ term favours perpendicular anchoring of $\p$ at the droplet interface (see figure \ref{fig:elastic-modes}(b)): if $\beta_1>0$, $\p$ tends to point outwards (from the polar to the isotropic phase) whereas if $\beta_1<0$, $\p$ tends to point inwards. Finally, the $\beta_2$ term promotes parallel anchoring of $\p$ at the interface (see  figure \ref{fig:elastic-modes}(c)).

For the case of nematic (rather than polar) emulsions, the second rank tensor $\Q(\r,t)$ takes the place of $\p(\r,t)$ as a hydrodynamic variable, alongside the composition $\phi(\r,t)$ and the fluid velocity $\v(\r,t)$.
The free energy can be written as $F[\phi,\Q]=F_{\phi}[\phi]+F_{\Q}[\phi,\Q]$,
where $F_{\phi}$ is of the same form as Eq.~(\ref{eq:F-phi}).
$F_{\Q}$ comprises the so-called Landau--de Gennes free energy functional for bulk nematics \citep{DeGennes02}, augmented with appropriate couplings to $\phi$. In three dimensions this reads
 \citep{Sulaiman06}
\begin{equation}
\begin{split}
F_{\Q}[\phi,{\Q}] = \int\bigg\{ &\frac{A_0}{2}\left(1-\frac{\gamma'(\phi)}{3}\right)Q_{ij}Q_{ij} 
											 - \frac{A_0}{3}\gamma'(\phi)Q_{ij}Q_{jk}Q_{kl} 
											 + \frac{A_0}{4}\gamma'(\phi)(Q_{ij}Q_{ij})^2 \\
											 &+ \frac{K}{2}(\partial_iQ_{jk})^2 
											 + \beta_0 (\partial_i\phi)Q_{ij}(\partial_j\phi) \bigg\}\d\r.\end{split}
\label{eq:F-Q}
\end{equation}
where for simplicity we have taken a single elastic constant $K>0$. The thermodynamic parameter $A_0>0$ is a scale factor for bulk free energies, while $\gamma'(\phi)$ controls the isotropic to nematic transition; the nematic phase is stable for 
$\gamma'>2.7$ and the isotropic phase stable for $\gamma'<2.7$. This transition is discontinuous (thanks to the term in \ref{eq:F-Q} cubic in $\Q$) with a finite hysteresis window within which the phase of higher $F_\Q$ remains metastable.
(These statements can be checked by parameterizing ${\Q}=\lambda(\hat{z}\hat{z}-{\I}/{3})$ and 
then minimizing $F_{\Q}(\lambda)$ with respect to $\lambda$.)
Choosing $\gamma'(\phi)=2.7+\gamma_1\phi$ with some positive constant $\gamma_1$ ensures a nematic phase at compositions $\phi>0$ and isotropic phase at $\phi<0$.
For nematics, the physics of orientational anchoring at the surface of a droplet can be captured by a single term, proportional to $\beta_0$.
For $\beta_0<0$, the director field $\nhat$ will tend to align perpendicular to the interface, whereas for $\beta_0>0$,
$\nhat$ will tend to align parallel to the interface.
This is similar to figure~\ref{fig:elastic-modes}(b) and (c) respectively except that one must replace the vector with a headless vector in each case.

\begin{figure}
\centering
\includegraphics[height=3cm]{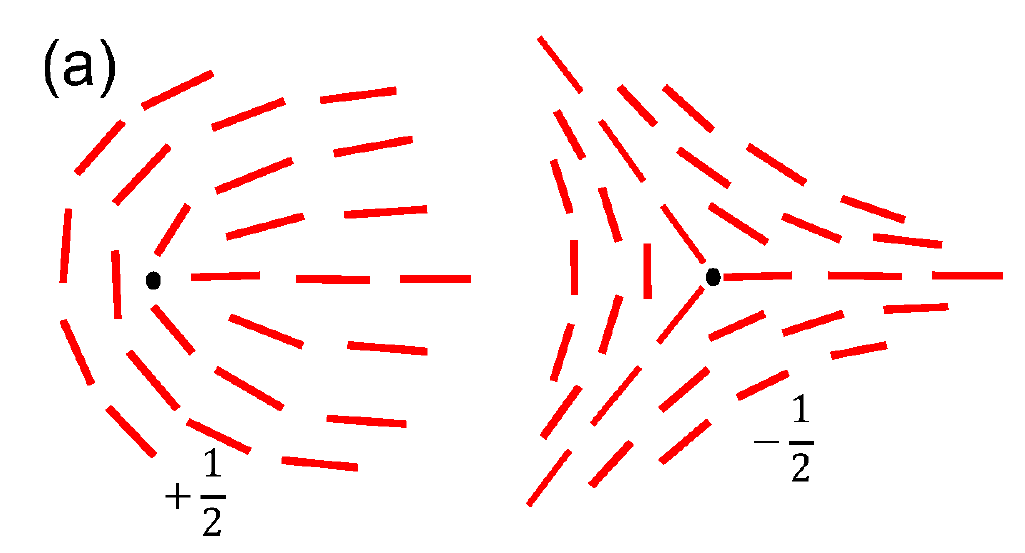}
\includegraphics[height=3cm]{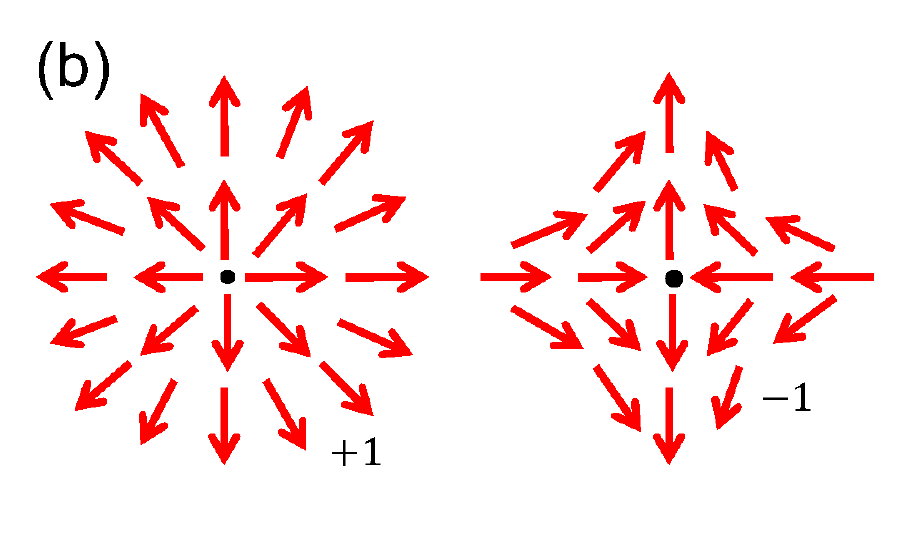}
\includegraphics[height=3cm]{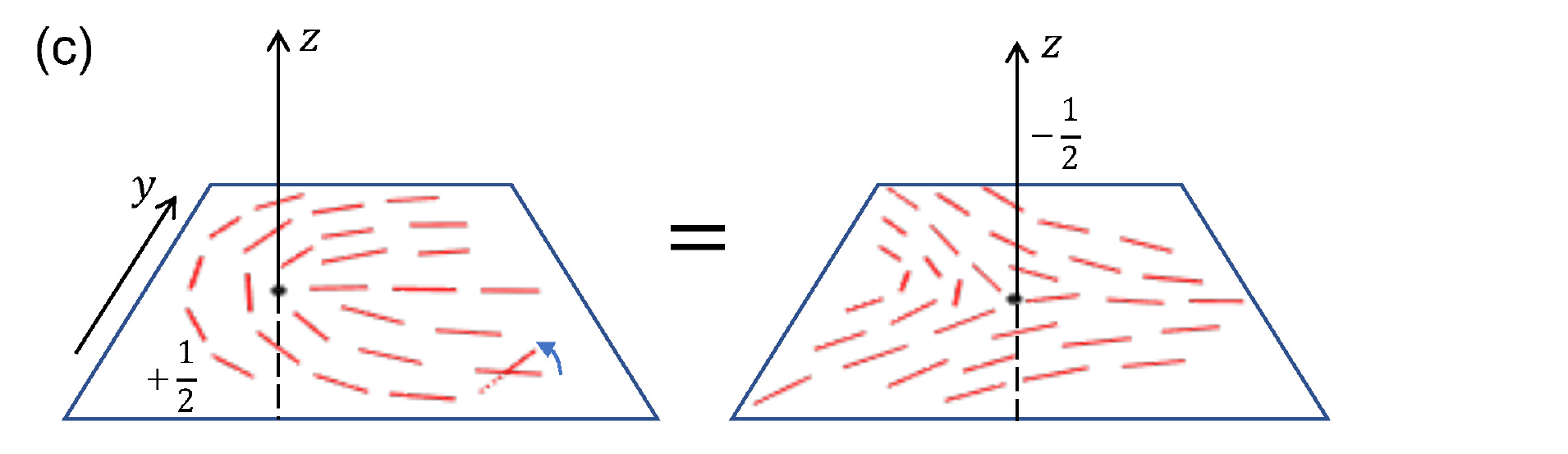}
\caption{(a) Topological defects in 2D nematic liquid crystals.
(b) In polar liquid crystals half-integer defects are forbidden.te
(c) In 3D, $+\frac{1}{2}$ and $-\frac{1}{2}$ point defects of the nematics become line defects.
Furthermore, $+\frac{1}{2}$ and $-\frac{1}{2}$ line defects are then topologically equivalent (one can see this by flipping each molecule $180^\circ$ about the $y$--axis) \citep{Chaikin95}.}
\label{fig:defects}
\end{figure}

\subsection{Topological defects}
If we ignore the anchoring term $\beta_1$ in the polar free energy $F_\p$,
we notice that the free energy is invariant under a \emph{global} inversion $\p\rightarrow-\p$.
On the other hand for nematics, the free energy is invariant under a \emph{local} inversion $\nhat\rightarrow-\nhat$.
As long as we are in a region without topological defects -- which are places where $\p$ or $\nhat$ are undefined -- there should be no distinction between a global or local symmetry since any attempt to make different sign choices
in different neighbourhoods will lead to $\p$ or $\nhat$ being undefined wherever these neighbourhoods meet.
In practice, therefore, in defect-free regions the static and dynamic properties for both $\p$ and $\nhat$ are found to be very similar, so long as we ignore terms which break the $\p\leftrightarrow-\p$ symmetry in $F_\p$ such as the $\beta_1$ term in (\ref{Fp}).

However, differences arise when we do have topological defects. Such a defect is present when,  if the state of order in some region is mapped onto a state of uniform orientation, this map cannot be made smooth everywhere in real space \citep{Chaikin95}.
In 2D, the lowest order topological defects for nematics ($\nhat$) are pointlike and have `topological charge' $+\frac{1}{2}$ or $-\frac{1}{2}$, as shown in Fig.~\ref{fig:defects}(a). This charge denotes the number of full turns of the director along a path that makes one full turn around the defect. Because the vector $\nhat$ is headless, a rotation comprising a whole number of half-turns in either direction brings it back to the same state. (Defects of charge $\pm n/2$ with $n>1$ are also possible, but in practice these rapidly dissociate for energetic reasons into $n$ half-integer defects of the appropriate sign.)
On the other hand, the lowest order defects for polar fluids have charge $\pm 1$ since a full rotation of $\p$ is needed to recover the same state, as shown in figure \ref{fig:defects}(b). (Again, defects of charge $\pm n$ with $n>1$ dissociate to reduce the elastic energy.)
In both the polar and nematic cases, defects of the same sign repel one another whereas those of opposite charge attract and then annihilate. Conversely, during the isotropic to liquid crystal transition, triggered typically by a quench which alters $\gamma$ or $\gamma'$, pairs of defects with opposite signs are created.

In 3D, the $\pm\frac{1}{2}$ point defects of the nematics become line defects, as shown in figure \ref{fig:defects}(c).
Furthermore, in 3D, $+\frac{1}{2}$ and $-\frac{1}{2}$ line defects are topologically equivalent.
To see this, consider the $+\frac{1}{2}$ line defect shown in figure \ref{fig:defects}(c) left.
We can then continuously rotate the director field everywhere through $180^\circ$ about an axis perpendicular to $\hat{\bf z}$ (in this example, the $y$ axis) to get the $-\frac{1}{2}$ line defect shown in Fig.~\ref{fig:defects}(c) right. Similar reasoning establishes that any two defects whose charge differs by an integer are equivalent, so that all integer defect lines are equivalent to no defect at all, and all half integer defects are equivalent to each other. Thus in 3D nematics there is only one type of line defect; these are called `disclinations'. Any two disclination lines can annihilate. Note that disclinations can also form closed loops \citep{Chaikin95}.

For polar liquid crystals there are no line defects in 3D. This is because cylindrical versions of the structures shown in figure \ref{fig:defects}(b) can be converted into a defect-free state by smoothly rotating all the arrows to point out of (or into) the page. The basic defects are instead the obvious 3D equivalents of the $\pm1$ point defects shown in~\ref{fig:defects}(b), left and right. These are called the radial hedgehog and hyperbolic hedgehog defects, respectively \citep{Lubensky98}.
These two types of point defects are also possible in nematics, but are then topologically equivalent to one another, and also equivalent, at distances much larger than its radius, to a closed disclination loop. Note that their topological equivalence does not mean that two structures are {\em freely} interconvertible; energetic considerations often favour one over the other.

We also note in passing that at the centre of a line defect in a polar liquid crystal, $\p$ passes through zero 
such that its direction is undefined. For a 2D nematic, $\Q$ likewise becomes zero at the defect core, meaning that the medium is isotropic there and $\nhat$ is undefined. However at the core of a disclination line in 3D, $\nhat$ is undefined not because $\Q$ is zero, but because its two largest eigenvalues are degenerate so that the orientational distribution of molecules is isotropic in the plane normal to the defect line.

\subsection{Defects in and around emulsion droplets}

\begin{figure}
\centering
\includegraphics[height=4.5cm]{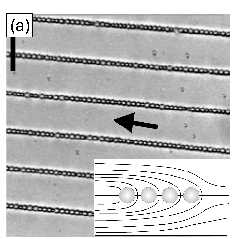}
\includegraphics[height=2.5cm]{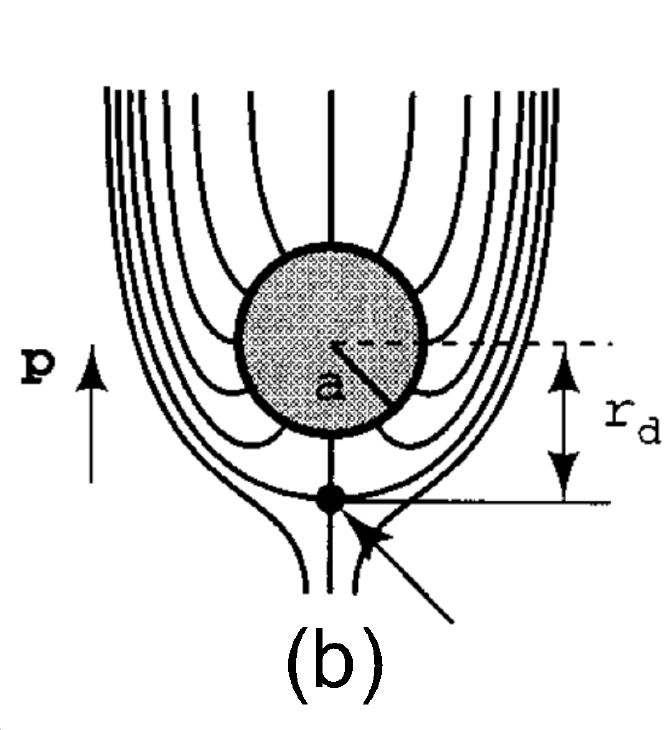}
\includegraphics[height=2.5cm]{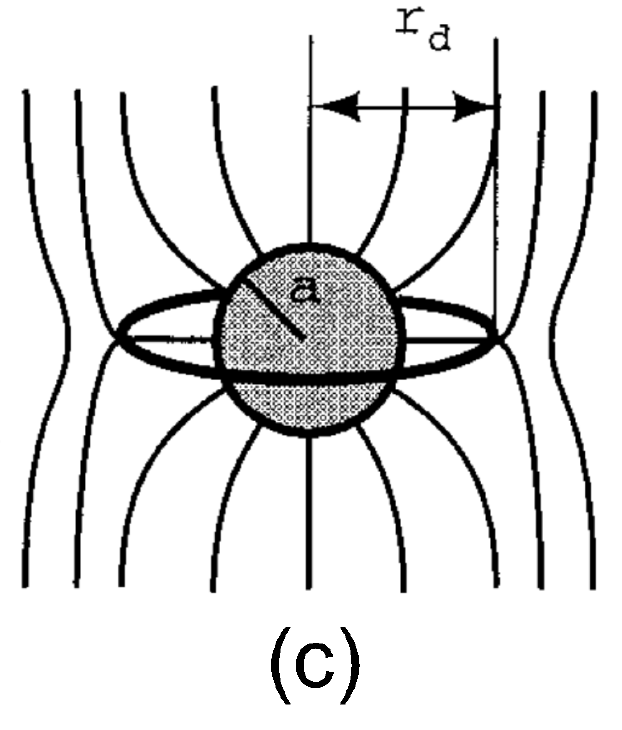}
\includegraphics[height=2.5cm]{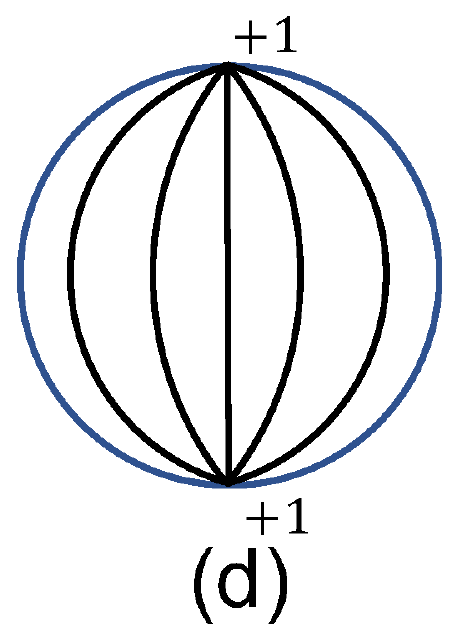}
\includegraphics[height=2.5cm]{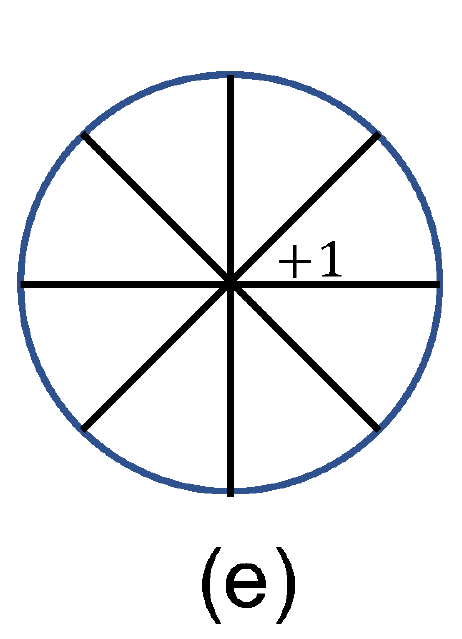}
\caption{(a) Long chains of isotropic droplets in a bulk nematic fluid; the chains have roughly equal separation.
(b) A hyperbolic hedgehog (point defect) formed by the nematics liquid crystal just outside an isotropic droplet.
(c) Half-integer ring defect can also form around an isotropic droplet (saturn ring).
(d) Boojum defects at the poles of nematic droplet.
(e) Radial hedgehog in the middle of nematic droplet.
(Images of (a) are adapted from \cite{Loudet00} and image of (b,c) are adapted from \cite{Lubensky98}. [Permissions not yet requested.])}
\label{fig:defects2}
\end{figure}

Emulsions comprising droplets of isotropic fluid in a bulk nematic (sometimes called `inverted' nematic emulsions)
can display interesting states of organization such as the one shown in figure \ref{fig:defects2} (\cite{Loudet00}).
Here, isotropic droplets can form parallel and very long chains, separated by roughly equal distance.
This phase is obtained by quenching the mixture from an initial high temperature phase that is uniform and isotropic, so that it phase separates into isotropic and nematic phases along the lines discussed in section \ref{PSK} for binary mixtures of simple fluids.

These long chains of isotropic droplets are stabilized in 3D by topological defects. In this system, surface anchoring requires $\nhat$ to be radial at the droplet edge, creating in effect a radial hedgehog that cannot be annihilated because its core is effectively inside the droplet. To restore uniform ordering at large scales, there needs to be either another point defect just outside the droplet, for which the lowest energy choice is a hyperbolic hedgehog
(see figure \ref{fig:defects2}(b)), or a disclination loop (which is equivalent at large distances as discussed above)  \citep{Lubensky98}. 
This loop can be off-centred or equatorial, as shown in figure \ref{fig:defects2}(c).
For the asymmetric arrangements such as that with the hyperbolic hedgehog, long-ranged elastic distortion in the director field creates an effective dipolar attraction \citep{Poulin97} and this causes the droplets to come together to form chains (figure \ref{fig:defects2}(a) inset). The interaction is quadrupolar for a symmetric (equatorial) disclination loop, with attraction only when the separation between particles is in a range of oblique angles to the far-field director.
If perpendicular anchoring is replaced by parallel anchoring, the exterior nematic instead develops a pair of $+1$ defects at opposite poles of the included droplet. Note that half of each defect, which would lie interior to the droplet, is effectively absent. Such surface defects are called `boojums'. The leading order interaction is again quadrupolar \citep{Poulin97}.

Interesting structures also arise in `direct' nematic emulsions comprising nematic droplets immersed in an isotropic fluid.
Again the director field can be aligned either parallel to the droplet interface ($\beta_0>0$ in the free energy equation (\ref{eq:F-Q})) or
perpendicular to it ($\beta_0<0$).
In the case of parallel anchoring, the director field typically forms a pair of boojum defects at each pole of the droplet  
(now with the missing part of the defect outside rather than inside the droplet, see figure \ref{fig:defects2}(d)). 
These point defects create a cusp at each pole of the droplet \citep{Prinsen03}.
On the other hand, in the case of perpendicular anchoring, the director field will typically form a single radial hedgehog in the centre of the droplet \citep{Lopez11}, 
at least within the single elastic constant approximation (figure \ref{fig:defects2}(e)). Importantly, because nematic order exists only in the droplet interior, these effects do not lead to any long-range interaction between droplets through the isotropic continuous phase. For this reason, the self assembly properties of direct nematic emulsions are broadly the same as for conventional emulsions of two isotropic liquids. They can be functionally useful, however, because the discrete nematic compartments can have faster switching times in response to external fields than a continuous phase in which defects can move slowly across large distances. Materials where the continuous phase is a polymer solution, from which the solvent is then evaporated to give nematic droplets in a solid matrix, are referred to as polymer-dispersed liquid crystals and widely used in technology \citep{Bouteiller96}. A related structure is obtained when a binary fluid is quenched into isotropic+nematic coexistence in the presence of colloidal particles. To minimize anchoring constraints these tend to segregate into the isotropic phase. When the final phase volume of that phase is small, a biliquid foam resembling that in figure \ref{twelve} is formed, with particles compressed into films that now lie between nematic cells with different directors \citep{Anderson01}. 

%
%

\section{Dynamics of liquid-crystalline emulsions \label{sec:LC-dynamics}}
So far, we have only studied the static properties of liquid-crystalline emulsions, which amounts to finding configurations that minimize the free energy. Although many such static properties of nematic emulsions have been studied, 
the literature on their dynamics is sparse. \cite{Fernandez07} studied experimentally a nematic droplet under pipe flow. At rest, with a parallel anchoring condition, the director field inside the droplet forms a $+1$ boojum defect at each pole of the droplet (figure \ref{fig:defects2}(d)). Under pipe flow, fluid circulation inside the droplet can bring these two point defects together to form, momentarily, a $+2$ defect. \cite{Tiribocchi16} studied numerically the effect of a shear flow in an inverted nematic emulsion, finding that the flow not only distorts the droplet into an ellipsoidal shape but causes its equatorial disclination loop to be displaced. 

Despite the relative lack of dynamical studies so far, we present below the theoretical machinery for addressing the dynamics of polar and nematic emulsions, restricting attention for simplicity to the hydrodynamic level where noise terms are neglected. In part we do so because this framework is needed to address the behaviour of {\em active} liquid-crystalline droplets, to which we shall turn in section \ref{sec:active-LC}. In what follows, we outline the derivation of dynamical equations for $\p,\phi,\v$ in the polar case (section \ref{pdynamics}), and then quote without derivation the corresponding equations for $\Q,\phi,\v$ in nematics (section \ref{Qdynamics}).

\subsection{Equations of motion: polar liquid crystals} \label{pdynamics}

\begin{figure}
\centering
\includegraphics[height=3.3cm]{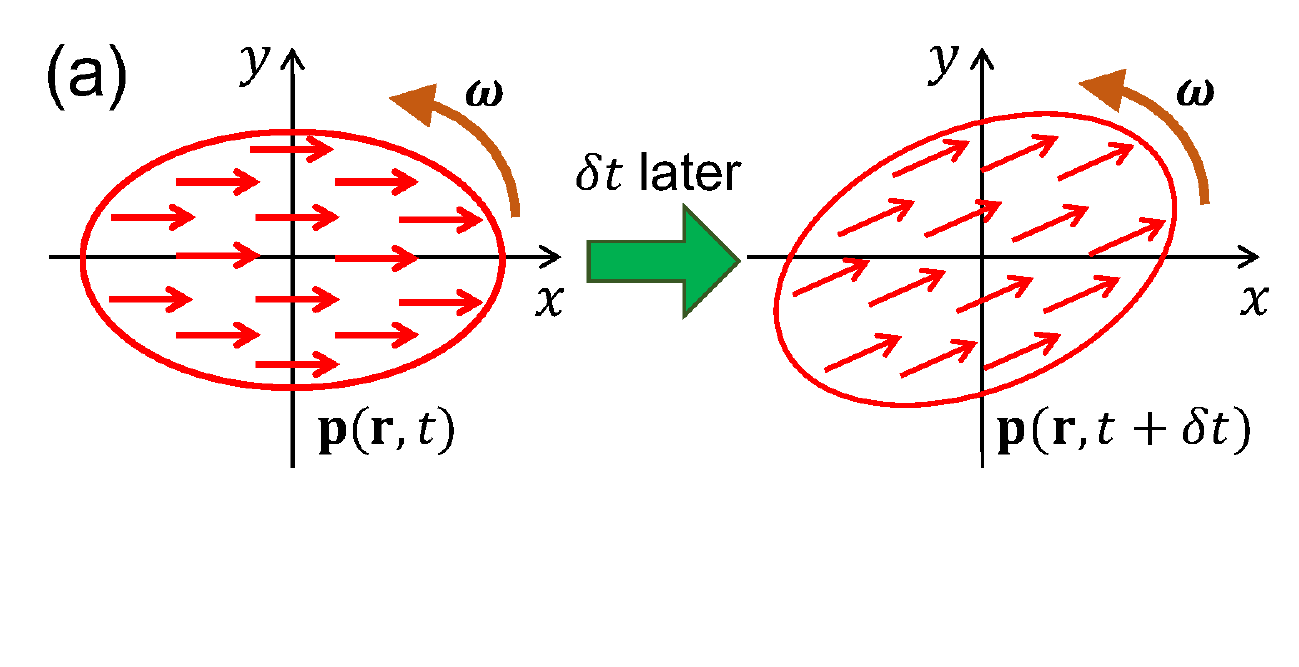}
\includegraphics[height=3.3cm]{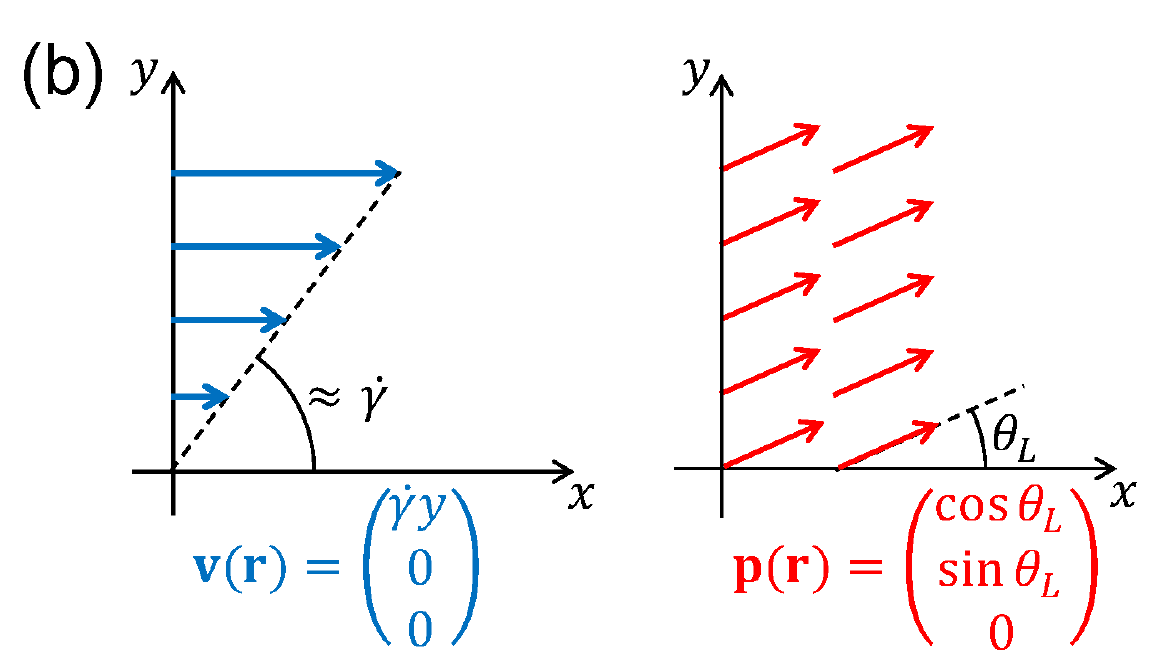}
\caption{(a) Rigid body rotation in a patch of liquid-crystalline material; (red) arrows indicate $\p$. 
(b) However most liquid crystals do not rotate like a rigid body, they tend to align with shear flow at steady state.
Here $\dot{\gamma}$ is the strain rate and $\theta_{L}$ is the Leslie angle, 
which is related to the phenomenological parameter $\xi$.
Horizontal (blue) arrows (left) represents fluid velocity $\v(\r)$ and inclined (red) arrows (right) represent the orientational field $\p(\r)$ at steady state.}
\label{fig:blob}
\end{figure}

Consider a patch of polar liquid-crystalline material as shown in figure \ref{fig:blob}(a). Let us initially assume that this patch rotates as a rigid body with some angular velocity $\boldsymbol{\omega}$.
In other words, the fluid velocity $\v(\r,t)$ at position $\r$ at time $t$ is given by $\v=\boldsymbol{\omega}\times\r$. 
At time $\delta t$ later, the $\p$-field can then be written as:
\begin{equation}
\underbrace{\p(\r,t+\delta t)=\p(\r-\v\delta t,t)}_{\text{advection}}+\underbrace{\boldsymbol{\omega}\delta t\times\p(\r-\v\delta t,t)}_{\text{rotation}}.
\end{equation}
The first term in this equation is simple advection;
we displaced the material by $\v\delta t$ in the time interval $\delta t$. 
However, as we can see from figure \ref{fig:blob}(a), 
the advective term alone is not enough; we also have to rotate $\p$.
This is given by the second term. 
Expanding to first order in $\delta t$ we obtain for the rigid rotation of $\p$:
\begin{equation}
\frac{\partial\p}{\partial t}+\v\bdot\del\p=\boldsymbol{\omega}\times\p=-\bOmega\bdot\p.\label{eq:p-dynamics1}
\end{equation}
The second equality above follows from the fact that the angular velocity can be expressed as 
$\omega_{i}=\frac{1}{2}\epsilon_{ijk}\Omega_{jk}$,
where $\Omega_{ij}\equiv\frac{1}{2}\left(\partial_{i}v_{j}-\partial_{j}v_{i}\right)$ is the anti-symmetric part of the velocity gradient tensor. 

In general flows, such as shear flows, liquid-crystalline materials do not rotate like a rigid body; typically $\p$ tends to align with the streamlines of the flow. This results in an additional contribution in the full equation of motion for $\p$; since it is advective, this term is bilinear in $\del\v$ and $\p$. The resulting full equation reads:
\begin{equation}
\frac{\partial\p}{\partial t} +\v\bdot\del\p =
								-\bOmega\bdot\p
								+ \xi{\D}\bdot\p
								- \frac{1}{\Gamma}\h.\label{eq:p-dynamics2}
\end{equation}
Here we have introduced $\xi$ as the flow-alignment parameter in the second term on the right. In this term, $D_{ij}\equiv\frac{1}{2}\left(\partial_{i}v_{j}+\partial_{j}v_{i}\right)$ is the symmetric part of the velocity gradient tensor. The parameter $\xi$ is set by the molecular geometry and is independent of the terms in the free energy $F$.
If $\left|\xi\right|>1$, the orientation of the liquid crystal tends to align in shear with the flow direction, with a steady-state orientation set by the so-called Leslie angle, $\theta_{L}=\tan^{-1}(\frac{\xi-1}{\xi+1})^{1/2}$ (see figure \ref{fig:blob}(b)). 
On the other hand if $\left|\xi\right|<1$, the orientation $\p$ is not stationary in a shear flow but instead shows tumbling behaviour \citep{Larson99}.  Importantly, there is no corresponding molecular influence on the coefficient of the $\bOmega$ term: this is always unity, otherwise $\p$ would fail to evolve properly when the only motion is the slow rigid rotation of the entire sample \citep{Beris94}. 

The final ingredient in the dynamics for $\p$ is the relaxation term proportional to ${1}/{\Gamma}$ in (\ref{eq:p-dynamics2}).
Here $\h(\r,t)={\delta F}/{\delta\p}$ is called the molecular field, and is analogous to the chemical potential $\mu={\delta F}/{\delta\phi}$ for $\phi$. 
This term, which relaxes $\p$ towards the minimum of the free energy, differs in form from the corresponding term in the equation of motion for the composition $\phi$ since $\p$ is not a locally conserved quantity. The equation for $\phi$ itself is the same as in Model H for a simple binary fluid (see (\ref{Hphi}) above), as is the NSE (see (\ref{HNSE})) for $\v$ except that a different expression, detailed below, is now needed for the thermodynamic stress $\bsigma$.

To summarise, for polar liquid crystals, at hydrodynamic (deterministic) level, the dynamics is governed by the following set of equations \citep{Kung06}:
\begin{eqnarray}
\rho(\dot{\v} + \v\bdot\del\v) &=& \eta\nabla^2\v 
											- \del P 
											+ \del\bdot{\bsigma}[\phi,\p], \label{eq:p1}\\
\del\bdot\v &=& 0, \label{eq:p2}\\
\dot{\phi} +\del\cdot(\phi\v)&=&  \del\bdot (M\del\mu),\label{eq:p3}\\
\dot{\p}  +\v\bdot\nabla\p &=&  
			- \bOmega\bdot\p
			+ \xi\D\bdot\p
			- \h/\Gamma. \label{eq:p4}
\end{eqnarray}
Here $M$ is the mobility and $\mu[\phi,\p] = \delta F/\delta\phi$ is the chemical potential. By construction these reduce to the noise-free version of Model H in the limit where $\p = \0$ everywhere.

It only remains to derive the total elastic stress ${\bsigma}[\phi,\p]$ which enters the NSE (\ref{eq:p1}).
We know already from section \ref{sec:stress} that even in the absence of $\p$ there an interfacial stress contribution. This takes the form (compare (\ref{stress3}))
\begin{equation}
\sigma_{ij}^\phi = ({\mathbb{F}}_\phi-\phi\mu)\delta_{ij} - \frac{\partial {\mathbb{F}}}{\partial(\partial_j\phi)}\partial_i\phi ,\label{eq:sigma-phi}
\end{equation}
where ${\mathbb{F}}_\phi = f(\phi)+\kappa(\del\phi)^2/2$ is the local free energy density 
as defined in (\ref{eq:F-phi}). In the final term we have generalized (\ref{stress3}) to allow for the addition of anchoring terms, which also can contribute to the chemical potential and the interfacial stress; hence the full free energy density ${\mathbb{F}}= {\mathbb{F}}_\phi+{\mathbb{F}}_\p$ appears here. This can be confirmed by a straightforward analogue of the following argument, in which for simplicity we calculate the elastic stress contribution from $\p$ only \citep{Beris94}.

To find this stress contribution, we consider the rate of change in $F_\p$:
\begin{equation}
\frac{dF_{\p}}{dt}=\int\frac{\delta F_\p}{\delta p_{i}}\frac{\partial p_{i}}{\partial t}\d\r = \int h_i\dot p_{i}\,\d\r.
\end{equation}
Substituting (\ref{eq:p4}) for $\dot p_i$, we obtain
\begin{equation}
\frac{dF_{\p}}{dt}=\int\left(\underline{(p_{j}\partial_{i}h_{j})v_{i}}-\Omega_{ij}h_{i}p_{j}+\xi D_{ij}h_{i}p_{j}-\frac{\left|\h\right|^{2}}{\Gamma}\right)\d\r,\label{eq:dFdt}
\end{equation}
where the underlined term has been integrated by parts. The resulting surface contribution vanishes for periodic boundary conditions which we choose here without loss of generality (just as we did in section \ref{sec:stress} when discussing Model H). The underlined term in (\ref{eq:dFdt})  is reminiscent of (\ref{stress1}) for $\delta F(\phi)$ in the case where only a composition variable is present. 
Thus $-p_j\partial_i h_j$ is effectively a force density, playing a similar role to $-\phi\nabla_i\mu$ there. Allowing for the fact that the small displacement $\u$ is now $\v dt$, this term gives a contribution $\delta F_\p' = \int(p_{j}\partial_{i}h_{j})u_{i}\d\r$ (where the prime on $F_\p'$ reminds us that this is not the only contribution). Following the same procedure using (\ref{stress2}) as in section \ref{sec:stress}, this translates into an elastic stress contribution that is the direct analogue of (\ref{eq:sigma-phi}):
\begin{equation}
\sigma_{ij}'=({\mathbb{F}}_\p-\p\bdot\h)\delta_{ij}-\frac{\partial {\mathbb{F}}}{\partial(\partial_{j}p_{k})}\partial_{i}p_{k}.
\end{equation}
It is simple to check from this that $\partial_j\sigma_{ij}'=-p_j\partial_i h_j$.

The remaining contribution $\sigma''_{ij}$ to the elastic stress has no counterpart in Model H; it stems from the rotation and alignment terms (\ref{eq:dFdt}) which may be written
\begin{equation}
\frac{dF''_{\p}}{dt}=\int(-\Omega_{ij}h_{i}p_{j}+\xi D_{ij}h_{i}p_{j})\,\d\r = \int\left(\Omega_{ij}\frac{1}{2}(p_{i}h_{j}-p_{j}h_{i})+\xi D_{ij}\frac{1}{2}(p_{i}h_{j}+p_{j}h_{i})\right)\d\r.\label{eq:anti-symmetric-sigma1}
\end{equation}
In the second equality, we have decomposed $h_{i}p_{j}$ into symmetric and anti-symmetric parts, and used 
the fact that $X_{ij}Y_{ij}=0$ for any symmetric tensor $X_{ij}$ and anti-symmetric tensor $Y_{ij}$. Once again using (\ref{stress2}) with $\u = \v dt$ we obtain
\begin{equation}
\frac{dF''_\p}{dt}=\int\sigma''_{ij}\partial_{j}v_{i}\,\d\r=\int\left({\sigma''_{ij}}^{S}D_{ij}-{\sigma''_{ij}}^{A}\Omega_{ij}\right)\d\r,\label{eq:anti-symmetric-sigma2}
\end{equation}
where the second form decomposes the rate of strain tensor $\partial_jv_i$ into its symmetric and antisymmetric parts.
Comparing Eq.~(\ref{eq:anti-symmetric-sigma1}) and Eq.~(\ref{eq:anti-symmetric-sigma2}) we identify 
${\sigma''_{ij}}^{S}=\frac{\xi}{2}(p_{i}h_{j}+p_{j}h_{i})$ and 
${\sigma''_{ij}}^{A}=-\frac{1}{2}(p_{i}h_{j}-p_{j}h_{i})$. The presence of an antisymmetric contribution is generic in liquid crystals in which rotational symmetry is spontaneously broken. 

Note finally that the
last term, proportional to $1/\Gamma$, in (\ref{eq:dFdt}) causes a purely dissipative loss of free energy and hence does not contribute to the elastic stress, which instead describes how the stored free energy changes with sample shape. This term is negative definite, ensuring that in the absence of a driving force the free energy steadily decreases towards its minimum value.

Adding the contributions $\bsigma^\phi$ and $\bsigma^\p = \bsigma' + \bsigma''^S + \bsigma''^A$ found above, we can assemble
the total elastic stress tensor ${\bsigma}[\phi,\p]$ in the NSE (\ref{eq:p1}). This reads
\begin{equation}
\sigma_{ij}=({\mathbb{F}}-\phi\mu-\p\bdot\h)\delta_{ij}-\frac{\partial {\mathbb{F}}}{\partial(\partial_{j}\phi)}\partial_{i}\phi-\frac{\partial {\mathbb{F}}}{\partial(\partial_{j}p_{k})}\partial_{i}p_{k}+\frac{\xi}{2}(p_{i}h_{j}+p_{j}h_{i})-\frac{1}{2}(p_{i}h_{j}-p_{j}h_{i}).\label{elasticstress}
\end{equation}
It is important to note that, alongside the rigid-rotation term (the final term on the right), the flow alignment parameter $\xi$, as defined via the equation of motion (\ref{eq:p4}) for $\p$, also enters the elastic stress in a non-negotiable fashion \citep{Beris94}. This is because it controls the response of $\p$ to an incremental strain and hence affects the resulting free energy increment which determines the stress.

\subsection{Equations of motion: nematic liquid crystals} \label{Qdynamics}

In principle, one can repeat the same calculation as above for nematic liquid crystals.
Here we shall just quote the results \citep{Beris94} which are included for completeness.
The dynamical equation for the order parameter $\Q$ is
\begin{equation}
\frac{\partial\Q}{\partial t} +(\v\bdot\del)\Q =
						  + \boldsymbol{S}(\del\v,\Q)
						  - \frac{1}{\Gamma}\boldsymbol{H}, \label{Qdynamics}
\end{equation}
where the first term on the right describes the combined effect of rigid body rotation and shear aligning.
This takes the following form
\begin{equation}
\boldsymbol{S}(\del\v,\Q) = (\xi\D + \boldsymbol{\Omega})(\Q + \I/d) 
							+ (\Q + \I/d)(\xi\D-\boldsymbol{\Omega})
							- 2\xi(\Q + \I/d)\Tr(\Q{\boldsymbol :}\del\v),
\end{equation}
where $\xi$ is a flow-aligning parameter analogous to that introduced previously for $\p$. This equation describes similar physics to the corresponding terms for $\p$. The quantity $\Q+\I/d$ enters (rather than just the traceless $\Q$) because the entire orientational distribution function, including its isotropic part, is stretched and rotated by the flow.
The final term in (\ref{Qdynamics}) is a local relaxation term similar to that used in (\ref{eq:p4}) for $\p$, with $\Gamma$ an analogous relaxation constant. The molecular field for nematics is defined as
\begin{equation}
\boldsymbol{H} = \frac{\delta F}{\delta \Q}  - \frac{1}{d}\Tr\left( \frac{\delta F}{\delta \Q} \right).
\end{equation}
This is traceless by construction, so as to maintain $\Tr\Q=0$ under time evolution.
The equations of motion for $\phi$ and $\v$ are the same as before, (\ref{eq:p1}), (\ref{eq:p2}) and (\ref{eq:p3}),
with the total elastic stress $\bsigma(\Q,\phi)$ now given by 
\begin{equation}
\begin{split}
\sigma_{ij} = &({\mathbb{F}}-\phi\mu)\delta_{ij} 
			- \frac{\partial {\mathbb{F}}}{\partial(\partial_j\phi)}(\partial_i\phi)
			- \frac{\partial {\mathbb{F}}}{\partial(\partial_jQ_{kl})}(\partial_iQ_{kl}) 
			+ Q_{ik}H_{kj} - H_{ik}Q_{kj} \\
			&- \xi H_{ik}(Q_{kj} + \delta_{kj}/d)
			- \xi (Q_{ik} + \delta_{ik}/d)H_{kj}
			+ 2\xi(Q_{ij} - \delta_{ij}/d)Q_{kl}H_{kl}.
\end{split}\label{FQ}
\end{equation}

This equation set was, for example, solved numerically by \cite{Sulaiman06}, who addressed droplet shapes and defect textures in equilibrium, and also switching behaviour in an applied external field ${\bf E}(t)$ (which adds a term  in ${\bf E}\bdot\Q\bdot{\bf E}$ to the free energy density $\mathbb{F}$, not needed here).
Note, however, that for both the nematic case and the polar one of the preceding section, only recently have the relevant numerical tools (primarily involving the lattice Boltzmann method) been developed to solve the equations of motion presented above \citep{Cates09,Sulaiman06}. With these methods in hand, it should be possible to understand more fully the unusual dynamical phenomena seen in direct and inverse liquid crystal emulsions, such as the kinetics of chain-formation among isotropic emulsion droplets in a nematic fluid \citep{Poulin97}. Conversely the recent numerical studies of rheology in such systems \citep{Tiribocchi16} may hopefully promote new experimental studies of their response to imposed flow.
%
%

\section{Active binary fluids} \label{active}
Most of the systems we have addressed so far above will, if left alone long enough, reach a state of thermal equilibrium at fixed volume, governed by the Boltzmann distribution or, if fluctuations are neglected, by minimising the free energy $F$. For example, a finite sample of phase-separating binary fluid will ultimately achieve a state with two large domains of the immiscible phases separated by an interface of minimal area consistent with the geometry of the container, modulo small thermal fluctuations of the interface itself. The main exceptions we have encountered are systems with particle-stabilized interfaces, where thermal energies are insufficient to detach particles and hence cannot achieve equilibration, and systems that are being continuously sheared in what is (experimentally at least) a boundary-driven flow. These exemplify two important ways in which a system can remain out of equilibrium: through kinetic arrest, and by being subject to continuous boundary driving. Recently however, a major focus of research has been systems that depart from thermal equilibrium because of continuous {\em microscopic} driving at the scale of the constituent particles \citep{Marchetti13}. For example, in a suspension of micro-organisms such as swimming bacteria, each `particle' moves through the surrounding solvent by self-propulsion, converting chemical energy (ultimately derived from a food source) into mechanical motion and thence viscous dissipation in the fluid. Such particles are called `motile'. Synthetic colloidal swimmers can be designed that also achieve motility, fuelled either by a chemical agent (such as dissolved hydrogen peroxide) or in some cases by the energy of light. In these cases, the colloids have surface chemistry that breaks rotational invariance, typically being Janus colloids on which each hemisphere is coated with a different material. If one of the coatings catalyse the breakdown of fuel, this creates local concentration gradients in reagents and/or products which in turn cause the Janus particle to move up or down those gradients in an autophoretic manner. The same gradients can also induce motion of other, neighbouring particles (cross-phoresis). There are also many systems, mostly biological in origin, where the activity is at a molecular rather than colloidal scale. Important examples include so-called actomyosin gels, in which molecular motors (myosin) crawl along polymeric filaments (actin). Such gels are a sub-cellular component of most multicellular (eukaryotic) organisms, forming part of the cytoskeleton which allows cells to change shape and move from place to place.

For active systems, equations of motion at continuum level (in which the active species are represented by a smooth density field rather than individually resolved particles) can be developed bottom-up by explicit coarse-graining of more detailed models in which motile particles enter as discrete, possibly point-like, objects \citep{Marchetti13,Cates15}. Below we follow a more phenomenological route, in keeping with the approaches developed above for passive systems. In this route, we adopt a suitable passive continuum model with appropriate symmetries, and add to it minimal extra terms to represent activity. The key properties of these additional terms are: (i) they are local -- reflecting the fact that activity is a local rather than global forcing of the system; and (ii) they break time-reversal symmetry (TRS).

Note, crucially, that although deterministic equations such as (\ref{ModelB}) are first order in time and therefore appear already to break time-reversal symmetry, in passive systems this symmetry is restored in thermal equilibrium by the noise terms; this is the content of the fluctuation dissipation theorem which fixes their form, as in (\ref{HJnoise}). This remark applies to both Models B and H discussed previously, and indeed (at least in the absence of magnetism), it is a general feature of thermal equilibrium that any movie of the fluctuating steady state is statistically indistinguishable running forwards from running backwards.
The role of the new time-reversal symmetry-breaking terms for active systems is to destroy this symmetry even in steady state. One way to do this is to introduce a mismatch between the noise and dissipative terms so that the fluctuation dissipation theorem no longer holds. However, bottom-up coarse graining instead suggest a slightly different structure in which TRS is broken through terms in the deterministic sector that are incompatible with existence of a free energy, meaning that no Boltzmann distribution is possible \citep{Marchetti13,Cates15}. Recent work suggests these two different modes of TRS breaking can be quite closely linked, in the sense that choosing one or the other microscopically can lead to essentially the same continuum equations \citep{Fodor16}. The generality or otherwise of this `duality' remains under current investigation. 

The simplest microscopic models of active matter address motile particles with isotropic inter-particle forces. This means that the angular degrees of freedom, responsible for liquid crystallinity among passive rodlike particles, do not need to enter the continuum description: the continuum variables are the fluid velocity $\v$ and the composition $\phi$, the latter now linearly related to the local number density of active particles. (There is still a unit vector attached to each particle which is body-fixed and determines the propulsive direction.) 

Possibly the most striking prediction of these microscopic models is Motility-Induced Phase Separation (MIPS). This differs from passive fluid-fluid phase separation in that it stems directly from activity; indeed MIPS arises in systems of active particles whose interactions with each other are purely repulsive, including active hard spheres. One way to understand this for synthetic Janus colloids is to note that two particles are more likely to collide if they are pointing in roughly opposite directions. Upon contact, in the absence of interparticle torques, the radial component of their relative velocity is then cancelled by the repulsive force, leaving a tangential component that is small or indeed zero for a head-to-head collision. The particles then remain in contact until a slow, typically diffusive, tangential motion allows them to separate. This contrasts with passive dynamics for which repulsive particles rapidly separate; instead it resembles passive attractive particles which linger in each others vicinity. Thus the combination of repulsion and activity can give an effective attraction \citep{Cates15}. Another view of MIPS is to note that the effectiveness of the particles' propulsive effort in producing forward motion is likely to be reduced at high density (for instance because of collisions as just described). In addition, active particles tend to accumulate in regions where they move more slowly, essentially because these regions are easy to enter but hard to get out of \citep{Schnitzer93}. This effect is similar to the accumulation of pedestrians in front of a distracting shop window where they slow down; but it is absent for the specific case of isothermal passive diffusers whose particle density is fixed by the free energy $F$ alone, independent of any choice of kinetics. The combination of density-induced slowdown and slowness-induced densification leads to the unstable growth of fluctuations by essentially the same spinodal instability as a phase-separating system of attractive particles \citep{Cates15}.

\subsection{Active Models B and H}\label{ABandH}
Continuum models for the description of motility-induced phase separation remain under development. Here we outline some of the interesting cases looked at so far. We start by suppressing the fluid velocity so that the relevant passive model is Model B, described by (\ref{ModelB},\ref{ModelBmu}). The simplest way to break time-reversal symmetry in this model is to retain (\ref{ModelB}) but add to the chemical potential in (\ref{ModelBmu}) a term that is {\em not} of the form $\delta F/\delta\phi$. To lowest order in gradients, this term is $|\del\phi|^2$. We therefore introduce a nonequilibrium chemical potential
\begin{equation}
\mu  = a\phi+ b\phi^3 -\kappa(\nabla^2\phi) +\lambda|\del\phi|^2. \label{ActiveB}
\end{equation}
Slightly more generally, one can consider the case where $\kappa(\phi)$ and $\lambda(\phi)$ are functions of composition; time-reversal symmetry is then broken ($\mu \neq \delta F/\delta\phi $ for any $F$) whenever $\lambda \neq d\kappa/\d\phi$. Microscopic models \citep{Stenhammar13} give exactly this structure, albeit with non-polynomial $F$ and nontrivial $\phi$-dependence in $\kappa$ and $\lambda$; suppressing the latter dependence and restoring the simplest form for the local part of $\mu$ gives (\ref{ActiveB}) as the prototypical model for motility-induced phase separation without fluid flow.

Active Model B has some interesting properties \citep{Wittkowski14}. First, the non-TRS term (which also breaks $\phi\to-\phi$ symmetry) alters the phase boundaries at coexistence, even at mean-field (zero noise) level. This was unexpected since in equilibrium problems the phase diagram is found by a common-tangent construction on $f(\phi)$ (see (\ref{uniform})) in which no gradient terms arise. However, it turns out that this construction is valid only if the gradient terms stem from a free energy functional. If one continues to define $F[\phi]$ as the functional arising when the active term is switched off ($\lambda = 0$), the effect of activity is to create an inequality between phases in the pressure-like quantity $P_{\rm Th}\equiv \mu\phi-f(\phi)$, causing a shift of the binodal compositions away from their values at $\lambda = 0$.
(This pressure, which is defined by the usual equilibrium relation between $P,\mu$ and $\phi$, should not be confused with the mechanical one defined as the force density acting on a wall; only in passive systems are these two definitions equivalent \citep{Solon15}.) The resulting ``pressure jump'' across the interface, which when small is linear in $\lambda$, effectively provides an anomalous active contribution to the Laplace pressure, which is finite even for a flat interface.

Because it interferes with Laplace pressure and hence with the driving force for coarsening dynamics, one might expect activity to have some important influence on the diffusive growth law (\ref{LSlaw}) that gave a domain size $L\sim t^{1/3}$. However, in Active Model B there is no conclusive numerical evidence for a change in exponent \citep{Wittkowski14}. Since coarsening is driven by Laplace pressure {\em differences}, this outcome can be rationalized by noting that the activity induced pressure-jump across interfaces is curvature-independent at leading order. For the same reason, diffusive coarsening continues indefinitely. The latter is also found to be true for the particular microscopic models (of so-called active Brownian particles, or ABPs) that inspired the form of (\ref{ActiveB}). In several experimental systems, however, coarsening appears to saturate while clusters are at a finite size of perhaps 60-100 particles; the reasons for this are not clear, and various explanations have been suggested, for instance involving cross-phoresis mechanisms \citep{Buttinoni13}.

This suggests that Active Model B, though appealingly simple, may not capture all we need to address the phase behaviour of scalar active matter. That view is confirmed by the observation that the mathematical structure of (\ref{ActiveB}), in combination with (\ref{ModelB}), enforces $\nabla\times \J = \0$. This rules out steady-state circulating particle currents in real space. These currents are a low dimensional projection of a circulating probability flux in the space of configurations $\phi(\r)$.  (The latter, more abstract, currents do remain present however: one finds that in the phase separated state there is continuous birth, in one phase, of droplets of the other, which then migrate to the interface and disappear; see  \cite{Stenhammar13}.) However, we know of situations where steadily circulating real-space currents do arise, at least in computer simulations -- for example when active Brownian particles are placed in a ratchet-like environment \citep{Stenhammar16}. Alongside the fact that Active Model B cannot explain cluster phases, this observation has motivated the recent introduction of an extended model, known as Active Model B+, in which additional time-reversal symmetry-breaking gradient terms are included in the expression for $\J$ \citep{Nardini17}. This model remains under investigation.

For systems in which the fluid velocity field $\v$ plays an important role, the natural starting point is Model H, to which we can again add minimal TRS-breaking terms. One such term, in the chemical potential, is the same as just described for Active Model B. (The additional terms arising in Active Model B+ are yet to be addressed in this context.) Another new term enters the NSE (\ref{HNSE}) whose passive version contains a thermodynamic stress obeying $\del{\bdot}\bsigma = -\phi\del\mu$. This form assumes a thermodynamic relation between stress and chemical potential which only holds for equilibrium systems (in which mechanical forces and thermodynamic ones stem from the same microscopic Hamiltonian and are not independent). But in a system undergoing motility-induced phase separation, for instance, even the fact that the local ``free energy density" $f(\phi)$ has two minima can arise purely from activity and not from attractive interactions. This means that, while the active contributions to $f(\phi)$ do not break time-reversal symmetry in themselves, they have no reason to feed through via thermodynamics into the stress term in the NSE.

What matters in an incompressible fluid is the deviatoric stress which is traceless and differs from the full stress by a pure pressure. From (\ref{stress3}), this is (in $d$ dimensions)
\begin{equation}
\sigma_{ij}^D = -\zeta((\partial_i\phi)(\partial_j\phi) - \frac{1}{d}|\del\phi|^2\delta_{ij}), \label{devio}
\end{equation}
in which $\zeta = \kappa$. In the absence of an external field that breaks rotational invariance, this form is in fact the only one possible to leading order in gradients, so that in passing from the passive to the active case, all that is lost is the connection between $\zeta$ and $\kappa$.  Active Model H thus
reads \citep{Tiribocchi15}
\begin{eqnarray}
\rho(\dot\v+\v\bdot\del\v) &=& \eta\delsq \v-\del P - \del\bdot\bsigma^D +\del\bdot\bsigma^n,\label{AHNSE}\\
\del\bdot\v &=& 0, \label{Aincomp}\\
\dot\phi + \v\bdot\del\phi &=& -\del\bdot(-M\del\mu+\J^n),\label{AHphi}\\
\mu(\r) &=& a\phi+ b\phi^3 -\kappa\delsq \phi+\lambda|\del\phi|^2.\label{AmumodelH}
\end{eqnarray}
where $\bsigma^D$ obeys (\ref{devio}), in which $\zeta$ is a parameter that depends on both interaction forces and activity. Because this is no longer linked to $\kappa$ (which is always postive), $\zeta$ can have either sign. It includes an active contribution that is positive for ``extensile" swimmers, which draw fluid in along the axis of motion and expel it equatorially, and negative for ``contractile" ones, which do the opposite. (For a further discussion of this distinction, see section \ref{sec:active-LC} and figure \ref{fig:force-dipole} below.)

\begin{figure}
\centering
\includegraphics[height=4cm]{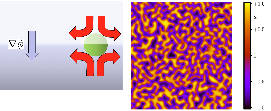}
\caption{Active Model H describes a fluid of active particles with conserved fluid momentum, but no orientational interactions so that a scalar order parameter $\phi$ is appropriate. Left: Schematic of stretching flow at an interface between low- and high-density regions ($\phi<0$ and $\phi>0$ respectively) of contractile swimmers. Right: Snapshot of a steady state in which this interfacial stretching balances the intrinsic coarsening dynamics of Model H. (Image courtesy of A. Tiribocchi.)}
\label{fig:AMH}
\end{figure}

Whenever $\zeta\neq\kappa$, there are effectively two interfacial tensions in Active Model H, one controlling the diffusive flux (set by $\kappa$) and one controlling the fluid flow driven by interfaces, now set by $\zeta$. This mismatch breaks time-reversal symmetry, but if both tensions are positive the consequences are relatively mild. (This statement is based on numerical studies in the simplest case which has $\zeta\neq\kappa$ but $\lambda = 0$.) In particular, the physics of coarsening in the viscous hydrodynamic regime remains broadly consistent with linear scaling, $L\sim t$. In the case when $\kappa$ is finite but $\zeta$ is zero, the order parameter field does not drive fluid motion and one recovers $L\sim t^{1/3}$ as for Model B. On the other hand, decreasing $\zeta$ below zero, as would be required to describe strongly contractile swimmers, one has in effect a negative interfacial tension in the mechanical sector (while that in the diffusive dynamics remains positive). Microscopically this arises because swimming particles tend to orient perpendicular to the interface between phases (there is a net polarization there proportional to $\del\phi$), where their contractile swimming action pulls fluid inwards normal to the interface and pushes out sideways in the interfacial (equatorial) plane causing the interface to stretch (figure \ref{fig:AMH}, left). The spontaneous stretching motion of the interface is mechanically equivalent to a negative tension.  Clearly, in this case one expects new and interesting effects to arise. One such effect is that the phase separation can arrest at a finite length scale where the diffusive shrinkage of the interfacial area is in balance with its contractile stretching (figure \ref{fig:AMH}, right). This offers a hydrodynamic, rather than microscopic, mechanism for the existence of cluster phases, but only in cases where the swimming is contractile \citep{Tiribocchi15}; its relation to earlier work in which the swimming particles are individually resolved \citep{Ishikawa08} is not yet established. The true character of cluster phases and their origins remains an active topic of current research \citep{Saha14}.

%
%

\section{Active liquid-crystalline emulsions \label{sec:active-LC}}

Above, we have introduced the concept of active binary fluids which are driven out of equilibrium by the irreversible dynamics of their constituent particles. We addressed the simplest case in which the active particles do not develop bulk orientational order. Such systems, which include spherical synthetic colloidal swimmers, can be described macroscopically by a scalar compositional order parameter alone.  More general active fluids include dilute and dense suspensions of rodlike bacteria, rodlike self-propelled colloids, and the active networks of fibres that arise in the cytoskeleton of living cells \citep{Marchetti13}.
Such systems can show mesoscopic or macroscopic orientational order, requiring additional liquid-crystalline order parameter fields. 

To describe them, we need a theory of 
{\em active liquid-crystalline} emulsions. This can be constructed by selectively adding non-equilibrium terms to the dynamical equations for their passive counterparts; the latter were derived in section~\ref{sec:LC-dynamics}. This procedure follows a similar philosophy to the development of Active Models B and H outlined above, but in fact preceded that work, albeit initially in the context of uniform bulk systems in which the compositional field $\phi$ is not also required. 
(For a comprehensive review of  active liquid crystals in bulk, see \cite{Marchetti13}.) 

\begin{figure}
\centering
\includegraphics[height=3.2cm]{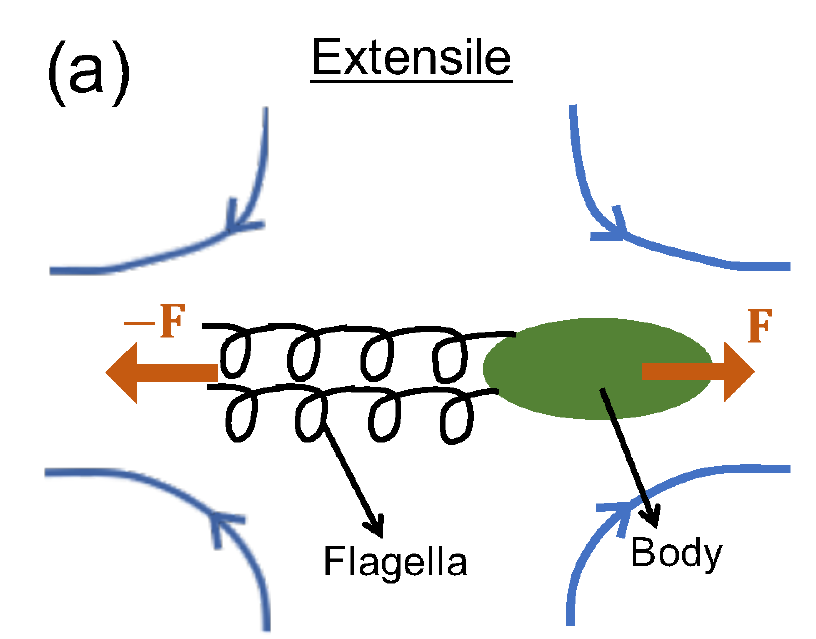}
\includegraphics[height=3.2cm]{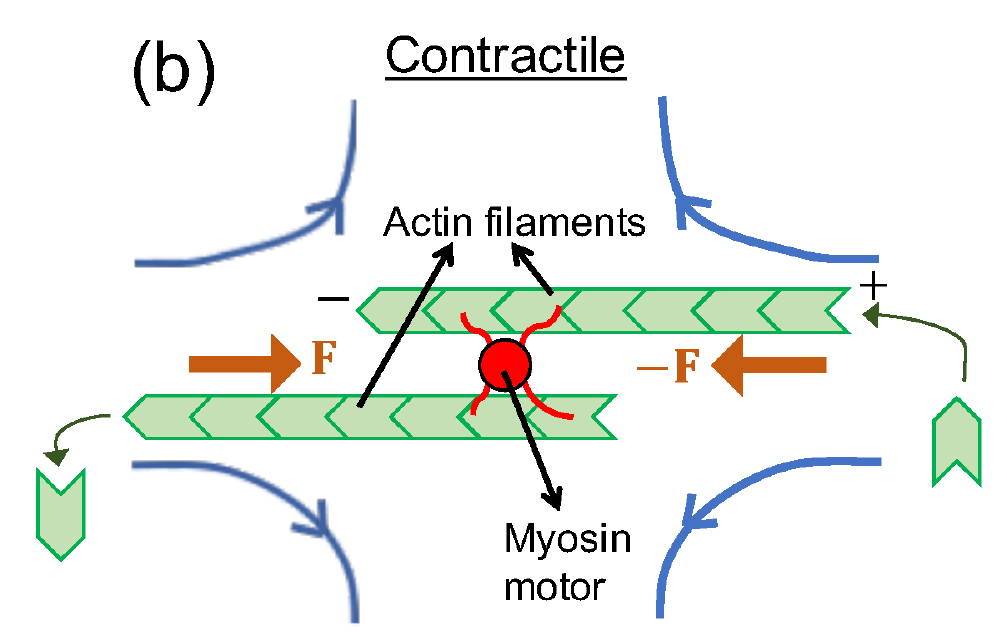}
\includegraphics[height=3.2cm]{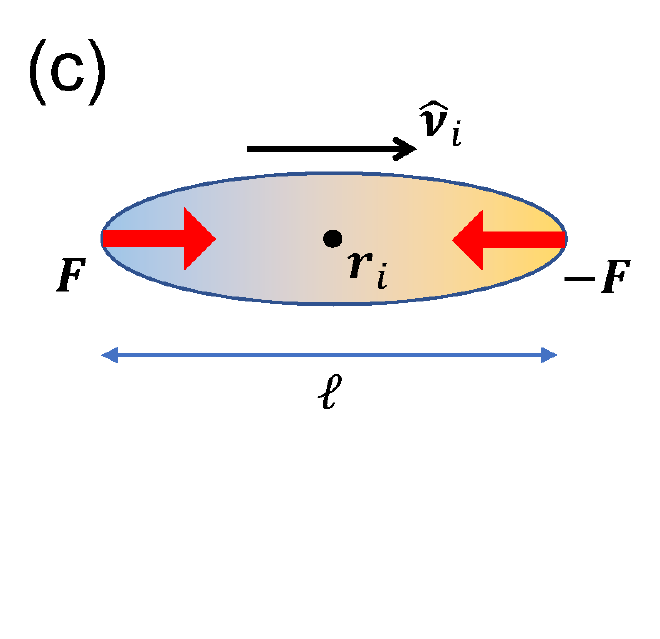}
\caption{Active stresses generated by active particles. 
(a) In bacteria, the flagella rotate anti-clockwise and the body rotates clockwise. 
This rotation expels the fluid fore-aft away from the bacteria and generates an extensile fluid flow. 
(b) In actomyosin contraction, the motor pulls the fluids inwards. 
This generates a contractile fluid flow. 
Actin filaments also tend to polymerise at the plus (barbed) end and de-polymerise at the minus (pointed) end. 
(c) Active particles can be modelled as force dipoles. \label{fig:force-dipole}}
\end{figure}

For liquid crystals there are two main types of non-equilibrium term that need to be added to represent activity. Consider, for example, the case of bacterial suspensions. Here the flagella of the bacteria (which have helical shape) rotate anticlockwise whereas 
the bacterial bodies (which have rodlike shape) rotate clockwise, 
resulting in self-propulsion forwards along a head-tail axis which can be described by a unit vector $\nuhat$. 
This propulsion leads to a `self-advection' in the dynamical equations whereby the bacterial concentration field $\phi(\r,t)$ is transported along $\langle\nuhat\rangle_{\rm meso}$ at a rate set by the mean propulsion speed, relative to a surrounding fluid that is stationary far away. Self-advection is important whenever $\langle\nuhat\rangle_{\rm meso}\neq\0$, that is, in polar phases.
Secondly, the propulsive motion creates a circulating flow pattern around each swimmer. The specific swimming mechanism of bacteria causes fluid to be expelled both forward and backward along the fore-aft axis, and drawn inwards
radially towards this axis, creating an extensile flow pattern; see figure \ref{fig:force-dipole}(a).
This action gives rise to an active stress contribution in the NSE in addition to the standard elastic stresses derived for liquid crystals in section~\ref{sec:LC-dynamics}. The form of the active stress, in both polar and nematic phases, is considered further below.

The cell cytoskeleton, on the other hand, is a network of various protein filaments, cross-linked by motor and/or linking proteins.
Such structures are found in the interior of all eukaryotic cells and play an important role in cellular shape-changes, motility and division. 
One family of protein filaments, called actins, are relatively thin and flexible. The actin filaments are cross-linked by motor proteins called myosins; see
figure \ref{fig:force-dipole}(b), which shows a pair of actin filaments linked by a single myosin motor.
The resulting `actomyosin network' is an active system because the motor proteins can pull the filaments together causing them to contract lengthwise. 
This creates a contractile fluid flow which is opposite to the extensile fluid flow found in the previous example of bacteria (compare figure \ref{fig:force-dipole}(a) to (b)). 
Note that if the myosin motor instead pushes the filaments outwards, as it would do eventually if the motion in figure \ref{fig:force-dipole}(b) were to continue, these will tend to buckle so that the time-averaged effect is a net contractile stress.
Actomyosin contraction has been shown to play an important role in the swimming motility of some tumour cells \citep{Poincloux11,Hawkins11}. 

Each actin filament is also polar (in our usual, geometrical sense), being comprised schematically of repeat units that are shaped like an arrowhead. It therefore has a `plus' end and a `minus' end (corresponding to the barbed end and the pointed end, respectively). 
We can then define a unit vector $\nuhat$ which is embedded in each filament and points from minus to plus. On average, actin monomers tend to polymerise at the plus end and de-polymerise at the minus end, creating an illusion of swimming in the direction of $\nuhat$. More precisely, this process of actin polymerisation and de-polymerisation (called `treadmilling') creates mass transport of the composition field $\phi(\r,t)$ by self-advection, just as swimming would do, so long as $\phi$ now refers to polymerized material as opposed to free monomeric actin.  Treadmilling plays an important role in the crawling motility of eukaryotic cell types
such as keratocyte cells \citep{Mogilner09}. 
These cells have been observed to crawl on a glass slide in the direction of polarization $\p=\left\langle \nuhat\right\rangle_{\rm{ meso}}$
\citep{Yam07}. 
Another class of protein filaments found in the cell cytoskeleton are microtubules. 
They are much stiffer and longer filaments than those made of actin. 
Microtubules are cross-linked by another class of motor proteins, called kinesins, which can create
extensile as well as contractile mean stresses, depending on physiological conditions.
Models of the microtubular network as an active liquid crystalline medium have recently shed light on the process of cell division \citep{Brugues14,Leoni17}.

In what follows we shall assemble the tools needed to describe the phenomenology of active polar emulsion droplets. 
Physically, these represent a minimal model for cellular motility (via the cytoskeleton) and also for emulsified bacterial swarms \citep{Aranson16,Marchetti13}. 
The hydrodynamic variables are (as usual) the scalar composition field $\phi(\r,t)$,  the average orientation $\p(\r,t)$ and the fluid velocity $\v(\r,t)$. 
As just discussed, the equations of motion will be similar to those of passive liquid-crystalline emulsions Eqs.~(\ref{eq:p1}--\ref{eq:p4}), but supplemented by some extra terms, chosen to break time-reversal symmetry.
\subsection{Active stress in liquid crystals}\label{ASILC}

As already described, the extensile swimming action of bacteria, or the contractile action of actomyosin, gives rise to an extra mechanical stress term in the NSE.
To find its form, consider a rod-shaped particle to represent a single bacterium or a pair of actin filaments cross-linked by a myosin motor as shown in figure \ref{fig:force-dipole}(c). A coordinate
$\r_{i}$ defines the centre of mass of the particle which has extent $\ell$ along the for-aft unit vector $\nuhat_{i}$. The activity creates a flow pattern that can be complicated in the near-field, but whose far-field is generically described by the lowest order spherical harmonic compatible with global momentum conservation. The fact that there is no external force on the swimmer rules out a Stokeslet contribution as would arise from a point force acting upon it. The lowest order term is therefore a stresslet, which is equivalent to the action of a force dipole and can be represented as such.
We therefore embed equal and opposite point forces at each end of the particle. 
The direction of the forces will determine whether the stress is contractile (forces pointing inwards, as shown in figure \ref{fig:force-dipole}(c)) or
extensile (forces pointing outwards).
Following \cite{Hatwalne04} (see also \cite{Saintillan07}) we write the force density acting on the fluid from $N$ such particles as:
\begin{align}
\mathbf{f}(\r) & =\sum_{i=1}^{N}\left\{ -F\nuhat_{i}\delta\left(\r-\r_{i}-\frac{\ell}{2}\nuhat_{i}\right)+F\nuhat_{i}\delta\left(\r-\r_{i}+\frac{\ell}{2}\nuhat_{i}\right)\right\} \nonumber \\
 & =\sum_{i=1}^{N}\left\{ \del\bdot\left(F\ell\nuhat_{i}\nuhat_{i}\delta(\r-\r_{i})\right)+O(\ell^{3}\nabla^{3})\right\}.\label{eq:force-dipole}
\end{align}
Here the summand is the force density from one active particle.
Note that $\ell$ is a microscopic length scale whereas $\del$ is the inverse of a mesoscopic or macroscopic length scale. 
Thus we Taylor expanded the Dirac delta functions in the first line above in powers of $\ell\nabla$, assumed to be small. 
Even in a polar fluid the nematic order parameter $\Q$ can be defined as usual:
$\Q=\left\langle \nuhat_{i}\nuhat_{i}\right\rangle _{\rm meso}-\mathbf{I}/d$.
More precisely, this can be expressed as:
\begin{equation}
\Q(\r)=\frac{1}{c}\left\langle \sum_{i=1}^{N}\left(\nuhat_{i}\nuhat_{i}-{\mathbf{I}}/{d}\right)\delta(\r-\r_{i})\right\rangle_{\rm meso}, \label{eq:Q-def}
\end{equation}
where $c(\r)=\langle\sum_{i}\delta(\r-\r_{i})\rangle_{\rm meso}$ is the number density of the particles and the angle brackets indicate ensemble averaging. 
Subsituting (\ref{eq:Q-def}) into (\ref{eq:force-dipole}), we obtain (ignoring terms $O(\ell^{3}\nabla^{3})$)
\begin{equation}
\mathbf{f}(\r)=\del\cdot\left[F\ell c(\phi)\left(\Q+{\mathbf{I}}/{d}\right)\right].
\end{equation}
Since the force density is related to the stress by $\mathbf{f}=\del\cdot\bsigma$,
we identify the active stress due to the force dipoles:
\begin{equation}
\bsigma^{\rm active}=\bar\zeta c(\phi)\left(\Q+{\mathbf{I}}/{d}\right),\label{eq:contractile1}
\end{equation}
where we have introduced an activity parameter $\bar\zeta=F\ell$, which is positive for contractile, negative for extensile and zero in equilibrium. (Choosing $\zeta = -\bar\zeta$ matches the notation to that of section \ref{ABandH} above.) Here
$c(\phi)$ is the concentration of the active particles and can be taken to obey $c=c_{0}\frac{\phi+1}{2}$ so that 
$c\simeq c_{0}$ in the active polar phase ($\phi=1$) and $c\simeq0$ in the passive isotropic phase ($\phi=-1$) for some positive constant $c_{0}$.
Also, for incompressible fluids, the isotropic part of $\bsigma^{\rm active}$ can be absorbed into the isotropic pressure $P$. Thus an equally good choice is 
\begin{equation}
\bsigma^{\rm active} = 
\bar\zeta c(\phi)\Q .\label{eq:contractile1a}
\end{equation}

The form (\ref{eq:contractile1a}) can be used directly in the equations for an active nematic, in which the particles (or more generally their orientational statistics) are symmetric with respect to inversion $\nuhat_i\rightarrow-\nuhat_i$. Self-propulsion breaks this symmetry at single-particle level, but nematic phases of self-propelled particles (known as movers) are possible in principle so long as there are equal numbers swimming up and down any chosen spatial axis. Alternatively, there are active particles (known as shakers) that set up a local circulation of fluid but do not self-advect. These can have the full nematic symmetry even at single-particle level.  On the other hand, polar active liquid crystals with nonzero $\p(\r,t) = \langle\nuhat\rangle_{\rm meso}$ break this head-tail symmetry by definition. 
In this case, one either has to carry two separate order parameter fields, $\Q$ and $\p$, or re-express $\Q$ locally as a function of $\p$ so that only $\p(\r,t)$ need be retained as a dynamical field variable. 

There is no general relation of this kind, as exemplified by the case where molecular orientations are distributed uniformly over the unit sphere and normal to it, but with arrows pointing outwards in the upper hemisphere and inwards in the lower. This state has nonzero $\p$ but zero $\Q$, because if the arrow heads are removed, the distribution of headless vectors is isotropic. Usually though,
both $\p$ and $\Q$ are nonzero and one can assume $\p$ to point along the major axis of $\Q$, which is the director $\nhat$. Thus $\p = \pm p\nhat$ while $\Q = S(\nhat\nhat-\I/d)$ so that the active stress can be written
\begin{equation}
\bsigma^{\rm active}=\bar\zeta(p) c(\phi)\p\p, \label{eq:contractile2}
\end{equation}
where $\bar\zeta$ has been redefined to absorb a factor of $S/p^{2}$. This expression only differs from (\ref{eq:contractile1}) or (\ref{eq:contractile1a}) by an isotropic term of the kind by which they already differ. Note that this form of stress cannot be derived from any free energy functional for our polar liquid crystal. In a quiescent system of nonzero uniform $\p$, the existence of such a free energy structure, from which $\p\neq \0$ arises by spontaneous breaking of rotational symmetry, demands that the deviatoric stress involves gradients of $\p$, not $\p$ itself, as in (\ref{elasticstress}). 

Absence of a free energy structure breaks time reversal symmetry, just as it did via the chemical potential contribution $\lambda|\del\phi|^2\neq\delta F/\delta \phi$ for Active Model B above. Nonetheless, just as the remaining chemical potential there was of the passive form, here we retain a free energy functional that generates (effectively passive) elastic stresses as in (\ref{elasticstress}), which we now denote as $\sigma^{\rm passive}$. The chosen form is
\begin{equation}
F[\phi,\p]=\int\left(-\frac{a}{2}\phi^{2}+\frac{a}{4}\phi^{4}+\frac{\kappa}{2}\left|\del\phi\right|^{2}+\frac{1}{2}\gamma(\phi)\left|\p\right|^{2}+\frac{\alpha}{2}\left|\p\right|^{4}+\frac{K}{2}\left|\del\p\right|^{2}+\beta_{1}\del\phi\bdot\p\right)\d\r.\label{eq:F-phi-p}
\end{equation}
Here $\gamma(\phi)=-\alpha\phi$, giving the same as (\ref{Fp}) for passive liquid-crystalline emulsions, except that we have taken the single elastic constant approximation, and set $\beta_2 = 0$ which restricts us to cases of perpendicular anchoring. As in the passive case, this free energy will stabilize a droplet of active polar phase ($\phi\simeq1$ and $\left|\p\right|\simeq 1$)
within a background fluid of the passive, isotropic phase ($\phi\simeq-1$ and $\left|\p\right|\to 0$), or vice versa.

\subsection{Self-advection and equations of motion}

Following the above arguments, we arrive at the dynamical equations for an active polar liquid-crystalline emulsion as follows \citep{Kruse05}:
\begin{align}
\rho\left[\dot{\v}+(\v\bdot\del)\v\right] & =\eta\nabla^{2}\v-\del P+\del\bdot\bsigma^{\rm passive}+\del\bdot\bsigma^{\rm active},\label{eq:active-p1}\\
\del\bdot\v & =0,\label{eq:active-p2}\\
\dot{\phi}+\del\bdot(\phi\v+\phi w\p) & =M\nabla^{2}\mu,\label{eq:active-p3}\\
\dot{\p}+(\v\bdot\del)\p+(w\p\bdot\del)\p & =-\bOmega\bdot\p+\xi\mathbf{D}\bdot\p-\mathbf{h}/\Gamma.\label{eq:active-p4}
\end{align}
This equation set differs from that of polar liquid-crystalline emulsions in (\ref{eq:p1}--\ref{eq:p4}), first via the active stress term in (\ref{eq:active-p1}) as already discussed, and second by the self-advection terms proportional to $w$ in (\ref{eq:active-p3},\ref{eq:active-p4}). 
These describe the fact that the active material is propelled through space with average local velocity $\langle w\nuhat\rangle_{\rm meso} = w\p$ where $w$ is the swim speed of a particle (or, for actin, a suitably defined treadmilling rate). 
This motion is additional to the mesoscopically defined fluid velocity $\v$. Accordingly we replace  $\v\rightarrow\v+w\p$ in the advective terms of (\ref{eq:p1}) and (\ref{eq:p4}) to get (\ref{eq:active-p3}) and (\ref{eq:active-p4}) above. 
Equations such as (\ref{eq:active-p1}-\ref{eq:active-p4}) are often referred to as `active gel theory'. Here the word `gel' is a slight misnomer, since liquid crystals are not strictly gels. Introducing additional polymeric degrees of freedom allows models of true active gels to be considered but this area of study remains in its infancy (e.g., \cite{Hemingway15}).

Turning to the case of active nematics, the dynamics for these is described by order parameter fields $(\Q,\phi,\v)$ instead of $(\p,\phi,\v)$. Since $\p$ is zero, the self-advective terms proportional to $w$ are absent, but we still have an active stress in the form (\ref{eq:contractile1}). This is added to $\bsigma^{\rm passive}$ obeying (\ref{FQ}) in the Navier Stokes equation for $\v$ with the equations of motion for $\phi$ and $\Q$ unchanged from the passive case.

\begin{figure}
\centering
\includegraphics[height=3.2cm]{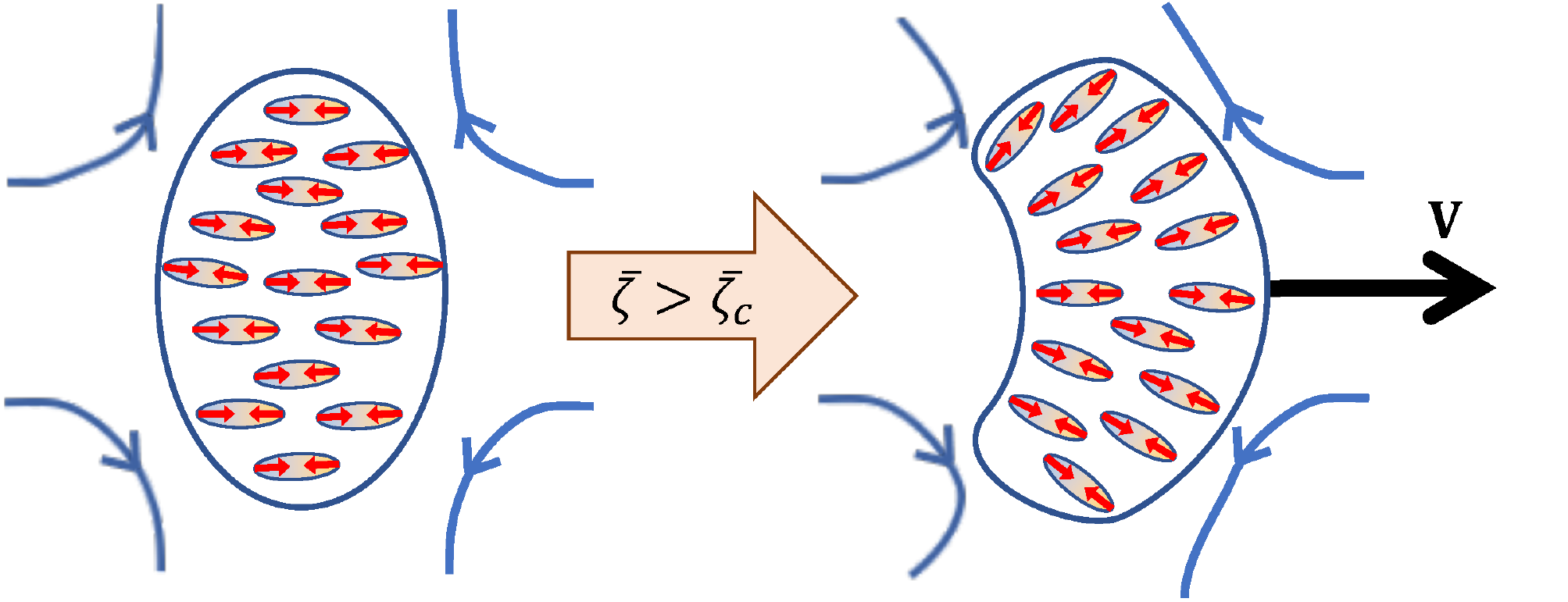}
\caption{For small activity $0<\bar\zeta<\bar\zeta_{c}$, an active polar droplet pumps the surrounding fluid to create a quadrupolar (force-dipolar, or stresslet) fluid flow which is left-right symmetric. 
At higher activities $\bar\zeta>\bar\zeta_{c}$, the polarization field $\p$ become unstable with respect to splay and 
this breaks the left-right symmetry resulting in translational motion in the direction of splay (here, to the right). 
(See figure \ref{fig:elastic-modes} for the definition of splay.)
\label{fig:motility}}
\end{figure}

\subsection{Spontaneous motion of active liquid-crystalline droplets}

For simplicity, we first set both the self-advection $w$ and the anchoring term $\beta_{1}$ to zero. 
In this case the equation of motion becomes symmetric with respect to a \emph{global }inversion $\p\rightarrow-\p$.
Figure \ref{fig:motility} shows what  happens when we increase the contractile stress ($\bar\zeta>0$) from some small values less than critical activity $\bar\zeta_{c}$ 
to a large value greater than $\bar\zeta_{c}$ \citep{Tjhung12}. 
For $\bar\zeta<\bar\zeta_{c}$, the active gel forms a uniform alignment inside the droplet.
The droplet also slightly contracts along $\p$ and pumps the surrounding fluid, creating a left-right symmetric fluid flow
of stresslet form (see figure \ref{fig:motility} left). 
Since the fluid flow is left-right symmetric, there is no reason for this droplet to move. 
In other words, the droplet as a whole behaves like a large ``shaker'' particle.

However when we increase the contractile activity $\bar\zeta$ above some threshold $\bar\zeta_{c}$, 
the liquid crystalline material inside the droplet become unstable with respect to splay deformation. 
This breaks the left-right symmetry in the fluid flow and causes the droplet to spontaneously swim either to the left or to the right (see figure \ref{fig:motility} right). These phenomena are unchanged by replacing $\p\rightarrow -\p$ globally: in particular, the direction of motion corresponds to $\p\del\bdot \p$ which is invariant under that replacement. For this reason the sense of $\p$ is not shown in figure \ref{fig:motility}, and indeed the same figure equally depicts the corresponding phenomena that arise in contractile nematics; the direction of motion is then set by $\nhat\del\bdot\nhat$.
Again the director field $\nhat$ inside the droplet becomes unstable with respect to splay and the droplet spontaneously swims for $\bar\zeta>\bar\zeta_{c}$. 
This mechanism for self-propulsion has been proposed for swimming motility in some eukaryotic cells \citep{Hawkins11}.
Here the source of the contractile stresses comes from the actomyosin contraction in the cytoskeletal bulk of the cell.
Note that extensile liquid crystals show instead an instability towards bending at large activities, resulting in banana-shaped droplets that also swim spontaneously \citep{Tjhung12}.

In striking (and somewhat related) experimental studies, \cite{Sanchez12} looked at the dynamics of suspensions of microtubules and kinesin motors in a droplet; under the conditions used, these form an extensile nematic phase.
They observed spontaneous motion of the droplet driven by the activity of the kinesin motors, albeit with complex dynamics resulting in non-Brownian diffusion at the droplet scale. A further complication in this system is that the active gel phase separates into a thin layer near the droplet surface; indeed these experiments have fuelled theoretical and numerical studies of active nematics in quasi-2D geometries. Both experiments and simulations show complex flow patterns within the 2D film, driven primarily by defect motion. Intriguingly, defects of topological charge $-1/2$ are advected by the flow field $\v$ in a quasi-passive manner, whereas those of charge $+1/2$ have additionally an active ballistic motion, somewhat resembling the self-propulsion of contractile emulsion droplets described in the previous paragraph \citep{Giomi13}. This stems from the fact that the $+1/2$ defect structure breaks spatial symmetry because it looks like an arrow, unlike the $-1/2$ case which is 3-fold symmetrical (see figure \ref{fig:defects}(a)). The ballistic separation of defect pairs, which are created by active flow, counters the normal passive evolution in which $\pm 1/2$ defects are attracted to one another and gradually annihilate to give an increasingly ordered state; this competition allows the active system to attain a stationary chaotic flow.

Active polar and nematic emulsions models have also been used to address the biological process  of cell division, known as mitosis. 
During mitosis, the filaments of the cytoskeleton self-organize themselves to create 
a structure called the `mitotic spindle' that in liquid-crystalline language can be viewed as a pair of  $+1$ aster defects.
Experimental measurements of the fluctuations in the mitotic spindle shows very good agreement with those predicted from the active gel theory outlined above \citep{Brugues14}.
Once the two asters are created, subsequent division of the droplet can be explained mainly by equilibrium free energy minimization.  Specifically, if the anchoring term $\beta_{1}$ is much larger than the elastic constant $K$, 
division into two equally-sized droplets is energetically favourable. 
However, if the elastic constant $K$ is much larger than $\beta_{1}$, 
the droplet only elongates, without dividing \citep{Leoni17}. This is one of several examples where a complex and highly regulated biophysical process can be impersonated by models involving minimal physical ingredients, prompting speculation that the biochemical machinery of the cell is sometimes used to control and exploit the autonomous function of generic active-matter building blocks, rather than to create cellular functionality from scratch. We next address another example of this type.

\begin{figure}
\begin{centering}
\includegraphics[height=4cm]{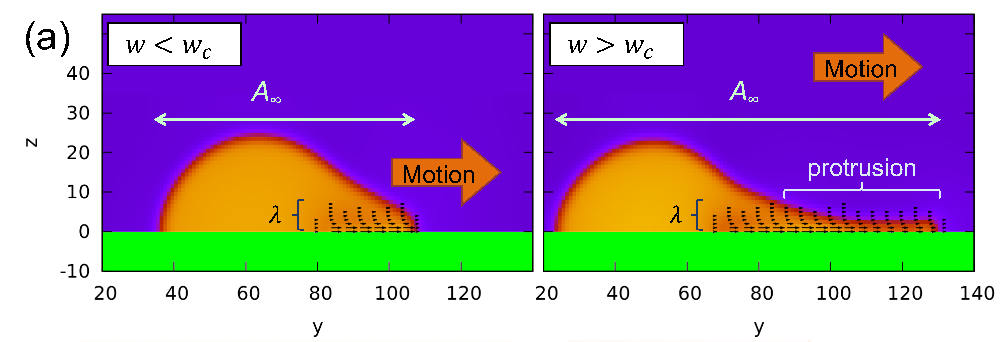}
\includegraphics[height=2.5cm]{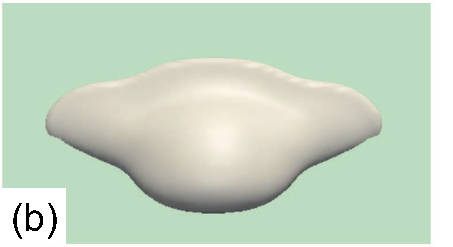}
\includegraphics[height=2.5cm]{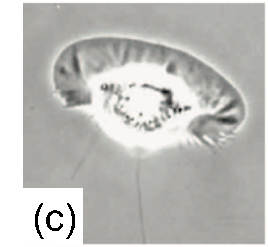}
\par\end{centering}
\caption{(a) Simple model of cell crawling: a 2D active polar droplet on a substrate.
When the actin treadmilling rate $w$ is larger than some critical $w_{c}$,
a thin protrusion layer is formed at the leading edge of the droplet.
Black arrows indicate the orientation of actin filaments $\p$. In this 2D model, the polarization field is confined to a thin layer near the wall and near the front of the droplet.
(b) A similar model in 3D can also capture the shape of the crawling cell, showing a fan-shaped lamellipodium, compared
to experimental observation (c). In this case there is polarization $\p$ throughout the droplet but treadmilling occurs only in a thin layer near the wall.
(See \cite{Tjhung15}, image of (c) is taken from \cite{Barnhart11}.) [Permission for (c) not yet requested.] \label{fig:crawling}}
\end{figure}

\subsection{Active emulsion droplets and the physics of cell crawling}

Some eukaryotic cell types, such as keratocytes, are able to crawl on a solid substrate such as glass slide \citep{Verkhovsky99}. 
The crawling motility is driven mainly by the actin treadmilling process. 
The actin filaments are found mostly at the leading edge of the crawling cell where they have
polarization $\p=\left\langle \nuhat\right\rangle_{\rm meso}$.
The filaments polymerise at their front (plus) ends and therefore self-advect in the direction of $\p$, creating forward motion of the entire cell. An important feature is that the actin filaments communicate chemo-mechanically with the supporting wall through structures called `focal adhesions'. These provide anchorpoints for the filaments so that their treadmilling is converted into a tank-treading motion of the cellular perimeter, driving the cell forward. This communication takes place across the plasma membrane, a lipid bilayer that encloses the cell. 

To model this situation within active gel theory, we consider an active polar droplet on a solid substrate as shown in figure \ref{fig:crawling}(a). For simplicity the confining plasma membrane is not modelled directly; instead there is, as usual, an interfacial tension between the interior of the droplet (or now cell) and the exterior. Focal adhesions are accounted for via a no-slip or partial-slip boundary condition on the fluid flow at the wall. 
We introduce a free energy functional similar to (\ref{eq:F-phi-p}), but assume either (a) that the
polarization field $\p$ is confined to a thin layer of thickness $\lambda$ close to the wall (and away from the rear of the droplet), with fixed treadmilling rate $w$ or that (b) the polarization $\p$ extends throughout the droplet but $w$ is nonzero only within a thin layer of thickness $\lambda$ near the wall. Such assumptions may be implemented by having explicit spatial dependence of the model parameters, for example $w = w(0)\exp[-z/\lambda]$ where $z$ is a coordinate normal to the wall. Assumption (a) is probably closer to the real behaviour, but assumption (b) is found to give very similar results in a 2D version of the model \citep{Tjhung15}, and is much simpler to implement in 3D. The confinement of actin and/or self-advection to a thin layer arises in part because the actin network not only adheres to the wall via focal adhesions, but also receives biochemical signals from these that alter its concentration and treadmilling rate. 

The equations of motion of the various fields are unchanged from (\ref{eq:active-p1}--\ref{eq:active-p4}), although some models of cell crawling also neglect hydrodynamics by putting $\v=\0$, in the manner of Model B \citep{Ziebert13}.
Since we retain the NSE, we need to specify the boundary condition for $\v$ at the solid substrate.
We choose a partial slip boundary condition whereby $v_z(z=0)=0$ and $(1-s)\v_\parallel(z=0) = 3\eta s  \left({\partial \v_\parallel}/{\partial z} \right)\rvert_{z=0} $. Here $\v_\parallel$ is the tangential velocity and $s$ is the slip parameter \citep{Wolff12}; $s\rightarrow0$ recovers the familiar no-slip boundary condition whereas $s\rightarrow1$ allows full slip.
We assume a boundary condition on $\p$ at the substrate such that it lies parallel to the wall: $p_z(z=0)=0$ and $\left({\partial \p_\parallel}{\partial z} \right)\rvert_{z=0} = \0$, whereas the anchoring term $\beta_1$ favours perpendicular alignment at the fluid-fluid interface. 

Typical results for both 2D and 3D models are shown in figure \ref{fig:crawling}. 
In 2D, for $w$ less than some critical value $w_c$, the droplet crawls forward in the direction of $\p$.
Interestingly, for larger values of $w>w_c$, a thin layer of protrusion is formed at the leading edge of the droplet, 
which is suggestive of the lamellipodium, a fanlike protrusion that is frequently seen in real crawling cells such as keratocytes \citep{Verkhovsky99}; see figure \ref{fig:crawling}(c).
The onset of this protrusion involves a dynamic transition which becomes increasingly discontinuous when we increase the slip parameter \citep{Tjhung15}.
This prediction might be tested experimentally by measuring the distribution of the projected areas of crawling cells ($A_{\infty}$ in figure \ref{fig:crawling}(a)). 
In the case of a discontinuous transition, this should show a bimodal distribution.

Physically, the slip parameter $s$ in the model represents the inverse adhesion strength between the filaments and the substrate:
if focal adhesions are few, or weak, the actin filaments are more likely to slip backward instead of being propelled forward.
For the same treadmilling rate $w$, the protrusion size (or projected area $A_{\infty}$) and the crawling speed are both found to decrease with increasing $s$, that is, with decreasing adhesion. This agrees qualitatively with experimental observations of \cite{Barnhart11}.

In 3D simulations, it is found that the active gel model supports a wide range of cell shapes. These include a `fried egg' structure, with a radially symmetric protrusion and no motion; a fully three dimensional representation of a crawling lamellipodium as shown in  figure \ref{fig:crawling}(b); and a crawling finger-like protrusion, resembling a structure known to biologists as a filopodium. Transitions between these shapes are controlled in part by an interplay between the contractile activity, which promotes spontaneous splay, and the anchoring term $\beta_1$, which favours states with $\p$ normal to the perimeter of the cell. From the active liquid crystal viewpoint, the lamellipodium is the result of a contractility-induced spontaneous splay in the interior, which then demands a fan-shaped morphology to maintain normal anchoring at the leading edge. This viewpoint is complementary to a more conventional biological one, pointing as it does to mechanistic elements that may stem from the generic properties of active fluids, rather than specific biochemical mechanisms within the cell.

Cellular motility and morphology remain active areas of research where active gel theory  and its extensions can provide a physical modelling framework that captures many phenomenologies.
In two recent examples, \cite{Camley14} modelled a pair of interacting cells that can spontaneously rotate inside a confined geometry, while \cite{Lober15} studied the collective dynamics of multiple motile cells and predicted that collective motion is inhibited with increasing cell-cell adhesion.

\section{Conclusion and outlook}

In this paper we have presented a modelling framework for several types of binary fluid system. This framework starts from mesoscopic order-parameter fields describing composition (a scalar order parameter), fluid velocity (a vector) and, where also present, polar (vector) or nematic (tensor) orientational order. All order parameters are controlled by evolution equations for their conservative (currents) and non-conservative (relaxation) dynamics. Among these is the familiar Navier Stokes equation for the fluid velocity, augmented by additional stress terms arising from the coupling between fluid momentum and the remaining order parameters. These stresses can be purely interfacial (for simple fluids) or also elastic (for liquid crystals) and can be derived from an underlying free energy expressed as a functional of the relevant order parameters. In some situations, particularly involving defects in liquid crystals, free-energy considerations allow aspects of the dynamics to be predicted without explicit consideration of the equations of motion. To capture processes such as nucleation of droplets from a metastable phase, and also the diffusion of disconnected fluid droplets through a quiescent fluid, one must add noise terms to the evolution equations for the order parameters, including the fluid velocity. In addition to interfacial and elastic stresses, we considered active stresses that can also drive a fluid flow but (by definition) do not derive from a free energy. These arise in systems whose fundamental constituents are maintained far from equilibrium by a continuous conversion of fuel into work, such as suspensions of self-propelled particles.

Our use of mesoscopic order parameters allows relatively sharp structural features, such as the interface between fluids or the cores of topological defects in liquid crystals, to be handled within the framework of continuous fields.  This offers advantages, both numerically and conceptually, over alternative descriptions in which interfaces (say) are discontinuities in composition with singular interfacial stresses. Our choices of free energy functional, typically involving polynomial local terms plus square-gradient interfacial contributions, were deliberately kept simple: each of these was not intended as an accurate descriptions of a particular material, but as a prototype of some general class of system. (For the same reason, we ignored the dependence of viscosity on concentration and similar kinetic complications.) Those general classes include binary mixtures of two simple fluids (describing conventional emulsions); binary mixtures of one simple fluid and a polar or nematic liquid crystal (liquid crystalline emulsions); and for both cases, recently studied extensions to analogues in which one or more component is active.

To give examples of its use, we have tried to connect the general modelling framework presented here to specific scientific questions. One classical area concerns the issue of how to maintain stability of emulsions over long periods. In that context we used a generic model for simple binary fluids to explore the kinetics of phase separation and to discuss the origins of the diffusive coarsening (the Ostwald process) that, alongside coalescence, is typically responsible for instability. We discussed remedies including the use of surfactants, trapped species and interfacial particles. Readers may have noticed that large parts of the latter discussion departed from our announced framework of continuous order parameter fields -- particularly when addressing interfacial colloids, surfactant micellization, bending elasticity and lipid bilayers. This is an acknowledgement that in many real situations the multi-scale character of these interfacial problems requires additional tools to be brought to bear before the full physics can be identified.

This multi-scale limitation applies equally to liquid-crystalline emulsions, our second main class of materials. In this field, the required detailed studies of interfacial phenomena in the presence of surfactants, interfacial particles, {\em{etc.}}, are mostly yet to be initiated. In contrast, the role of topological defects in these systems is relatively well explored, at least for static situations. However, much of the interfacial physics is controlled by anchoring, in which the interface sets a preferred direction for orientational order parameters; this is already captured in the chosen free energy functionals for this case. Surfactants will then influence the strengths, but not the form, of the anchoring terms. Nonetheless, there are many open questions involving interfacial microstructure, for example in ternary systems comprising colloidal particles in a mixture of liquid-crystalline and simple fluids. Here the structural analogues of Pickering emulsions and bijels could lead to a range of functionalities that remain largely unexplored. In addition, the unusual interactions present in liquid-crystalline emulsions have many dynamical consequences such as the kinetics of chain formation among isotropic droplets in a nematic matrix. The exploration of these dynamics, enabled by the kinds of description presented here combined with recently introduced numerical methods, remains a promising area for future work.

The active analogues of simple binary fluids include suspensions of spherical, self-propelled colloids whose mesoscopic description -- now at a scale larger than the colloids themselves -- requires a scalar concentration field only. Here activity can lead to new types of `motility-induced' phase separation. These can be partly described at continuum level by equations of motion that differ from those studied historically (Models B and H) by active contributions to the compositional current and the stress. One experimental mystery is the observation that such active phase separation is often incomplete, culminating in a state of dynamical clusters whose average size no longer increases with time. Various system-specific microscopic interpretations of this behaviour have been offered; continuing work on active continuum models aims to clarify whether fully generic mechanisms also exist. 

For rodlike swimmers and filamentary cytoskeletal materials the corresponding theory is that of active liquid crystals, which are commonly, but not always, polar rather than nematic. (The opposite is true for the passive case.) Within our general modelling framework, theories for binary systems involving these materials are again obtained by selectively adding active terms, that break time-reversal symmetry and do not therefore stem from any free energy, to the equations of motion of an otherwise passive system. For active polar media the main terms are an active stress and self-advection of the polarization and concentration fields. The latter encodes the mean effect of either self-propulsion along the polarization vector of individual molecules, or, in the case of actin filaments, polymerization at one end and depolymerization at the other. These self-advection contributions are absent in active nematics, whose mean polarization is zero by definition.

When such active terms are included in theories of polar liquid-crystal emulsions, some surprising phenomena emerge which appear similar to observations of contractility-induced swimming and cell-division in biological cells. Allowing also for spatial variations of the active terms within an emulsion droplet next to a solid wall, these similarities extend to cell-crawling, where the occurrence of a fan-shaped protrusion at the front of the crawling cell emerges from an interplay of a splay instability caused by activity and the anchoring of polarity at the droplet surface. These are among the first findings of a new and developing field in which continuum theories of active fluids form a platform for physics-inspired models of biological function. The aim of these models is not to replace the biologists' understanding of cellular motility and related functionality, which is largely based on a relatively detailed analysis of underlying biochemical process. Rather, the aim is to determine the extent to which these processes are necessary to sustain even basic modes of functionality, and the extent to which they instead exert close control over a set of pre-existing autonomous behaviours that are generically present in active fluids, whether closely controlled or not.

\section*{Acknowledgements}
Sections \ref{Intro}--\ref{PE} of this article are based in part on previous lecture notes by one of us \citep{Cates12}. We acknowledge the contributions of numerous colleagues, students and collaborators with whom we have discussed the topics covered in this article. MEC acknowledges funding from the Royal Society in the form of a Research Professorship. We thank Kevin Stratford, Paul Clegg and Adriano Tiribocchi for the images in figures \ref{four}, \ref{twelve} and \ref{fig:AMH} (right) respectively.



\begin{thebibliography}{14}
\expandafter\ifx\csname natexlab\endcsname\relax\def\natexlab#1{#1}\fi

\bibitem[Batchelor(1971)]{Batchelor59}
{\sc Batchelor, G.~K.} 1971 Small-scale variation of convected quantities like
  temperature in turbulent fluid. part 1. general discussion and the case of
  small conductivity. {\em J.~Fluid Mech.\/} {\bf 5}, 113--133.

\bibitem[Brownell \& Su(2004)]{Brownell04}
{\sc Brownell, C.~J. \& Su, L.~K.} 2004 Planar measurements of differential
  diffusion in turbulent jets. {\em AIAA Paper 2004-2335\/}.

\bibitem[Brownell \& Su(2007)]{Brownell07}
{\sc Brownell, C.~J. \& Su, L.~K.} 2007 Scale relations and spatial spectra in
  a differentially diffusing jet. {\em AIAA Paper 2007-1314\/}.

\bibitem[Dennis(1985)]{Dennis85}
{\sc Dennis, S. C.~R.} 1985 {Compact explicit finite difference approximations
  to the Navier--Stokes equation}. In {\em Ninth Intl Conf. on Numerical
  Methods in Fluid Dynamics\/} (ed. Soubbaramayer \& J.~P. Boujot), {\em
  Lecture Notes in Physics\/}, vol. 218, pp. 23--51. Springer.

\bibitem[Hwang \& Tuck(1970)]{Hwang70}
{\sc Hwang, L.-S. \& Tuck, E.~O.} 1970 On the oscillations of harbours of
  arbitrary shape. {\em J.~Fluid Mech.\/} {\bf 42}, 447--464.

\bibitem[Koch(1983)]{Koch83}
{\sc Koch, W.} 1983 Resonant acoustic frequencies of flat plate cascades. {\em
  J.~Sound Vib.\/} {\bf 88}, 233--242.

\bibitem[Lee(1971)]{Lee71}
{\sc Lee, J.-J.} 1971 Wave-induced oscillations in harbours of arbitrary
  geometry. {\em J.~Fluid Mech.\/} {\bf 45}, 375--394.

\bibitem[Linton \& Evans(1992)]{Linton92}
{\sc Linton, C.~M. \& Evans, D.~V.} 1992 The radiation and scattering of
  surface waves by a vertical circular cylinder in a channel. {\em Phil.\
  Trans.\ R. Soc.\ Lond.\/} {\bf 338}, 325--357.

\bibitem[Martin(1980)]{Martin80}
{\sc Martin, P.~A.} 1980 On the null-field equations for the exterior problems
  of acoustics. {\em Q.~J. Mech.\ Appl.\ Maths\/} {\bf 33}, 385--396.

\bibitem[Miller(1991)]{Miller91}
{\sc Miller, P.~L.} 1991 Mixing in high schmidt number turbulent jets. PhD
  thesis, California Institute of Technology.

\bibitem[Rogallo(1981)]{Rogallo81}
{\sc Rogallo, R.~S.} 1981 Numerical experiments in homogeneous turbulence. {\em
  Tech. Rep.\/} 81835. NASA Tech.\ Mem.

\bibitem[Ursell(1950)]{Ursell50}
{\sc Ursell, F.} 1950 Surface waves on deep water in the presence of a
  submerged cylinder i. {\em Proc.\ Camb.\ Phil.\ Soc.\/} {\bf 46}, 141--152.

\bibitem[{van Wijngaarden}(1968)]{Wijngaarden68}
{\sc {van Wijngaarden}, L.} 1968 On the oscillations near and at resonance in
  open pipes. {\em J.~Engng Maths\/} {\bf 2}, 225--240.

\bibitem[Worster(1992)]{Worster92}
{\sc Worster, M.~G.} 1992 {The dynamics of mushy layers}. In {\em In
  Interactive dynamics of convection and solidification\/} (ed. S.~H. Davis,
  H.~E. Huppert, W.~Muller \& M.~G. Worster), pp. 113--138. Kluwer.

\end{thebibliography}


\begin{thebibliography}{99}

\bibitem[Andelman \emph{et al.} (1987)]{Andelman87} 
\textsc{Andelman D., Cates, M. E., Roux D. \& Safran S. A.} 1987 
Structure and phase equilibria of microemulsions,
J. Chem. Phys. {\bf 87}, 7229--7241.

\bibitem[Anderson \emph{et al.} (2001)]{Anderson01}
\textsc{Anderson V. J., Terentjev E. M., Meeker S. P., Crain J. \& Poon W. C. K.} 2001
Cellular solid behaviour of liquid crystal colloids -1. Phase separation and morphology,
Eur. Phys. J. E {\bf 4} 11--20.

\bibitem[Aranson (2016)]{Aranson16} 
\textsc{Aranson I. S.} 2016
Physical Models of Cell Motility,
\emph{Springer}.

\bibitem[Aveyard (2012)]{Aveyard12} 
\textsc{Aveyard R.} 2012 
Can Janus particles give thermodynamically stable Pickering emulsions?
\emph{Soft Matt.} {\bf 8}, 5233--5240.

\bibitem[Barnhart \emph{et al.} (2011)]{Barnhart11} 
\textsc{Barnhart E. L., Lee K.-C., Keren K., Mogilner A. \& Theriot J. A.} 2011 
An adhesion-dependent switch between mechanisms that determine motile cell shape,
\emph{PLoS Biol.} {\bf 9}, e1001059.

\bibitem[Beris \& Edwards (1994)]{Beris94} 
\textsc{Beris A. N.  \& Edwards B. J.} 1994 
Thermodynamics of Flowing Systems with Internal Microstructure,
\emph{Oxford University Press}.

\bibitem[Bibette \emph{et al.} (2002)] {Bibette02} 
\textsc{Bibette, J., Leal-Calderon, F., Schmitt ,V. \& Poulin P.} 2002 
Emulsion Science,
\emph{Springer}.

\bibitem[Binks \& Horozov (2006)]{Binks06} 
\textsc{Binks B. P. \& Horozov, T. S., Eds.} 2006 
Colloidal Particles at Liquid Interfaces,
\emph{Cambridge}. 

\bibitem[Bouteiller \& LeBarney (1996)]{Bouteiller96}
\textsc{Bouteiller, L \& LeBarny, P.} 1996 
Polymer-dispersed liquid crystals: Preparation, operation and application,
\emph{Liquid Crystals} {bf 21}, 157--174.

\bibitem[Brady \& Bossis (1988)]{Brady88}
\textsc{Brady, J. F. \& Bossis, G.} 1988 
Stokesian dynamics, 
\emph{Ann. Rev. Fluid Mech.} {\bf 20}, 111--157.

\bibitem[Bray (1994)]{Bray94} 
\textsc{Bray A. J.} 1994 
Theory of phase-ordering kinetics, 
{\em Advances in Physics}, {\bf 43}, 357--459.

\bibitem[Brugues \& Needleman (2014)]{Brugues14}  
\textsc{Brugues J. \& Needleman D.} 2014 
Physical basis of spindle self-organization,
{\em Proc. Natl. Acad. Sci. USA}, {\bf 111}, 18496--18500.

\bibitem[Buttinoni {\emph et al.} (2013)]{Buttinoni13}
\textsc{Buttinoni I., Bialke J., Kummel F., Lowen H., Bechinger C. \& Speck T.}
Dynamic clustering and phase separation in suspensions of self-propelled colloidal particles,
{\em Phys. Rev. Lett.} {\bf 110} 238301.

\bibitem[Camley \emph{et al.} (2014)]{Camley14} 
\textsc{Camley B. A., Zhang Y., Zhao Y., Li B., Ben-Jacob E., Levine H. \& Rappel W.-J.} 2014 
Polarity mechanism such as contact inhibition of locomotion regulate persistent rotational motion of mammalian cells on micropatterns,
{\em Proc. Natl. Acad. Sci. USA} {\bf 111}, 14770--14775. 

\bibitem[Cates \emph{et al.} (2009)]{Cates09} 
\textsc{Cates, M. E., Henrich, O., Marenduzzo, D. \& Stratford, K.,} 2009 
Lattice Boltzmann simulations of liquid crystalline fluids: Active gels and Blue Phases, 
{\em Soft Matter} {\bf 5}, 3791--3800. 

\bibitem[Cates (2012)]{Cates12}
\textsc{Cates, M. E.} 2012 
Complex fluids: The physics of emulsions, arXiv preprint, 1209.2290; 
Ch. 10 in Soft Interfaces (proceedings of les Houches 2012 Summer School, Session XCVIII), 
Bocquet L., Qu\'er\'e D., Witten T. A. \& Cugliandolo L. F., et al, Eds., 
\emph{Oxford University Press} 2017.  

\bibitem[Cates \& Clegg (2008)]{Cates08}
\textsc{Cates M. E. \& Clegg P. S.} 2008 
Bijels: a new class of soft materials, 
{\em Soft Matter} {\bf 4} 2132--2138.

\bibitem[Cates \& Tailleur (2015)]{Cates15}
\textsc{Cates, M. E. \& Tailleur, J.}  2015 
Motility-induced phase separation, 
\emph{Ann. Rev. Cond. Mat. Phys.} {\bf 6}, 219--244.

\bibitem[Cavallaro \emph{et al.} (2011)]{Cavallaro11} 
\textsc{Cavallaro M., Botto L., Lewandowski E. P.,1, Wang M. \&  Stebe K. J.} 2011 
Curvature-driven capillary migration and assembly of rod-like particles,
\emph{Proc. Nat. Acad. Sci. USA}, {\bf 108}, 20923--20928.

\bibitem[Chaikin \& Lubensky (1995)]{Chaikin95} 
\textsc{Chaikin, P. M. \& Lubensky, T. C.} 1995 
Principles of Condensed Matter Physics
\emph{Cambridge University Press}.

\bibitem[Clegg \emph{et al.} (2016)]{Clegg16}
\textsc{Clegg P. S., Tavacoli J. W. \& Wilde P. J.} 2016 
One-step production of multiple emulsions: microfluidic, polymer-stabilized and particle-stabilized approaches,
\emph{Soft Matter} {\bf 12}, 998--1008.

\bibitem[David (2004)]{David04} 
\textsc{David, F.} 2004
Geometry and field theory of random surfaces and membranes, in 
Nelson D. R., Piran T. and Weinberg S., Eds., Statistical Mechanics of Membranes and Surfaces, 
\emph{World Scientific}.

\bibitem[de Gennes \& Taupin (1982)]{DeGennes82} 
\textsc{de Gennes P.-G. \& Taupin C.} 1982 
Microemulsions and the flexibility of oil-water interfaces,
\emph{J. Phys. Chem.} {\bf 86}, 2294--2304.

\bibitem[de Gennes \& Prost (2002)]{DeGennes02} 
\textsc{de Gennes P. G. \& Prost J.} 2002 
The Physics of Liquid Crystals, 2nd Edition,
\emph{Oxford Science Publication, Oxford}.

\bibitem[Doi \& Ohta (1991)] {Doi91} 
\textsc{Doi M. \& Ohta T.} 1991 
Dynamics and rheology of complex interfaces,
\emph{J. Chem. Phys.} {\bf 95}, 1242--1248.

\bibitem[Fernandez-Nieves \emph{et al.} (2007)]{Fernandez07}
\textsc{Fernandez-Nieves A., Link D. R., Marquez M. \& Weitz D. A.} 2007
Topological changes in bipolar nematic droplets under flow,
\emph{Phys. Rev. Lett.} {\bf 98}, 087801.

\bibitem[Fielding (2008)]{Fielding08}
\textsc{Fielding S. M.} 2008
Role of inertia in nonequilibrium steady states of sheared binary fluids,
\emph{Phys. Rev. E} {\bf 77},  021504. 

\bibitem[Fodor \emph{et al.} (2016)]{Fodor16}
\textsc{Fodor E., Nardini C., Cates M. E., Tailleur J., Visco P., van Wijland F.} 2016
How far from equilibrium is active matter?
\emph{Phys. Rev. Lett.} {\bf 117}, 038103.

\bibitem[Fryd \& Mason (2012)] {Fryd12}
\textsc{Fryd M. M., Mason T. G.} 2012 
Advanced nanoemulsions,
\emph{Ann. Rev. Phys. Chem.} {\bf 63}, 493--518.


\bibitem[Furukawa (1985)]{Furukawa85} 
\textsc{Furukawa H.} 1985 
Effect of inertia on droplet growth in a fluid, 
\emph{Phys. Rev. A} {\bf 31}, 1103--1108.


\bibitem[Giomi \emph{et al.} (2013)]{Giomi13}
\textsc{Giomi L., Bowick M. J., Ma X. \& Marchetti M. C.} 2013
Defect annihilation and proliferation in active nematics,
\emph{Phys. Rev. Lett} {\bf 110} 228101.


\bibitem[Gompper \& Schick (1994)] {Gompper94} 
\textsc{Gompper G. \& Schick M.} 1994  
Self Assembling Amphiphilic Systems, Phase Transitions and Critical Phenomena Vol 16, Domb C. and Lebowitz J. L., Eds.,
\emph{Academic Press, NY}.

\bibitem[Gonnella \emph{et al.} (1999)]{Gonnella99} 
\textsc{Gonnella G., Orlandini E. \& Yeomans J. M.} 1999 
Phase separation in two-dimensional fluids: The role of noise,
\emph{Phys. Rev. E} {\bf 59}, R4741--R4744.

\bibitem[Hatwalne \emph{et al.} (2004)]{Hatwalne04}  
\textsc{Hatwalne Y., Ramaswamy S., Rao M., \& Simha R. A.} 2004 
Rheology of active-particle suspensions,
\emph{Phys. Rev. Lett.} {\bf 92}, 118101.

\bibitem[Hawkins \emph{et al.} (2011)]{Hawkins11}  
\textsc{Hawkins R. J., Poincloux R., Benichou O., Piel M., Chavrier P. \& Voituriez R.} 2011 
Spontaneous contractility-mediated cortical flow generates cell migration in three-dimensional environments,
\emph{Biophys. J.} {\bf 101}, 1041--1045.

\bibitem[Hemingway \emph{et al.} (2015)]{Hemingway15}
\textsc{Hemingway E. J., Maitra A., Banerjee S., Marchetti M. C., Ramaswamy S., Fielding S. M. \& Cates M. E.} 2015
Active viscoelastic matter: From bacterial drag redulction to turbulent solids,
\emph{Phys. Rev. Lett} {\bf 114} 098302.

\bibitem[Herzig \emph{et al.} (2007)]{Herzig07} 
\textsc{Herzig E. M., White K. A., Schofield A. B., Poon W. C. K. \& Clegg P. S.} 2007 
Bicontinuous emulsions stabilized solely by colloidal particles,
\emph{Nat. Mater.} {\bf 6}, 966--971.

\bibitem[Hohenberg \& Halperin (1977)] {Hohenberg77}
\textsc{Hohenberg P. C. \& Halperin B. I.}1977
Theory of dynamic critical phenomena,
\emph{Rev. Mod. Phys.} {\bf 49}, 435--479.

\bibitem[Huse \& Leibler (1988)]{Huse88} 
\textsc{Huse D. A. \& Leibler S.} 1988 
Phase behaviour of an ensemble of nonintersecting random fluid films.
\emph{J. Phys. France} {\bf 49}, 605--621.

\bibitem[Ishikawa \emph{et al.} (2008)]{Ishikawa08}
\textsc{Ishikawa T., Locsei J. T. \& Pedley T. J.} 2008
Development of coherent structures in concentrated suspensions of swimming model micro-organisms.
\emph{J. Fluid Mech} {\bf 615} 401--431.

\bibitem[Kendon \emph{et al.} (2001)]{Kendon01} 
\textsc{Kendon, V. M., Cates, M. E., Pagonabarraga, I., Desplat, J.-C. \& Bladon, P.} 2001 
Inertial effects in three-dimensional spinodal decomposition of a symmetric binary fluid Mixture: A lattice Boltzmann study, 
{\em J. Fluid. Mech.} {\bf 440} 147--203.

\bibitem[Kruse \emph{et al.} (2005)]{Kruse05}  
\textsc{Kruse K., Joanny J. F-, Julicher F., Prost J. \& Sekimoto K.} 2005 
Generic theory of active polar gels: a paradigm for cytoskeletal dynamics,
{\em Eur. Phys. J. E} {\bf 16} 5--16.

\bibitem[Kung \emph{et al.} (2006)]{Kung06}
\textsc{Kung W., Marchetti M. C. \& Saunders K.} 2006
Hydrodynamics of polar liquid crystals,
{\em Phys. Rev. E} {\bf 73} 031708.  

\bibitem[Landau \& Lifshitz (1959)]{Landau59} 
\textsc{Landau L. V. \& Lifshitz I. M.} 1959 
Fluid Mechanics,
\emph{Pergamon, Oxford}.

\bibitem[Landau \& Lifshitz (1986)]{Landau86} 
\textsc{Landau L. V. \& Lifshitz I. M.} 1986 Theory of Elasticity 3rd Edition,
\emph{Pergamon, Oxford}.

\bibitem[Landfester (2003)]{Landfester03} 
\textsc{Landfester K.,} 2003 
Miniemulsions for nanoparticle synthesis, 
\emph{Topics in Current Chem.} {\bf 227}, 75--123.

\bibitem[Larson (1999)]{Larson99}
\textsc{Larson R.G,} 1999 
The Structure and Rheology of Complex Fluids,
\emph{Oxford University Press, New York}.

\bibitem[Lattuada \& Hatton (2011)]{Lattuada11} 
\textsc{Lattuada M. \& Hatton T. A.} 2011
Synthesis, properties and applications of Janus nanoparticles,
\emph{Nano Today} {\bf 6}, 286--308.

\bibitem[Lee \emph{et al.} (2013)] {Lee13} 
\textsc{Lee M. N. , Thijssen J. H. J., Witt J. A. \& Clegg P. S.} 2013 
Making a robust interfacial scaffold: Bijel rheology and its link to processability,
\emph{Adv. Funct. Mater.} {\bf 23}, 417--423.

\bibitem[Leoni \emph{et al.} (2017)] {Leoni17}  
\textsc{Leoni M., Manyuhina O. V., Bowick M. J. \& Marchetti M. C.} 2017 
Defect driven shapes in nematic droplets: analogies with cell division,
\emph{Soft Matter} {\bf 13}, 1257--1266.

\bibitem[Lober \emph{et al.} (2015)]{Lober15}  
\textsc{Lober J., Ziebert F. \& Aranson I. S.} 2015
Collisions of deformable cells lead to collective migration,
{\em Sci. Rep.}, {\bf 5}, 9172.

\bibitem[Lopez-Leon \& Fernandez-Nieves (2011)]{Lopez11}  
\textsc{Lopez-Leon T., \& Fernandez-Nieves A.} 2011
Drops and shells of liquid crystal,
{\em Colloid Polym. Sci.}, {\bf 289}, 345--359.

\bibitem[Loudet \emph{et al.} (2000)]{Loudet00}  
\textsc{Loudet J. C., Barois P. \& Poulin P.} 2000
Colloidal ordering from phase separation in a liquid-crystalline continuous phase,
{\em Nature}, {\bf 407}, 611--613.

\bibitem[Lubensky \emph{et al.} (1998)]{Lubensky98}  
\textsc{Lubensky T. C., Pettey D., Currier N. \& Stark H.} 1998
Topological defects and interactions in nematic emulsions,
{\em Phys. Rev. E}, {\bf 57}, 610--625.

\bibitem[Marchetti \emph{et al.} (2013)]{Marchetti13} 
\textsc{Marchetti, M. C., Joanny, J.-F., Ramaswamy, S., Liverpool T. B., Prost, J. Rao, M. \& Simha, R. A.} 2013 
Hydrodynamics of soft active matter, 
\emph{Rev. Mod. Phys.} {\bf 85} 1143.

\bibitem[Mogilner (2009)]{Mogilner09} 
\textsc{Mogilner A.} 2009 
Mathematics of cell motility: have we got its number?
\emph{J. Math. Biol.} {\bf 58} 105--134.

\bibitem[Nardini \emph{et al.} (2017)] {Nardini17} 
\textsc{Nardini C., Fodor E., Tjhung E., van Wijland F., Tailleur J. \& Cates M. E.} 2017
Entropy production in field theories without time reversal symmetry: Quantifying the non-equilibrium character of active matter,
\emph{Phys. Rev. X} {\bf 7}, 021007.

\bibitem[Nazarenko \emph{et al.} (2001)]{Nazarenko2001} 
\textsc{Nazarenko V. G., Nych, A. B. \& Lev, B. I.} 2001 
Crystal structure in nematic emulsion,
\emph{Phys. Rev. Lett.} {\bf 87}, 075504.

\bibitem[Onuki (2002)] {Onuki02} 
\textsc{Onuki, A.} 2002  
Phase Transition Dynamics,
\emph{Cambridge University Press}.

\bibitem[Poincloux \emph{et al.} (2011)]{Poincloux11}  
\textsc{Poincloux R., Collin O., Lizarraga F., Romao M., Debray M., Piel M. \& Chavrier P.} 2011
Contractility of the cell rear drives invasion of breast tumor cells in 3D Matrigel,
{\em Proc. Natl. Acad. Sci. USA}, {\bf 108}, 1943--1948.

\bibitem[Poulin \emph{et al.} (1997)]{Poulin97}  
\textsc{Poulin P., Stark H., Lubensky T. C. \& Weitz D. A.} 1997
Novel colloidal interactions in anisotropic fluids, 
{\em Science}, {\bf 275}, 1770--1773.

\bibitem[Poulin (1999)]{Poulin99} 
\textsc{Poulin, P.} 1999 
Novel phases and colloidal assemblies in liquid crystals, 
{\em Current Opinion in Colloid and Interface Science}, {\bf 4}, 66--71.

\bibitem[Prinsen \& van der Schoot (2003)]{Prinsen03} 
\textsc{Prinsen P. \& van der Schoot P.} 2003 
Shape and director-field transformation of tactoids, 
\emph{Phys. Rev. E}, {\bf 68}, 021701.

\bibitem[Roux \emph{et al.} (1992)]{Roux92} 
\textsc{Roux D, Coulon C. \& Cates M. E.} 1992 
Sponge phases in surfactant solutions,
\emph{J. Phys. Chem.} {\bf 96}, 4174--4187.

\bibitem[Safran \& Turkevich (1983)] {Safran83} 
\textsc{Safran S. A. \& Turkevich L. A.} 1983 
Phase diagrams for microemulsions,
\emph{Phys. Rev. Lett.} {\bf 50}, 1930--1933.

\bibitem[Safran (2003)]{Safran03} 
\textsc{Safran S. A.} 2003  
Statistical Thermodynamics of Surfaces, Interfaces and Membranes, 
\emph{Westview Press}.

\bibitem[Saha \emph{et al.} (2014)]{Saha14}
\textsc{Saha S., Golestanian R. \& Ramasawmy S.} 2014
Clusters, asters and collective oscillations in chemotactic colloids,
\emph{Phys. Rev. E} {\bf 89} 062316.

\bibitem[Saintillan \& Shelley (2007)]{Saintillan07}
\textsc{Saintillan D., \& Shelley M. J.} 2007
Orientational order and instabilities in suspensions of self-locomoting rods,
\emph{Phys. Rev. Lett.} {\bf 99} 058102.

\bibitem[Sanz \emph{et al.} (2009)]{Sanz09} 
\textsc{Sanz E., White K. A., Clegg P. S. \& Cates M. E.} 2009 
Colloidal gels assembled via a temporary interfacial scaffold,
\emph{Phys. Rev. Lett.} {\bf 103}, 255502.

\bibitem[Sanchez \emph{et al.} (2012)]{Sanchez12} 
\textsc{Sanchez T., Chen D. T. N., DeCamp S. J., Heymann M. \& Dogic Z.} 2009 
Spontaneous motion in hierarchically assembled active matter,
\emph{Nature} {\bf 491}, 431--435.

\bibitem[Schnitzer (1993)]{Schnitzer93}
\textsc{Schnitzer M. J.} 1993
Theory of continuum random walks and application to chemotaxis,
{\emph Phys. Rev. E} {\bf 48} 2553--2568.

\bibitem[Shimuzu \& Tanaka (2015)]{Tanaka15} 
\textsc{Shimuzu, R. \&  Tanaka, H.} 2015 
A novel coarsening mechanism of droplets in immiscible fluid mixtures,  
{\emph Nature Communications} {\bf 6}, 7407.

\bibitem[Siggia (1979)]{Siggia79} 
\textsc{Siggia E.} 1979 
Late stages of spinodal decomposition in binary mixtures,
{\emph Phys. Rev. A} {\bf 20}, 595--605.

\bibitem[Solon \emph{et al.} (2015)]{Solon15}
\textsc{Solon A. P., Fily Y., Baskaran A., Cates M. E., Kafri Y., Kardar M. \& Tailleur J.} 2015
Pressure is not a state function for generic active fluids,
\emph{Nature Physics} {\bf 11}, 673--678.

\bibitem[Stansell \emph{et al.} (2006)]{Stansell06} 
\textsc{Stansell P., Stratford K., Desplat J.-C., Adhikari R. \& Cates M. E.} 2006 
Nonequilibrium steady states in sheared binary fluids,
\emph{Phys. Rev. Lett.}  {\bf 96}, 085701.

\bibitem[Stenhammar \emph{et al.} (2013)]{Stenhammar13}
\textsc{Stenhammar J., Tiribocchi A., Allen R. J., Marenduzzo D. \& Cates M. E.} 2013
Continuum theory of phase separation kinetics for active Brownian particles,
\emph{Phys. Rev. Lett.} {\bf 111}, 145702.

\bibitem[Stenhammar \emph{et al.} (2016)]{Stenhammar16}
\textsc{Stenhammar J., Wittkowski R., Marenduzzo D. \& Cates M. E.} 2016
Light-induced self-assembly of active rectification devices,
\emph{Science Advances} {\bf 2}, e1501850.

\bibitem[Stratford \emph{et al.} (2007)] {Stratford07} 
\textsc{Stratford K., Desplat J.-C., Stansell P. \& Cates M. E.} 2007 
Binary fluids under steady shear in three dimensions,
\emph{Phys. Rev. E} {\bf 76}, 030501(R).

\bibitem[Stratford \emph{et al.} (2005)]{Stratford05} 
\textsc{Stratford K., Adhikari R., Pagonabarraga I., Desplat J.-C., Cates M. E.} 2005 
Colloidal jamming at interfaces: A route to fluid-bicontinuous gels,
\emph{Science} {\bf 309}, 2198--2201.

\bibitem[Subramanian \emph{et al.} (2005)]{Subramanian05} 
\textsc{Subramaniam A. B., Abkarian M. \& Stone H. A.} 2005
Controlled assembly of jammed colloidal shells on fluid droplets,
\emph{Nat. Mater.} {\bf 4}, 553--556.

\bibitem[Sulaiman \emph{et al.} (2006)]{Sulaiman06} 
\textsc{Sulaiman N., Marenduzzo D \& Yeomans J.} 2006 
Lattice Boltzmann algorithm to simulate isotropic-nematic emulsions,
{\emph Phys. Rev. Lett.} {\bf 74}, 041708.

\bibitem[Tiribocchi \emph{et al.} (2015)] {Tiribocchi15} 
\textsc{Tiribocchi A., Wittkowski R., Marenduzzo D. \& Cates M. E.} 2015
Active Model H: Scalar active matter in a momentum-conserving fluid,
\emph{Phys. Rev. Lett.} {\bf 115}, 188302.

\bibitem[Tiribocchi \emph{et al.} (2016)] {Tiribocchi16} 
\textsc{Tiribocchi A., Da Re M., Marenduzzo D. \& Orlandini E.} 2016
Shear dynamics of an inverted nematic emulsion,
\emph{Soft Matter} {\bf 12}, 8195--8213.

\bibitem[Tjhung \emph{et al.} (2012)] {Tjhung12} 
\textsc{Tjhung E., Marenduzzo D. \& Cates M. E.} 2012
Spontaneous symmetry breaking in active droplets provides a generic route to motility,
\emph{Proc. Natl. Acad. Sci. USA.} {\bf 109}, 12381--12386.

\bibitem[Tjhung \emph{et al.} (2015)] {Tjhung15} 
\textsc{Tjhung E., Tiribocchi A., Marenduzzo D. \& Cates M. E.} 2015
A minimal physical model captures the shapes of crawling cells,
\emph{Nat. Comms.} {\bf 6}, 5420.

\bibitem[Verkhovsky \emph{et al.} (1999)]{Verkhovsky99}  
\textsc{Verkhovsky A.B., Svitkina T.M. \& Borisy. G.G. } 1999
Self-polarization and directional motility of cytoplasm,
\emph{Curr. Biol.} {\bf 9}, 11--20.

\bibitem[Wagner \& Cates (2001)]{Wagner01}  
\textsc{Wagner A. \& Cates M. E.} 2001
Phase ordering of two-dimensional symmetric binary fluids: A droplet scaling state, 
\emph{Europhys. Lett.} {\bf 56}, 556--562.

\bibitem[Weaire \& Hutzler (1999)] {Weaire99} 
\textsc{Weaire D. \& Hutzler S.} 1999 
The Physics of Foams, 
\emph{Oxford University Press.}

\bibitem[Webster \& Cates (1998)] {Webster98}
\textsc{Webster A. J. \& Cates M. E.} 1998 
Stabilization of emulsions by trapped species,
\emph{Langmuir} {\bf 14}, 2068--2079 .

\bibitem[Wittkowski \emph{et al.} (2014)] {Wittkowski14} 
\textsc{Wittkowski R., Tiribocchi A., Stenhammar J., Allen R. J., Marenduzzo D. \& Cates M. E.} 2014
Scalar $\phi^4$ field theory for active-particle phase separation,
\emph{Nature Communications} {\bf 5}, 4351.

\bibitem[Wolff \emph{et al.} (2012)] {Wolff12} 
\textsc{Wolff K., Marenduzzo D. \& Cates M. E.} 2012
Cytoplasmic streaming in plant cells: the role of wall slip,
\emph{J. Roy. Soc. Int.} {\bf 71}, 1398.

\bibitem[Yam \emph{et al.} (2007)] {Yam07} 
\textsc{Yam P. T., Wilson C. A., Ji L. Hebert B., Barnhart E. L., Dye N. A., Wiseman P. W., Danuser G. \& Theriot J. A.} 2007
Actin--myosin network reorganization breaks symmetry at the cell rear to spontaneously initiate polarized cell motility,
\emph{J. Cell Biol.} {\bf 178}, 1207-1221.

\bibitem[Ziebert \& Aranson (2013)] {Ziebert13} 
\textsc{Ziebert F. \& Aranson I. S.} 2013
Effects of adhesion dynamics and substrate compliance on the shape and motility of crawling cells,
\emph{PLoS One} {\bf 8}, e64511.





\end{thebibliography}
\end{document}